%% file: main.tex
\documentclass{article} %
\usepackage{iclr2027_conference_named,times}
\input{math_commands.tex}

\usepackage{hyperref}
\hypersetup{colorlinks=true, linkcolor=blue, citecolor=blue, urlcolor=blue}
\usepackage{url}
\usepackage{booktabs}
\usepackage{graphicx}
\usepackage{float}
\usepackage{subcaption}
\usepackage{multirow}
\usepackage{soul}
\usepackage{longtable}

\usepackage{pmboxdraw}
\usepackage{wrapfig}
\usepackage{placeins}

\newcommand{\lowtilde}{\raisebox{-0.7ex}{\textasciitilde}}

\newcommand{\brainpfn}{BrainPFN}
\newcommand{\Logauc}{Log-AULC}
\newcommand{\logauc}{log-AULC}
\newcommand{\std}[1]{\scriptsize{$\pm$#1}}

\title{BrainPFN: Amortized Brain-Behaviour Prediction via Prior-Data Fitted Networks}

\author{Carlo Alberto Barbano \& Demian Wassermann \\ %
MIND, Université Paris-Saclay, Inria, CEA, Palaiseau, 91120, France \\
\texttt{\{carlo.barbano,demian.wassermann\}@inria.fr} \\
}

\begin{document}

\maketitle

\begin{abstract}
The standard approach to predicting behavioral phenotypes from functional MRI is represented by Kernel Ridge Regression (KRR) applied on functional connectivity (FC) matrices. However, this approach is limited: separate models must be fit and tuned for every score and cohort, they do not transfer across datasets, and accuracy degrades sharply when only a handful of subjects are available, as is typical of clinical or cognitive studies. Despite an increasing interest in developing brain foundation models, none has consistently outperformed KRR on behaviour prediction; most reported gains come from easier targets such as age, sex, or diagnosis rather than continuous behavioural scores. We introduce BrainPFN, a Prior Data Fitted Network for brain behaviour regression. BrainPFN is meta-trained on a domain-specific prior built from \lowtilde7k real FC matrices, from which we sample one billion synthetic training samples. At inference, a labelled context is simply passed through the network in a single forward pass, with no per-score or per-cohort refitting required. Across four independent cohorts, BrainPFN beats KRR on small sample prediction, and leads recent foundation models pre-trained at four to nine times larger scale. Our meta-training pool, aggregated from open repositories, is a fraction of the size of the corpora behind competing models, most of which sit behind restrictive data use agreements such as UK Biobank. We release this pool publicly alongside code and checkpoints, showing that a modest, openly available pool can match or beat models trained at Biobank scale. 
\end{abstract}

\input{1_introduction}

\input{2_related}

\input{3_method_reframed}

\input{4_setup}

\input{5_results_reframed}

\input{6_conclusions}

\input{statements}

\bibliography{references}
\bibliographystyle{iclr2027_conference}

\clearpage
\newpage
\input{appendix}

\end{document}

%% file: math_commands.tex
\usepackage{amsmath,amsfonts,bm}

\def\eqref#1{equation~\ref{#1}}

\def\1{\bm{1}}

\DeclareMathAlphabet{\mathsfit}{\encodingdefault}{\sfdefault}{m}{sl}
\SetMathAlphabet{\mathsfit}{bold}{\encodingdefault}{\sfdefault}{bx}{n}

%% file: 1_introduction.tex
\section{Introduction}

\begin{figure}[t]
    \centering
    \includegraphics[width=\linewidth]{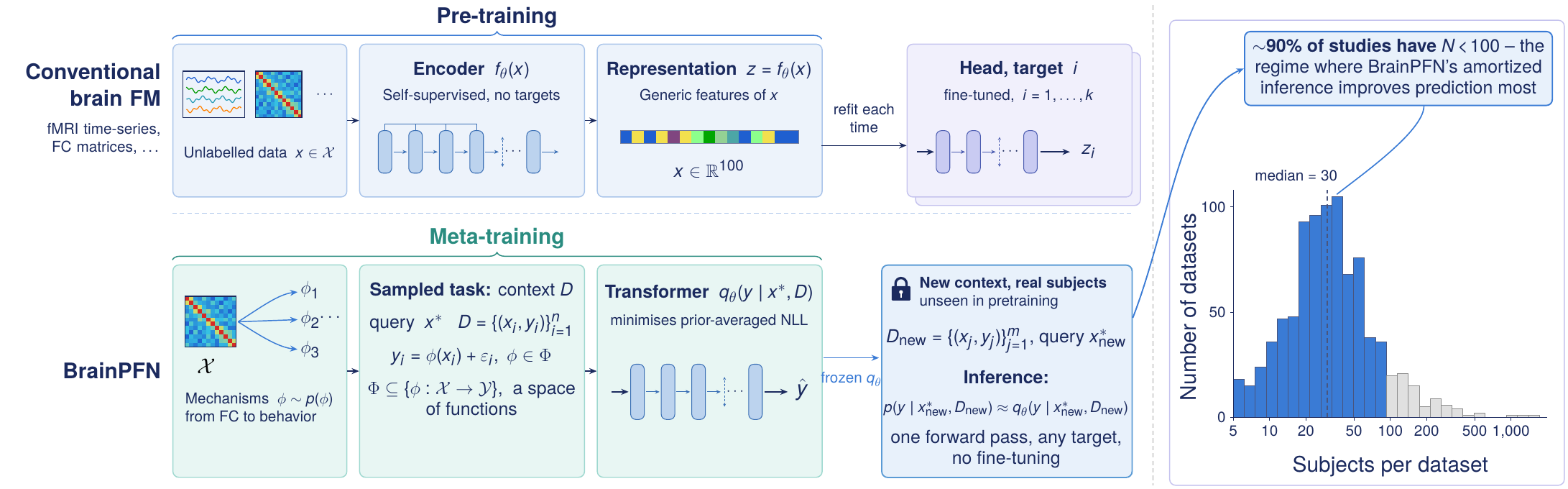}
    \caption{\textbf{BrainPFN overview:}  BrainPFN replaces the per-target, per-cohort refitting of conventional brain foundation models with a single frozen forward pass. Conventional brain FMs learn a generic representation via self-supervision, then require fitting a separate prediction head for each target. BrainPFN instead meta-trains a transformer $q_\theta$ entirely on synthetic tasks generated from a prior $\phi \in \Phi$ over brain-behavior to directly approximate the posterior predictive. At deployment, the same frozen $q_\theta$ is given a new context from real, previously unseen subjects and predicts any target in one forward pass, with no retraining. The right panel reports the distribution of cohort size across \lowtilde1000 neuroimaging datasets on OpenNeuro with fMRI data for at least 5 subjects: \lowtilde90\% have fewer than 100 subjects (highlighted in blue), precisely the small-cohort regime where BrainPFN's amortized inference improves prediction most.}
    \label{fig:dataset-histogram}
\end{figure}

Predicting behavioral and cognitive phenotypes from functional MRI (fMRI) is a central goal of cognitive neuroimaging, with applications ranging from mapping individual differences in cognition to identifying biomarkers of psychiatric and neurological disease. The dominant approach fits a Kernel Ridge Regression (KRR) model on functional connectivity (FC) matrices for each target score and cohort~\citep{ooi2022comparison}. This baseline is simple and hard to beat~\citep{he2020deep}, but comes with a structural cost: a new model must be fit and tuned from scratch for every phenotype and dataset, and its accuracy degrades sharply when only a handful of labeled subjects are available~\citep{ooi2025longer}. This is precisely the regime of most clinical and cognitive studies~\citep{szucs2020sample}, where cohorts number in the tens or low hundreds rather than thousands, as shown in Fig.~\ref{fig:dataset-histogram}.

A growing line of work has sought to replace this per-dataset fitting with brain foundation models~\citep{dong2024brain, thomas2022self,gijsen2026brain,caro2024brainlm,yang2024brainmass,dong2026brain}, pre-trained once on large fMRI corpora and then adapted, in principle cheaply, to new tasks. Despite this effort, no such model consistently outperforms a simple KRR baseline in continuous behavioral prediction~\citep{marraffini2026variance}: most reported gains are concentrated on coarser targets such as age, sex, or categorical diagnosis, and evaluations on genuine behavioral scores in the small-sample regime remain scarce. The result is a puzzling gap between the promise of large-scale pre-training and its demonstrated benefit for the settings where a transferable prior should matter most: small, heterogeneous, hard-to-pool cohorts.

We breach the fMRI-cognitive prediction gap with \textit{Prior-data Fitted Networks}~\citep[PFNs,][]{muller2022transformers}, a model class trained to perform Bayesian inference in a single forward pass. PFNs are meta-trained on synthetic datasets drawn from a task-specific prior, rather than on a fixed pre-training corpus. This design replaces per-dataset optimization with amortized inference: at test time, a labeled context of any size is simply passed through the network alongside query samples, with no gradient steps, refitting, or hyperparameter search required.
We instantiate this idea for brain-behavior prediction by constructing a domain-specific prior directly from a modest pool of \lowtilde7,000 real FC matrices, and meta-training on one billion synthetic regression problems sampled from it. Notably, our meta-training step is fully self-supervised and does not require any real behavioral label.

Our contribution is three-fold:
\begin{itemize}
    \item We introduce \emph{\brainpfn{}}, a Prior-data Fitted Network for brain-behavior prediction that amortizes inference across scores and cohorts in a single forward pass. This approach excels in the small-sample regime where existing approaches struggle most.
    \item We show that \brainpfn{} consistently outperforms KRR and existing brain foundation models -- several of them pre-trained on corpora four to nine times larger -- in low-data settings, while remaining competitive with the strongest baselines as more labeled data becomes available.
    \item We publicly release the meta-training pool underlying \brainpfn{}, a fraction of the size of, and free of the restrictive data-use agreements attached to, corpora such as UK Biobank, together with code and checkpoints, demonstrating that a small, openly available pool can match or exceed models trained at UK~BioBank scale.
\end{itemize}

%% file: 2_related.tex
\section{Related work}
\label{sec:related}

\paragraph{Classical brain-behaviour prediction and the small-sample regime.} Connectome-based predictive modelling predicts individual scores from functional connectivity (FC) with per-score regularized regression \citep{finn2015functional,rosenberg2016neuromarker,shen2017using}, and systematic benchmarks find simple linear models remarkably competitive on such data~\citep{dadi2019benchmarking,he2020deep,schulz2020different}. Recent studies on fMRI foundation models show evidence that larger datasets can improve downstream prediction~\citep{wang2026omnifmri}, however most studies do not typically focus on prediction on small cohorts. This matters practically because most clinical and cognitive neuroimaging studies never reach the large-$N$ regime: highly-cited fMRI studies have a median sample size around a dozen participants \citep{szucs2020sample}. With standard approaches, this results in underpowered designs that inflate and destabilize reported effects \citep{button2013power}, whereas brain-wide association studies require samples in the thousands to reproduce reliably \citep{marek2022reproducible}: a scale most single-site and, especially, clinical acquisitions never reach, where limited sample size is repeatedly identified as the main bottleneck to single-subject prediction \citep{arbabshirani2017single}. %
A predictive model whose accuracy is concentrated at small $N$ would therefore target the regime where most neuroimaging and clinical cohorts actually sit. %

\paragraph{Foundation models for brain imaging.} Foundation models pre-trained on large fMRI corpora include BrainLM \citep{caro2024brainlm}, BrainMass \citep{yang2024brainmass} (masked modelling on FC matrices across 30 datasets, \lowtilde70k samples), Brain-JEPA \citep{dong2024brain} and Brain-Semantoks \citep{gijsen2026brain} on ROI time series. These are typically deployed as frozen encoders with a per-target linear probe; Sec.~\ref{sec:results} compares \brainpfn{} against three of them under that protocol. All four
pre-train, in whole or in part, on UK Biobank \citep{bycroft2018uk}: Brain-JEPA on \lowtilde32k and Brain-Semantoks on \lowtilde39k UKB scans, BrainLM on a further \lowtilde61k, and BrainMass's 30-dataset pool includes it alongside HCP and Cam-CAN. \brainpfn{}'s meta-training pool aggregates \lowtilde7.7k FC matrices from public OpenNeuro releases \citep{markiewicz2021openneuro} with no access agreement, $4$--$9\times$ smaller than the corpora above, and we release it publicly alongside the model and code.

\paragraph{Prior-data Fitted Networks and In-Context Learning.} 
PFNs~\citep{muller2022transformers} have been popularized in the tabular domain, with TabPFN~\citep{hollmann2022tabpfn,hollmann2025accurate} and TabICL~\citep{qu2025tabicl}, with strong results in the small-data regime. %
Beyond tabular data, PFNs have been applied to time-series forecasting~\citep{dooley2023forecastpfn}, Bayesian optimization~\citep{muller2023pfns4bo}, learning-curve extrapolation~\citep{adriaensen2023efficient}, causal effect estimation~\citep{robertson2026pfn, balazadeh2026causalpfn} and genomics~\citep{ubbens2025gpfn}, and in-context learners have also reached neuroimaging~\citep{czolbe2023neuralizer, yu2025metalearning}.
In general, PFNs rely on fully synthetic priors: both inputs and targets come from randomly sampled generative processes such as structural causal models. Our prior is instead grounded in real data. We do not sample FC matrices; we draw them from a meta-training pool of real connectomes.

%% file: 3_method_reframed.tex
\section{Method: BrainPFN}

\subsection{Amortized inference for Brain-Behavior Prediction}
\label{sec:background}

We represent each subject by their resting-state FC
matrix parcellated with the Schaefer-400 atlas~\citep{schaefer2018local}. FC matrices are symmetric positive-definite (SPD) matrices, thus we map them to log-Cholesky coordinates \citep{lin2019riemannian}, which respects the curved geometry
of the SPD manifold. The log-Cholesky representation is then reduced by PCA to
$d = 100$ components, yielding feature vectors $x \in \mathbb{R}^d$. The target
$y \in \mathbb{R}$ is a continuous behavioral score. Given a labelled
training set $D_\text{train} = (X_\text{train}, y_\text{train})$, where $X_{train} \in \mathbb{R}^{n \times d}$ is composed of $n$ FC feature vectors, %
and a new test subject $x$, the goal is to predict $y$ without
refitting any model parameters, i.e.\ purely in-context. The training set $D_\text{train}$ is also called \textit{context}.

Suppose the data in $D_\text{train}$ are generated by some unknown mechanism $\phi \colon \mathcal{X} \to \mathbb{R}$, drawn from a prior $\phi \in \Phi \sim p(\phi)$ that encodes everything we know about how FC relates to behaviour. 
Each mechanism induces a conditional $p(y \mid x,\phi)$. Given $D_\text{train}$, we seek the predictor $q(y \mid x)$ that minimizes the Bayes risk under log loss: %
\begin{equation}
    \mathcal{R}(q) = \mathbb{E}_{\phi \sim p(\phi \mid D_{\text{train}})}\, \mathbb{E}_{y \sim p(y \mid x, \phi)} \bigl[-\log q(y \mid x)\bigr] = -\!\int p(y \mid x, D_{\text{train}}) \log q(y \mid x)\, dy\, .
\end{equation}
This is cross-entropy, so by Gibbs' inequality it is minimised by the posterior predictive:
\begin{equation}
    \label{eq:bayesian-inference}
    p(y \mid x, D_\text{train}) = \int_{\Phi} p(y \mid x,\phi)\, p(\phi\mid  D_\text{train})\, d\phi\,.
\end{equation}
A Bayes-optimal predictor therefore integrates over \textit{every}
possible mechanism weighted by how well it explains the context.
This integral is generally intractable, but it can be approximated with a Prior-Fitted Network~\citep[PFN,][]{muller2022transformers}, a Transformer trained to approximate:
\begin{equation}
\label{eq:objective-function}
q^* := \arg\min_{q}\; \mathbb{E}_{\phi \sim p(\phi)}\, \mathbb{E}_{D_\text{train},\,x,\,y \,\sim\, p(\cdot \mid \phi)} \bigl[-\log q(y \mid x, D_\text{train})\bigr] \;\underset{\text{PFN}}{=}\; \int_{\Phi} p(y \mid x, \phi)\, p(\phi \mid D_\text{train})\, d\phi \,,
\end{equation}

Approximating Eq.~\ref{eq:objective-function} is called \textit{amortized inference}. In practice, at each training step a mechanism $\phi$ is sampled, and a synthetic dataset ${D} = \{(x_i, y_i)\}_{i=1}^n$ is drawn from it. %
Specifically, once we sampled $\phi$, we produce $y_i=\phi(x_i)$ where the feature vectors $x_i$ come from a set of real FC matrices, i.e. the \textit{meta-training pool}.
${D}$ is then split into a context ${D}_{\text{train}}$ and a held-out test point $(x_{\text{test}}, y_{\text{test}})$. %
Taking this into account the network is trained by minimizing $L_\phi(\theta)$:
\begin{equation}
    L_\phi(\theta) = \mathbb{E}_{{D} \sim p({D} \mid \phi)} \left[ -\log q_\theta(y_{\text{test}} \mid x_{\text{test}}, {D}_{\text{train}}) \right] \, .
    \label{eq:pfnloss}
\end{equation}
Because a different $\phi$ is sampled at each step, a sufficiently expressive network trained in this way approximates $q_\theta$, a parametric minimizer for $q^*$ of Eq.~\ref{eq:objective-function}, as $L(\theta) = \mathbb E_{\phi\sim p(\phi)}L_\phi(\theta)$
without ever having to compute the posterior predictive explicitly~\citep{muller2022transformers}.

The key choice in this framework is therefore how to build the prior $p(\phi)$, which in our case must %
capture the structure of FC-behaviour associations.

\subsection{Brain-Behavior Association Prior}
\label{sec:brainpfn}

Regularised kernel methods, and kernel ridge regression (KRR) in
particular, remain among the strongest and most reproducible
predictors of behavior from FC data, matching or
exceeding deep learning architectures on this exact task
\citep{he2020deep,shen2017using,dadi2019benchmarking,ooi2022comparison}.
In KRR, predictions take the form $\hat{y} = K(x, X_\text{train})\,\alpha$, where $K$ is a kernel matrix computed from FC vectors, $\alpha = (K(X_\text{train}, X_\text{train}) + \lambda I)^{-1} y_\text{train}$ are the dual coefficients, and $\lambda$ is a regularisation parameter.
BrainPFN is accordingly built to approximate the posterior predictive 
distribution over the space of KRR functions mapping FC to behavior, 
with prior $p(\phi)$ defined over mechanisms $\phi = (K, \alpha, \sigma^2_\epsilon)$ 
parametrising the kernel choice, the dual coefficients $\alpha$, %
and the noise %
$\epsilon \sim \mathcal{N}(0, \sigma^2_{\epsilon})$.
For each synthetic task, we randomly select a set of FC vectors $X_\text{train}$ from our meta-training pool and generate synthetic continuous targets as:
\begin{equation}
    y = K(X_\text{train}, X_\text{train})\,\alpha + \epsilon, \qquad (K, \alpha, \sigma^2_{\epsilon}) = \phi \sim p(\phi),
\end{equation}
where the kernel $K$ is sampled from the linear, RBF, and Laplacian families, together with the kernel coefficients. %
A held-out test set $X_\text{test}$ is drawn from the same pool, the targets $y_\text{test} = K(X_\text{test}, X_\text{train})\,\alpha + \epsilon$ are generated from the same mechanism $\phi$, and the network $q_\theta$ 
is trained to predict $y_\text{test}$ from $(x_\text{test}, D_\text{train})$ by minimising the loss in Eq.~\ref{eq:pfnloss}.
We draw  $X_\text{train}$ and $X_\text{test}$ from real FC matrices (the meta-training pool) rather than simulated ones, as the statistical structure of functional connectivity is not trivial to capture with generative models. It is worth noting that for a single kernel with fixed hyperparameters the posterior predictive of Eq.~\ref{eq:objective-function} is available in closed form (App.~\ref{sec:appendix-gp}). In our prior, however, the kernel bandwidth is itself random so the exact posterior predictive is a mixture of GP posteriors weighted by their marginal likelihoods, with no closed form. The PFN learns to perform this averaging in-context.

\paragraph{Sampling dual coefficients.}
In the dual formulation of KRR, predictions take the form $f(x) = K(x, X_\text{train})\,\alpha$, and the regularisation term in the KRR objective penalises the squared reproducing kernel Hilbert space (RKHS) norm of $f$:
$\|f\|^2_\mathcal{H} = \alpha^\top K_\text{train}\,\alpha = \sum_j {\hat{f}_j^2}/{\lambda_j}$
where $\lambda_j$ and $v_j$ are the eigenvalues and eigenvectors of $K_\text{train}$, and $\hat{f}_j = \lambda_j\,v_j^\top\alpha$ is the prediction component along $v_j$. The RKHS norm thus penalises directions with small eigenvalues: the same prediction amplitude $\hat{f}_j$ incurs a cost $1/\lambda_j$ times larger in low-variance directions. To sample dual coefficients that respect this geometry we draw them from:
\begin{equation}
    \alpha \sim \mathcal{N}\!\left(0,\, \sigma^2_\alpha K_\text{train}^{-1}\right),
\end{equation}
which concentrates the sampled functions along the high-eigenvalue (smooth) directions of the kernel. %
This ties their effective complexity to the spectrum of the FC manifold rather than producing arbitrarily rough targets. We show this outperforms an isotropic prior in Sec.~\ref{sec:ablation-study}.

\textbf{Task difficulty and calibration.}
Brain-behavior associations vary widely in strength across cognitive domains and datasets. We model behavior as $y = s + \epsilon$ (where $s$ is the KRR prediction), meaning that some part is explained by the actual FC data and some part is aleatoric or epistemic noise. By varying the amount of noise $\epsilon$, we can control the strength of the correlation between FC and behavior $r = \mathrm{Corr}({s}, y)$, which represents the task difficulty. With $\epsilon \sim \mathcal{N}(0, \sigma^2)$ independent of ${s}$ and $\mathrm{Var}({s}) = 1$, a short calculation gives (full derivation in Sec.~\ref{appendix:effect-size-derivation}):
\begin{equation}
    \label{eq:effect-size-sigma}
    r = \mathrm{Corr}({s}, y)
    = \frac{\mathrm{Cov}({s},\, {s}+\epsilon)}{\sqrt{\mathrm{Var}({s})\,\mathrm{Var}(y)}}
    = \frac{1}{\sqrt{1 + \sigma^2}},
\end{equation}
so we can solve for $\sigma^2_\epsilon = (1 - r^2)/r^2$, and the observed target is:
\begin{equation}
    y = {s} + \epsilon, \qquad \epsilon \sim \mathcal{N}\!\left(0,\,\frac{1-r^{2}}{r^{2}}\right).
\end{equation}
To avoid circularity, the raw signal $s = K(X, X_{\text{train}})\,\alpha$ must be standardized to unit variance using subjects not involved in fitting it. 
Reusing the context or test block to estimate this variance would bias it differently across kernel families, so we draw a third, disjoint block of subjects for this purpose only.
We therefore draw a third block of subjects from the pool, disjoint from both the context and the test points, and use it only to estimate the scale of $s$. As a result, $r$ acts as a single, kernel-agnostic difficulty knob across the whole prior.

%% file: 4_setup.tex
\section{Experimental Setup}

\subsection{Datasets}
\label{sec:datasets}

\paragraph{Meta-training pool}
BrainPFN is meta-trained entirely on \lowtilde7.7k real resting-state FC matrices from \lowtilde4.9k unique subjects across 300 public OpenNeuro studies \citep{markiewicz2021openneuro} (full list in Sec.~\ref{appendix:dataset-list}). %
No dataset in the pool requires a data-use agreement.
Every subject is processed with an identical fmriprep-based pipeline (full preprocessing parameters in App.~\ref{sec:appendix-setup-details}) and parcellated with the Schaefer-400 atlas \citep{schaefer2018local}. We collect no behavioral scores for this pool: BrainPFN is meta-trained purely on synthetic regression tasks built from real FC features.

\paragraph{Downstream datasets}
\label{sec:downstream-datasets}
We evaluate cross-dataset transfer of BrainPFN on four independent cohorts spanning different age ranges, acquisition sites, and scanning protocols, processed with the same FC pipeline as the meta-training pool: HCP Young Adult \citep[HCP-YA,][]{vanessen2013wu}, HCP-Aging \citep[HCP-A,][]{bookheimer2019lifespan}, Cam-CAN
\citep{shafto2014cambridge,taylor2017cambridge}, and AOMIC-ID1000
\citep{snoek2021amsterdam}, none of which overlaps with the meta-training pool (Tab.~\ref{tab:datasets}).
Target scores span cognition, personality, emotion, physical, and demographic (age, sex) measures, drawn from each dataset's standard behavioural battery.

\begin{table}[h]
\centering
\caption{\textbf{Evaluation datasets} used to benchmark BrainPFN generalization to external cohorts. \# scores is the number of target behavioral measures evaluated per cohort.}
\label{tab:datasets}
\begin{tabular}{lcccc}
\toprule
\textbf{Dataset} & \textbf{Train / Test} & \textbf{Age range} & \textbf{Acquisition} & \textbf{\# scores} \\
\midrule
{HCP-YA} & 1029 & \multirow{2}{*}{22--37 (healthy young adult)} & resting-state & {62} \\
$\,\,\rightarrow$ \textit{Task} & 1008 & & task-based & 2\\
HCP-Aging & 1198 & 36--90+ (healthy older adult) & resting-state & 35 \\
Cam-CAN & 597 & 18--88 (lifespan) & resting-state & 3 \\
AOMIC-ID1000 & 881 & 19--26 (general population) & movie-watching & 14 \\
\bottomrule
\end{tabular}
\end{table}

\subsection{Implementation Details}
\label{sec:implementation-details}

\paragraph{BrainPFN architecture and meta-training.}
BrainPFN uses the asymmetric-attention Transformer of TabPFN~\citep{hollmann2022tabpfn} with a single linear regression head in place of the original softmax classifier. We report two presets: \textit{BrainPFN-0.8M} (4 layers,
$d_{\text{model}}{=}128$) and
\textit{BrainPFN-4.7M} (6 layers, $d_{\text{model}}{=}256$); full architectural and optimization hyperparameters are given in App.~\ref{sec:appendix-setup-details}.
The model is trained for 10k steps on \lowtilde1B synthetic context/test $(x,y)$ pairs sampled online from the meta-training pool, using a Laplace kernel and a fixed task difficulty $r=0.5$ by default (App.~\ref{sec:appendix-setup-details}); Sec.~\ref{sec:ablation-study} ablates both choices.

\paragraph{Evaluation protocol.} For each dataset we perform 5-fold cross-validation, repeatedly sampling 20 contexts of size $N$ from each split's train fold; BrainPFN (only its PCA basis refit per context) and the Ridge/KRR baseline are evaluated on the same held-out test fold via the Pearson correlation $r$ between predictions and ground truth, at context sizes $N \in \{5, 10, 20, \dots\}$ up to the train-fold size.
We summarize each learning curve with the log-area under the learning curve, \textbf{log-AULC}, computed via the trapezoidal rule, so that a
higher value indicates a more data-efficient model across the whole
$N$ range rather than at one arbitrarily chosen size. We restrict analysis to \textit{real-signal} targets, those decodable with at least $r \geq 0.1$ by any of the models, to avoid diluting results with near-random-level targets. Full per-score results, significance-testing, and evaluation-grid details are given in App.~\ref{sec:appendix-perscore} and App.~\ref{sec:appendix-setup-details}.

%% file: 5_results_reframed.tex
\providecommand{\reviewnote}[1]{\textcolor{red}{\textbf{[REVIEW: #1]}}}

\section{Results}
\label{sec:results}

\begin{figure}[t]
    \centering
    \includegraphics[width=\linewidth]{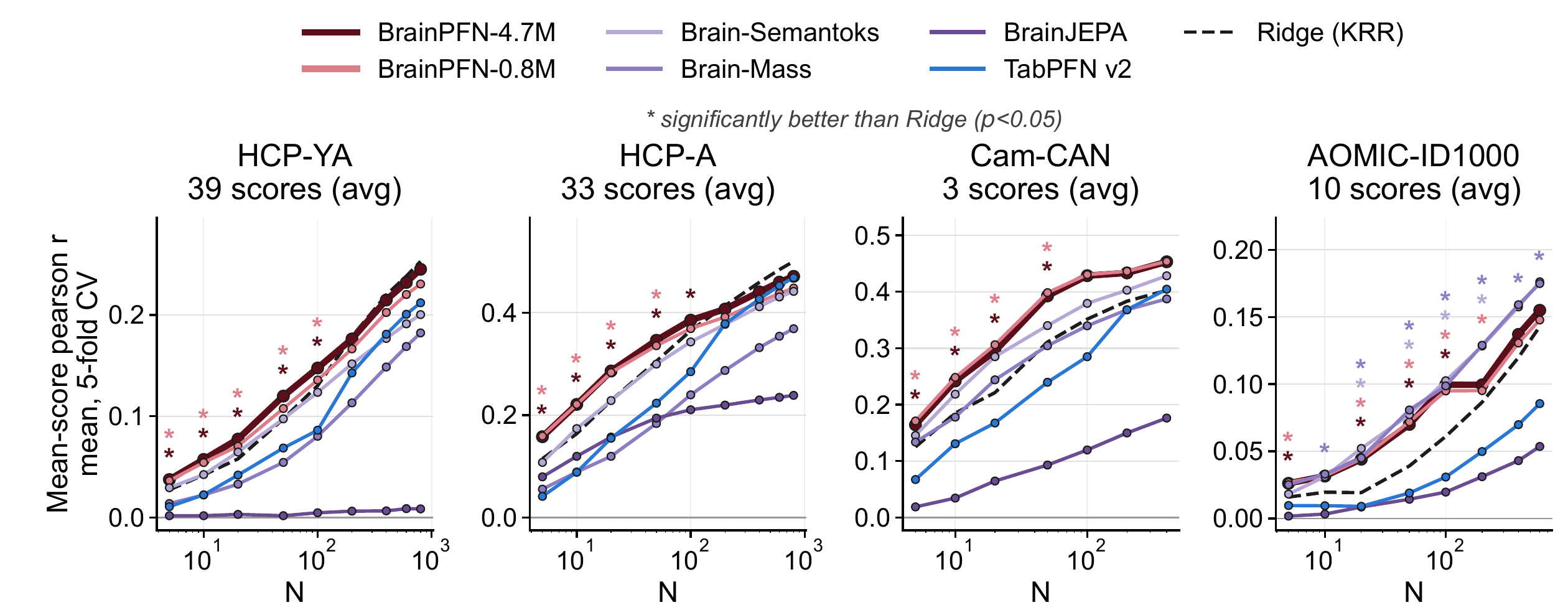}
    \includegraphics[trim=0 0 0 160px, clip,width=\linewidth]{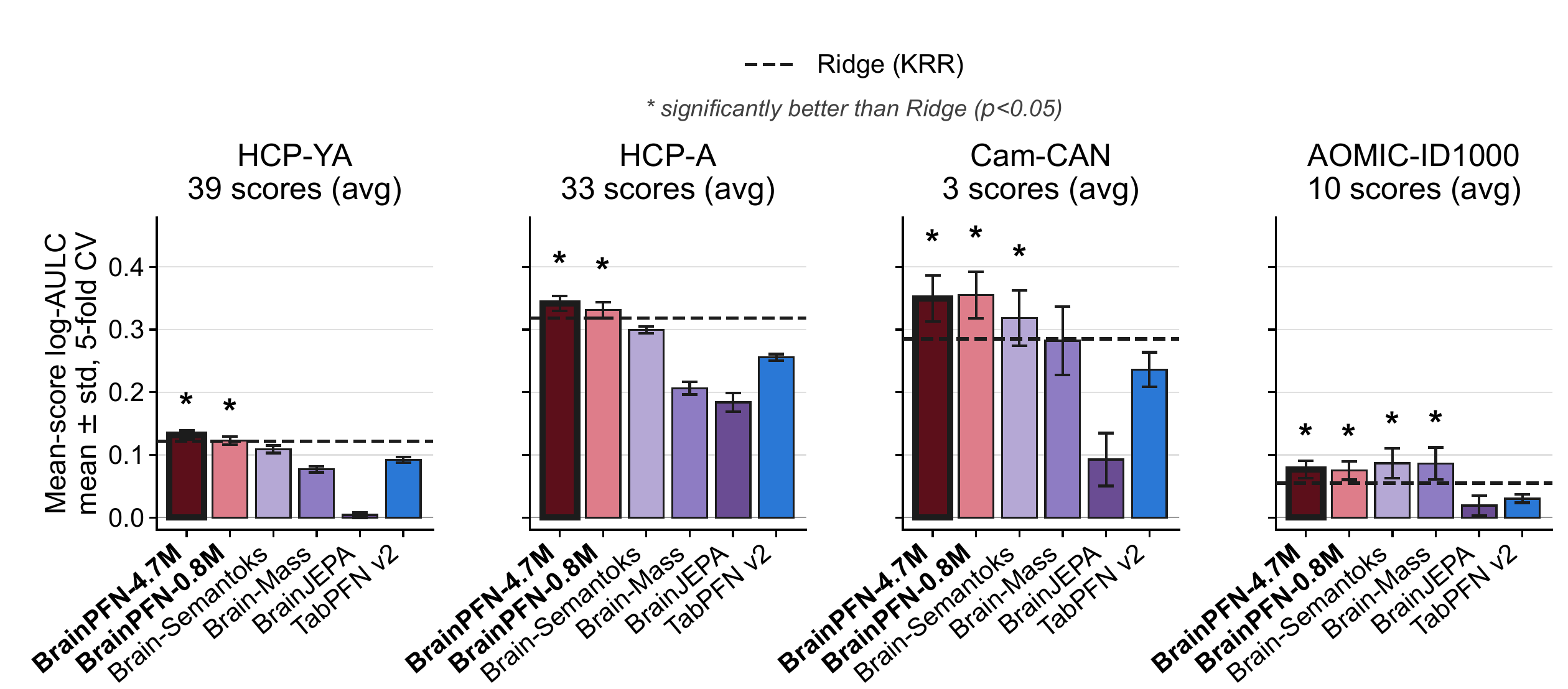}
    \caption{\textbf{Grand average across all behavioral tasks}. On a total of 85 tasks, BrainPFN beats Ridge and leads other foundation models on brain-behavior association. Top: mean Pearson $r$ vs. context size $N$; bottom: area under the learning curve (\logauc{}). BrainPFN is the only model consistently above Ridge (dashed; * $p < 0.05$) on all four cohorts.}
    \label{fig:headline-results}
\end{figure}

\subsection{BrainPFN leads the small N regime}
\label{sec:small-n}

Fig.~\ref{fig:headline-results} summarizes brain-behaviour prediction results across all four downstream cohorts on all 85 real-signal targets, plotting the learning curve of mean Pearson $r$ against context size $N$ (top) and its \logauc{} summary (bottom) against Ridge and four existing foundation models: Brain-JEPA, Brain-Mass, Brain-Semantoks, and the generic-prior TabPFN~v2. We restrict this comparison to \emph{real-signal} targets, those decodable with at least $r \geq 0.1$ by any of the compared models, so as not to dilute the average with near-random-level targets; results on the full, unfiltered set of targets are reported in App.~\ref{sec:appendix-full-results}. %
As a baseline we use linear-kernel KRR on log-Cholesky features, PCA with 800 components, and $\alpha=1$. This configuration was found to be the best during preliminary sweeps.

\begin{figure}[t]
    \centering
    \includegraphics[trim=0 0 0 110px, clip,width=\linewidth]{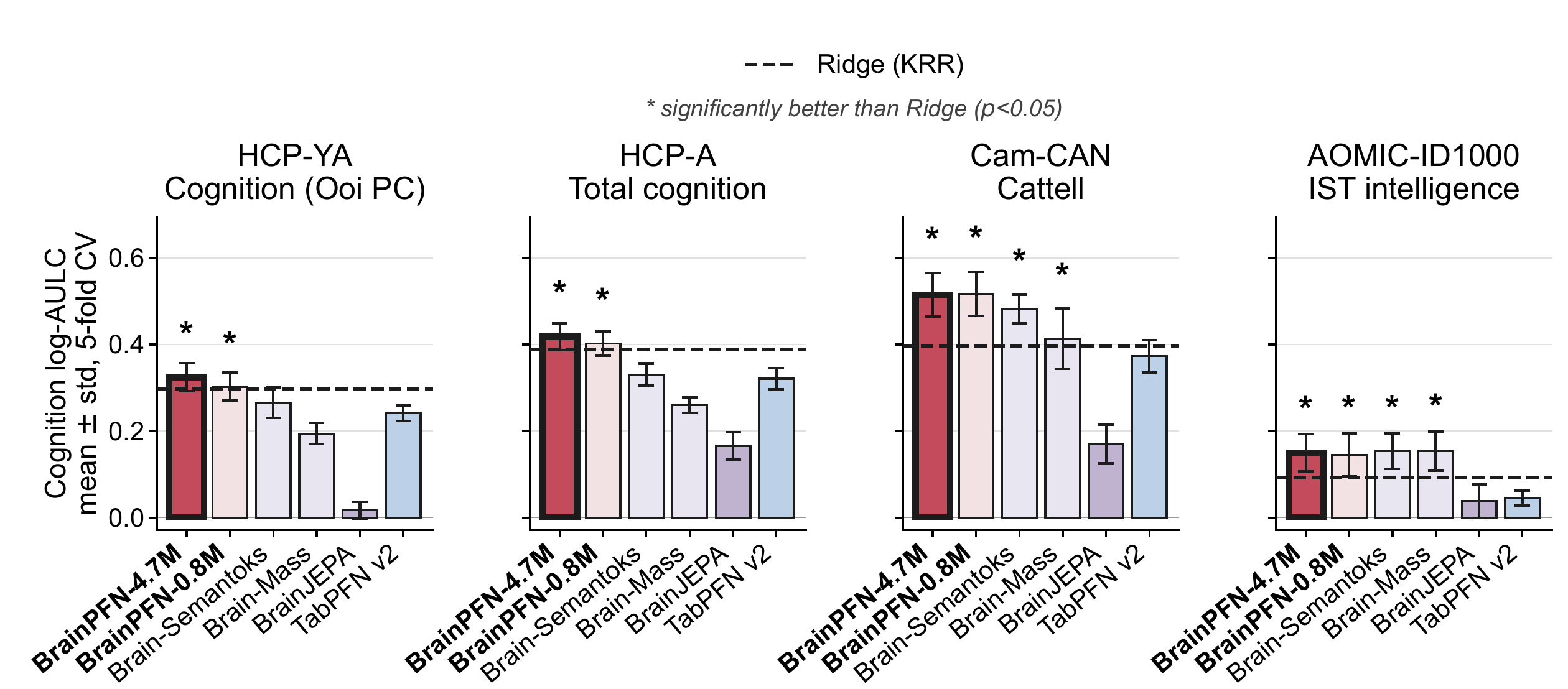}
    \caption{\textbf{BrainPFN beats ridge and leads other foundation models on cognition prediction.} \Logauc{} (mean $\pm$ std, 5-fold CV; dashed: Ridge; $^*$: $p<0.05$ vs.\ Ridge). BrainPFN is the only model significantly above Ridge on all four cohorts; other foundation models mostly fall below it.}
    \label{fig:headline-results-cognition}
\end{figure}

In the low-data regime this work targets, BrainPFN is the only model that beats Ridge consistently: both BrainPFN-4.7M and BrainPFN-0.8M are significantly above the Ridge baseline at $N \leq 20$ across all four cohorts (Fig.~\ref{fig:headline-results}, top, asterisks), and this advantage carries through to the \logauc{} summary, where BrainPFN is the only model to significantly outperform Ridge on HCP-YA, HCP-A, and Cam-CAN. None of the other foundation models matches this consistency. Brain-Semantoks is the strongest alternative, tracking Ridge closely and occasionally exceeding it, while Brain-Mass is competitive on Cam-CAN and AOMIC-ID1000 but noticeably weaker elsewhere; notably, Cam-CAN was part of Brain-Mass's pretraining corpus. %
Cognition is the behavioral domain where brain foundation models have so far
struggled most to improve on KRR~\citep{he2020deep}, so we examine it separately in Fig.~\ref{fig:headline-results-cognition}, using one general-cognition score per cohort: a cognition component in the style of~\cite{ooi2022comparison} on HCP-YA (more details in App.~\ref{sec:appendix-ooi}), the NIH Toolbox total cognition composite on HCP-A, the Cattell fluid-intelligence score on Cam-CAN, and the IST intelligence score on AOMIC-ID1000. Both BrainPFN presets are the only models whose \logauc{} is significantly above Ridge on all four cohorts. On HCP-YA and HCP-A, every competing foundation model falls below the Ridge baseline, with Brain-JEPA close to chance on HCP-YA. %

\begin{figure}[h]
 \centering
 \includegraphics[width=\linewidth]{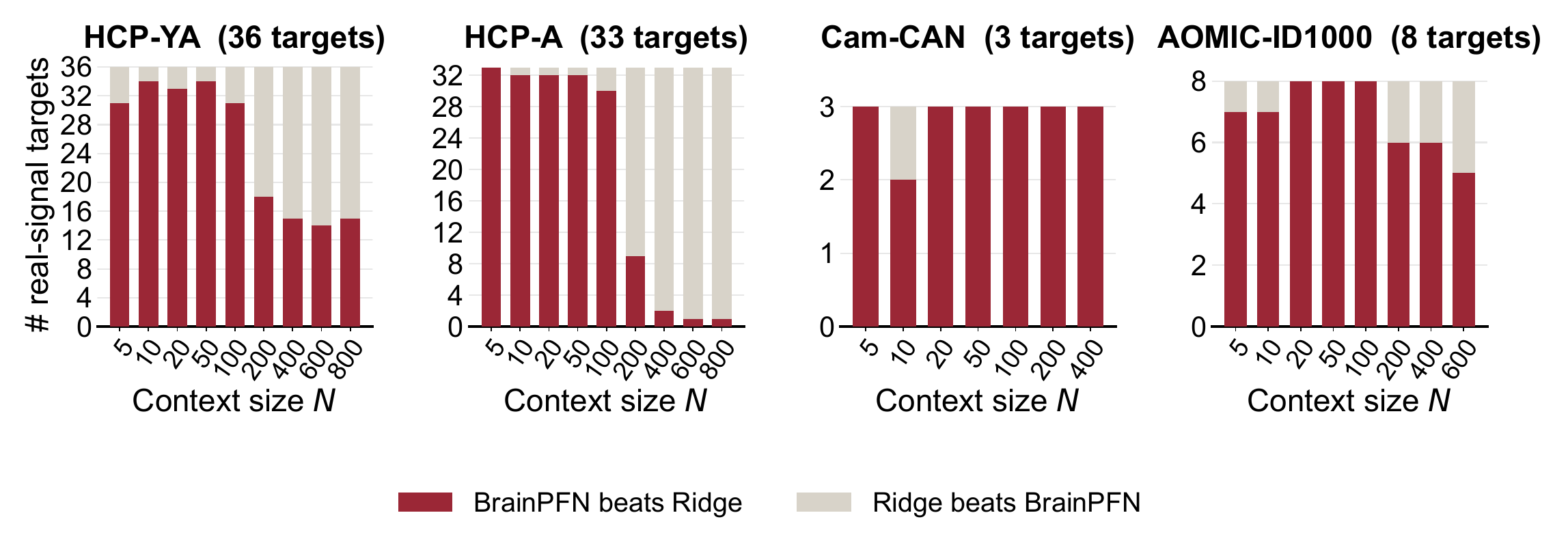}
 \caption{\textbf{Win-rate of BrainPFN versus Ridge} at each context size $N$. Bar height is the number of targets tested. BrainPFN's small-$N$ advantage is not carried by a few favorable targets: it wins on nearly every target on small contexts. 
 }
 \label{fig:winrate}
\end{figure}

The curves and \logauc{} summaries in Fig.~\ref{fig:headline-results} average over 85 real-signal targets in total. %
Hence, we check whether BrainPFN's average reflects a broad advantage or is carried by a few favourable scores. Fig.~\ref{fig:winrate} reports, per dataset and context size $N$, how many targets BrainPFN predicts more accurately than Ridge: BrainPFN wins on nearly every target on all four datasets up to $N{=}100$, the win rate then drops between $N{=}100$ and $N{=}200$ on HCP-YA and HCP-A specifically, and it never drops on Cam-CAN or AOMIC-ID1000. %
Cam-CAN is the one exception to the pattern above: BrainPFN's lead over Ridge does not shrink to zero as $N$ grows, and it is the one cohort where BrainPFN also stays ahead of the strongest competing foundation model at full context. %

BrainPFN's advantage is largest in the small-$N$ regime it was designed for. As $N$ grows, Ridge gradually closes the gap and overtakes BrainPFN between $N \approx 200$, several times the median OpenNeuro cohort size (Fig.~\ref{fig:dataset-histogram}), most visibly on HCP-A. On Cam-CAN and AOMIC-ID1000, no target crosses over within the available sample size. %
The same holds for every other foundation model we tested, none of which keeps a consistent advantage over Ridge at large $N$. BrainPFN remains competitive in this regime, while being the only model that consistently beats Ridge where data are scarce.

\subsection{Existing Brain Foundation Models fall short}
\label{sec:bfm-fall-short}

\paragraph{Impact of the domain-specific prior.} BrainPFN and TabPFN are the same kind of model: an in-context learner built on the same Transformer backbone. They differ in what each one is trained to predict: BrainPFN's synthetic KRR-over-real-FC prior versus TabPFN's generic, domain-agnostic prior. %
This leaves the prior itself as the source of BrainPFN's advantage. This evidence is further supported by the RKHS sampling ablation in Sec.~\ref{sec:ablation-study}, which removes the RKHS structure from the prior over dual coefficients and produces the largest degradation of any ablation tested.
Furthermore, simply training on domain-specific data is not by itself sufficient for good few-shot behavioral prediction: Brain-JEPA is the weakest model in every cohort despite being pretrained on fMRI data from more than 40k subjects, under-performing even the generic-prior TabPFN, which has seen no brain data at all. Although to a lesser extent, similar observations can also be made for the other foundation models. We read this as supporting context for the prior argument above: self-supervised fMRI pretraining and BrainPFN's explicit synthetic-function prior are different mechanisms, and Brain-JEPA's weak showing only demonstrates that the former is not sufficient on its own.

\paragraph{Model capacity, saturation, and transferability.} We find that model size is not predictive of performance: even though BrainPFN-4.7M gives a small, consistent edge over BrainPFN-0.8M, both models have far fewer parameters than the other foundation models. 
A compact model trained on a limited real-data footprint, augmented into \lowtilde1B synthetic training samples, is enough to learn a prior that outperforms models with substantially larger parameter counts (BrainJEPA has \lowtilde86M parameters, Brain-Semantoks \lowtilde62M, and BrainMass \lowtilde10M). %
We speculate that this gap could also reflect these larger models' pretraining corpora not scaling with their parameter count.
Furthermore, behavioural scores are defined differently across cohorts -- different instruments and scales, and sometimes different underlying constructs even when the name is similar (Sec.~\ref{sec:downstream-datasets}), so a per-score model like Ridge must be refit from scratch for every phenotype and every cohort, with nothing carried over from one dataset to the next. BrainPFN, in contrast, is used as-is, without retraining or refitting, on all four downstream cohorts and every target reported in this section.

\subsection{Generalization beyond resting-state FC}
\label{sec:generalization}

Beyond the results on naturalistic (movie-watching) fMRI from AOMIC-ID1000 (Figs.~\ref{fig:headline-results} and \ref{fig:headline-results-cognition}), where BrainPFN beats Ridge despite being pre-trained only on resting-state data, here we investigate whether this advantage also carries over to task-based fMRI.
Using the HCP-YA dataset, we test whether BrainPFN's small-context advantage extends beyond resting-state functional connectivity by predicting in-scanner working-memory (WM) task performance from FC estimated on the task fMRI run itself, compared against FC estimated from the resting-state run of the same subjects. Tab.~\ref{tab:wm_fc} reports 2-back and 0-back accuracy predicted from both FC sources. The small-context pattern from Sec.~\ref{sec:small-n} reproduces on both FC sources: BrainPFN beats Ridge on small contexts, and Ridge overtakes at larger $N$. Task FC is also simply a better predictor of task performance than resting FC, for either model, at every context size tested (Table~\ref{tab:wm_fc}). This is consistent with neuroscientific evidence in favor of task-based fMRI for behavioral prediction~\citep{zhao2023task}. %

\begin{table}[h]
 \centering
 \caption{\textbf{BrainPFN's small-context advantage carries over to task fMRI.} The table reports working-memory accuracy (2-back, 0-back) predicted from resting-state or task FC on the same HCP-YA subjects, as mean Pearson r ± std over 5 test folds. BrainPFN beats Ridge up to N = 100 on resting-state FC and up to N = 200 on task FC.}
 \label{tab:wm_fc}
 \small
 \setlength{\tabcolsep}{4pt}
 \resizebox{\textwidth}{!}{%
 \begin{tabular}{lcccccccc}
 \toprule
 & \multicolumn{4}{c}{\textbf{\texttt{WM\_Task\_2bk\_Acc} (2-back)}} & \multicolumn{4}{c}{\textbf{\texttt{WM\_Task\_0bk\_Acc} (0-back)}} \\
 \cmidrule(lr){2-5} \cmidrule(lr){6-9}
 & \multicolumn{2}{c}{\textbf{Resting-state FC}} & \multicolumn{2}{c}{\textbf{Task FC}} & \multicolumn{2}{c}{\textbf{Resting-state FC}} & \multicolumn{2}{c}{\textbf{Task FC}} \\
 \cmidrule(lr){2-3} \cmidrule(lr){4-5} \cmidrule(lr){6-7} \cmidrule(lr){8-9}
 \textbf{$N$} & \textbf{BrainPFN} & \textbf{Ridge} & \textbf{BrainPFN} & \textbf{Ridge} & \textbf{BrainPFN} & \textbf{Ridge} & \textbf{BrainPFN} & \textbf{Ridge} \\
 \midrule
 5   & \textbf{0.047\std0.009} & 0.029\std0.015 & \textbf{0.104\std0.039} & 0.081\std0.042 & \textbf{0.037\std0.034} & 0.023\std0.033 & \textbf{0.113\std0.064} & 0.083\std0.052 \\
 10  & \textbf{0.091\std0.017} & 0.055\std0.018 & \textbf{0.196\std0.043} & 0.140\std0.027 & \textbf{0.050\std0.028} & 0.025\std0.022 & \textbf{0.171\std0.029} & 0.103\std0.033 \\
 20  & \textbf{0.123\std0.031} & 0.091\std0.024 & \textbf{0.230\std0.019} & 0.162\std0.014 & \textbf{0.084\std0.028} & 0.061\std0.034 & \textbf{0.229\std0.037} & 0.148\std0.043 \\
 50  & \textbf{0.167\std0.038} & 0.131\std0.026 & \textbf{0.312\std0.077} & 0.241\std0.066 & \textbf{0.118\std0.041} & 0.096\std0.038 & \textbf{0.294\std0.061} & 0.223\std0.059 \\
 100 & \textbf{0.222\std0.052} & 0.200\std0.047 & \textbf{0.347\std0.083} & 0.298\std0.073 & \textbf{0.146\std0.016} & 0.129\std0.034 & \textbf{0.325\std0.063} & 0.268\std0.066 \\
 200 & 0.265\std0.057 & \textbf{0.269\std0.047} & \textbf{0.357\std0.075} & 0.352\std0.076 & 0.182\std0.050 & \textbf{0.188\std0.056} & \textbf{0.347\std0.086} & 0.324\std0.083 \\
 400 & 0.314\std0.049 & \textbf{0.331\std0.038} & 0.389\std0.088 & \textbf{0.409\std0.090} & 0.212\std0.062 & \textbf{0.223\std0.074} & 0.372\std0.082 & \textbf{0.375\std0.078} \\
 600 & 0.338\std0.054 & \textbf{0.362\std0.037} & 0.415\std0.093 & \textbf{0.441\std0.093} & 0.225\std0.062 & \textbf{0.240\std0.078} & 0.389\std0.080 & \textbf{0.399\std0.068} \\
 800 & 0.344\std0.055 & \textbf{0.375\std0.035} & 0.429\std0.089 & \textbf{0.461\std0.093} & 0.235\std0.060 & \textbf{0.256\std0.087} & 0.405\std0.077 & \textbf{0.421\std0.065} \\
 \bottomrule
 \end{tabular}%
 }
\end{table}

\subsection{Ablations}
\label{sec:ablation-study}

\begin{figure}
    \centering
    \includegraphics[width=\linewidth]{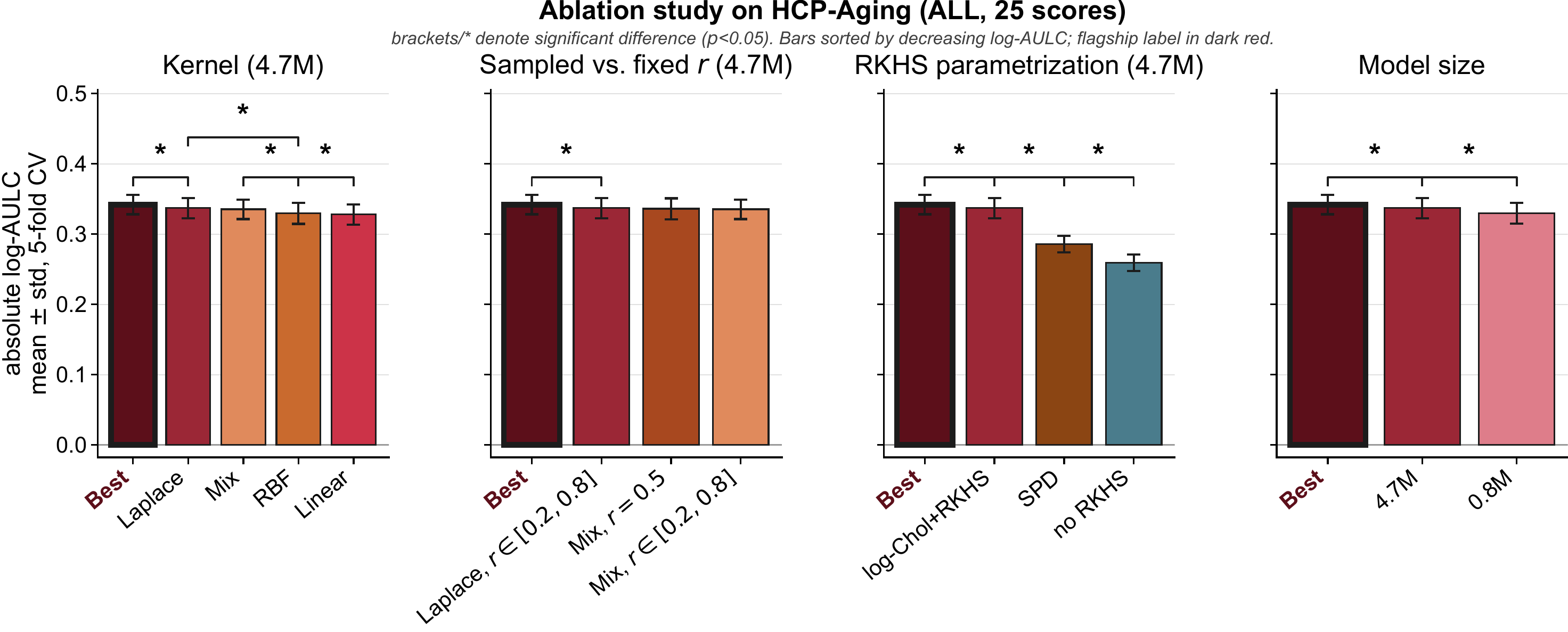}
    \caption{\textbf{Ablation study on HCP-Aging.} BrainPFN-4.7M (Laplace kernel family, fixed task difficulty) is the best configuration in every score group and significantly outperforms all variants. Removing the RKHS-sampling or working directly in SPD space causes the largest drops. \textit{Best} denotes BrainPFN-4.7M (Laplace, $r$=0.5).} %
    \label{fig:ablation-study-summary}
\end{figure}

We perform an ablation study on HCP-Aging in terms of \logauc{} to assess the contribution of key design choices in BrainPFN (Fig.~\ref{fig:ablation-study-summary}); App.~\ref{sec:appendix-ablations} extends the same comparison on the other three downstream cohorts. %
\textbf{Kernel:} sampling randomly or restricting the kernel family to linear or RBF kernels instead of the Laplace kernel likewise underperforms the best configuration. \textbf{Task-difficulty:} randomly sampling $r$ seems to slightly under-perform a fixed $r$, but not always significantly. \textbf{RKHS and log-Cholesky:} consistent with Sec.~\ref{sec:bfm-fall-short}, removing the RKHS prior on the $\alpha$ coefficients (``no RKHS'') produces the largest degradation of all ablations, and working directly in SPD space instead of using the log-Cholesky parameterization also reduces performance, though to a lesser extent. These are the two most impactful design choices tested. %
\textbf{Model size:} reducing model size from 4.7M to 0.8M parameters results in a modest decrease in performance. Hence, our preferred model is 4.7M because the computational overhead is minimal. %

%% file: 6_conclusions.tex
\section{Conclusions}

We presented \brainpfn{}, a Prior-data Fitted Network that recasts brain-behavior prediction as amortized inference: Meta-trained once on synthetic KRR tasks built on $\sim$7.7k real, unlabeled FC matrices from OpenNeuro, it predicts any continuous target from a labeled context in a single forward pass, with no refitting or hyperparameter search.
Across four independent cohorts and 85 targets, BrainPFN is the only model out of all the brain foundation models we compared that consistently beats KRR in the small-sample regime. That includes general cognition in every cohort. The advantage extends to movie-watching and task FC, even if BrainPFN was meta-trained using only resting-state data. BrainPFN closes for the first time, at least on small cohorts, the gap between the promise of pre-trained brain foundation models and simple linear baselines. Still, on large cohorts Ridge maintains the lead. Future work will focus on designing richer prior to extend BrainPFN's lead to larger $N$, other parcellations, and clinical cohorts.
Notably, our meta-training pool is a fraction of the size of UK BioBank data, typically used to train brain foundation models, and free of the access restrictions attached to it. By releasing this pool alongside code and checkpoints, we hope to lower the barrier to building and evaluating foundation models for neuroimaging, and to encourage a shift in how the field benchmarks such models: toward the low-data, high-heterogeneity settings where they are needed most.

%% file: statements.tex
\subsection*{AI use statement}

In this work, we used generative AI tools for: propose or refine hypothesis, design or provide feedback on research methodology or experiments, implement methods, clean and reformat dataset, interpret results.
We have not used generative AI tools for: generating synthetic dataset, help develop theoretical models or conceptual frameworks, formulate mathematical claims, provide critical ingredients for proving mathematical claims, assisting in the writing of proofs, support qualitative and thematic data analysis.
Additionally, we used generative AI tools for: creating or modifying scientific figures of images, suggesting experimental parameters, create or edit software code, creation of artifacts, drafting parts of a research paper, summarize or analyze existing literature, discover research topics or identify gaps, brainstorming, sourcing/searching for information, edit a research paper to improve readability, identify relevant literature, suggest a structure for a research paper. We have reviewed all AI-assisted work. We have not used AI to generate references, they were hand copied from the original sources. AI-generated code was verified and tested for correctness. We take responsibility for the final content of this work,
including text, claims or artifacts produced with the aid of generative AI.

\subsection*{Ethics statement}

This work uses recordings of human brain activity (resting-state and task fMRI) to predict behavioral and cognitive phenotypes. All data are either fully public or obtained under standard data use agreements (DUAs) from their original repositories; no new human data were collected, and we defer to the consent and IRB protocols of the original collectors. Our meta-training pool draws entirely from unrestricted OpenNeuro releases, and our downstream cohorts (HCP-YA, HCP-Aging, Cam-CAN, AOMIC-ID1000) are accessed under their standard DUAs.

Our aim is to advance cognitive and clinical neuroscience by making accurate brain-behavior prediction feasible in the small-sample regime typical of real research and clinical cohorts, ultimately improving our understanding of the brain and its diseases and supporting better diagnosis and treatment response.

We recognize a potential dual use of this technology: discriminating against or profiling individuals based on predicted cognitive ability or other traits, e.g., in employment, insurance, or education. We do not condone this use. BrainPFN is intended strictly for research and clinical purposes, and we release our pool, code, and checkpoints on that basis, encouraging downstream users to restrict use to appropriately consented research and clinical contexts.

\subsection*{Reproducibility statement}

BrainPFN's meta-training is fully reproducible: the meta-training pool is built entirely from public OpenNeuro data with no access restrictions, and we release the full source code, preprocessing pipeline, and model checkpoints, all of which will be made public upon acceptance. Evaluation on the four downstream cohorts (HCP-YA, HCP-Aging, Cam-CAN, AOMIC-ID1000) is likewise reproducible, provided the user agrees to the standard data use agreement (DUA) required by each dataset's distributor. Full details of the preprocessing pipeline, architecture, and training hyperparameters needed to reproduce our results are given in App.~\ref{sec:appendix-setup-details}, with dataset and evaluation-grid details in App.~\ref{sec:appendix-perscore}.

%% file: appendix.tex
\appendix
\section{Derivations}

\subsection{Task difficulty}
\label{appendix:effect-size-derivation}

We model the observed behavioral target as $y = s + \epsilon$, where $s = K(X, X_\text{train})\,\alpha$ is the KRR signal (standardised to unit variance, $\mathrm{Var}(s) = 1$) and $\epsilon \sim \mathcal{N}(0, \sigma^2)$ is additive noise independent of $s$. We derive the noise variance $\sigma^2$ required to achieve a target correlation $r = \mathrm{Corr}(s, y)$.

\paragraph{Covariance.}
Since $\epsilon \perp s$:
\begin{align}
    \mathrm{Cov}(s, y)
    &= \mathrm{Cov}(s,\, s + \epsilon) \notag \\
    &= \mathbb{E}[s(s+\epsilon)] - \mathbb{E}[s]\,\mathbb{E}[s+\epsilon] \notag \\
    &= \bigl(\mathbb{E}[s^2] - \mathbb{E}[s]^2\bigr) + \bigl(\mathbb{E}[s\epsilon] - \mathbb{E}[s]\,\mathbb{E}[\epsilon]\bigr) \notag \\
    &= \mathrm{Var}(s) + \mathrm{Cov}(s, \epsilon) \notag \\
    &= 1 + 0 = 1.
\end{align}

\paragraph{Variance of the target.}
Since $s$ and $\epsilon$ are independent:
\begin{equation}
    \mathrm{Var}(y) = \mathrm{Var}(s + \epsilon) = \mathrm{Var}(s) + \mathrm{Var}(\epsilon) = 1 + \sigma^2.
\end{equation}

\paragraph{Correlation.}
\begin{equation}
    r = \mathrm{Corr}(s, y)
    = \frac{\mathrm{Cov}(s, y)}{\sqrt{\mathrm{Var}(s)\,\mathrm{Var}(y)}}
    = \frac{1}{\sqrt{1 \cdot (1 + \sigma^2)}}
    = \frac{1}{\sqrt{1 + \sigma^2}}.
\end{equation}

\paragraph{Solving for $\sigma^2$.}
Setting the above equal to a target effect size $r$ and solving:
\begin{equation}
    \frac{1}{\sqrt{1+\sigma^2}} = r
    \;\implies\;
    1 + \sigma^2 = \frac{1}{r^2}
    \;\implies\;
    \sigma^2 = \frac{1 - r^2}{r^2},
\end{equation}
which gives Eq.~(\ref{eq:effect-size-sigma}).

\section{Additional Details}

\subsection{Additional experimental setup details}
\label{sec:appendix-setup-details}

\paragraph{Meta-training pool preprocessing.} Every subject in the meta-training pool (Sec.~\ref{sec:datasets}) is processed with the same pipeline based on fmriprep (25.2.3): resampling of volumes to the MNI152 template at 2mm isotropic resolution, followed by scrubbing of high-motion volumes (framewise displacement $>$ 0.5mm), 24-parameter confound regression including motion and tissue intensity parameters, and a 0.01--0.1Hz bandpass filter, before parcellation with the Schaefer-400 atlas \citep{schaefer2018local}.

\paragraph{Architecture and optimization hyperparameters.} Training tokens concatenate $x_i$ and the
scalar label $y_i$ before a linear projection; test tokens project
$x_j$ alone. Features and labels are normalised within each forward
pass using statistics computed from the training tokens and
de-normalised at output. \textit{BrainPFN-0.8M} uses 4 layers, $d_{\text{model}}{=}128$, 4 heads, $d_{\text{ff}}{=}256$, and \textit{BrainPFN-4.7M} uses 6 layers, $d_{\text{model}}{=}256$, 4
heads, $d_{\text{ff}}{=}512$; both parameter counts exclude the input
projection layers, whose size scales with the (data-dependent) feature
dimensionality $d$. In preliminary sweeps, further increasing depth
and width gave no accuracy improvement over BrainPFN-4.7M, so we
do not report larger presets.
The model is trained for 10k steps with AdamW
(learning rate $3\times10^{-4}$, weight decay $10^{-2}$, no warmup), batch size 512, and dropout 0.1. Each synthetic task in a batch draws $200$ context subjects, a disjoint calibration block of $200$ subjects (Eq.~\ref{eq:effect-size-sigma}), and $100$ held-out subjects from the meta-training pool, i.e. 500 subjects per synthetic task; over the full run this amounts to $10^4 \times 512 \times 300 \approx 1.5\times10^9$ individual context/test $(x,y)$ pairs seen by the network -- the
\lowtilde1B figure reported in the abstract. Training uses a single V100 or A100 GPU.

\paragraph{Significance testing and evaluation grid.} The other foundation models are evaluated at each context by linear probing on the same context, rather than by refitting Ridge/KRR. Significance against the Ridge baseline is computed on the per-fold averages: we first average across the 20 within-fold context draws, then compute significance across the 5 resulting fold-level values. The context-size grid $N \in \{5, 10, 20, \dots\}$ extends up to each cohort's train-fold size (Tab.~\ref{tab:datasets}).

\subsection{Data preprocessing for the other foundation models}
\paragraph{BrainJEPA.} Input is the raw resting-state BOLD timeseries (not a connectivity
matrix), parcellated into 450 regions (400 cortical + 50 subcortical) and reordered to match
BrainJEPA's own pretraining atlas convention. Each region's signal is robust-scaled --
$(x - \mathrm{median}) / \mathrm{IQR}$ -- using reference statistics computed once from the
training-fold subjects and then frozen and reused identically across every dataset (never
refit per dataset). The timeseries is cut into non-overlapping 160-timepoint windows (fixed
by the pretrained model's positional embeddings); every window from every run of a subject is
passed through the frozen pretrained encoder and the resulting window-level embeddings are
averaged into a single 768-dimensional subject embedding.

\paragraph{BrainMass.} Input is a functional-connectivity (correlation) matrix computed from
a 100-region parcellation, matching the resolution BrainMass was pretrained on (coarser than
this project's own 400-region atlas). The frozen pretrained transformer encoder processes
this matrix through its attention layers and a dimensionality-reduction step, yielding an
800-dimensional subject embedding (taken before the self-supervised pretraining projection
head).

\paragraph{Brain-Semantoks.} Input is again the raw BOLD timeseries, resampled to a fixed
2-second TR, over the same 450-region parcellation used for BrainJEPA plus a cerebellar atlas
that this project's data does not cover -- that channel is zero-filled and explicitly masked
out inside the model so it contributes no signal. The timeseries is split into non-overlapping
100-timepoint windows, encoded by the frozen pretrained model, and window/run embeddings are
averaged per subject into a 1536-dimensional embedding (a concatenation of pooled local tokens
and a global summary token).

\paragraph{TabPFN.} Unlike the three embedding models above, TabPFN is a general-purpose
tabular foundation model with no neuroimaging pretraining. Input is the same raw
functional-connectivity features used for BrainPFN. The PCA basis is refit separately for every few-shot
training-set size evaluated, rather than fixed once -- matching the same per-context
convention used for the ridge and BrainPFN baselines, so TabPFN always sees a feature
representation scaled appropriately to how much data it is being given.

\section{Additional Results}

\subsection{Ooi et al. (2022)-style behavioural components (HCP-YA and HCP-Aging)}
\label{sec:appendix-ooi}
\input{appendix_ooi.tex}
\clearpage

\subsection{Exact Gaussian-process baseline}
\label{sec:appendix-gp}

BrainPFN is meta-trained on synthetic tasks whose targets are kernel functions of FC plus
calibrated noise (Sec.~\ref{sec:brainpfn}). For a fixed kernel and fixed hyperparameters, the
posterior predictive of such a task is that of a Gaussian process (GP) and is available in
closed form; with random hyperparameters it is a mixture of GP posteriors weighted by their
marginal likelihoods. A GP with the same kernel, fitted on the same features, is therefore the
natural analytic counterpart of BrainPFN: it tests whether the network does more than
reproduce this Bayesian computation.

\paragraph{GP baselines.} Each target is standardised on the context set and modelled as
$y = f(x) + \epsilon$, with $f \sim \mathcal{GP}(0, r^2 k_\gamma)$ and
$\epsilon \sim \mathcal{N}(0, 1 - r^2)$, where $r$ is the signal--target correlation (the
task-difficulty parameter of the prior) and $k_\gamma(x, x') = \exp(-\gamma \lVert x - x'
\rVert_1)$ is the prior's Laplace kernel, with bandwidth $\gamma = s / m$ and $m$ the median
pairwise $\ell_1$ distance in the context. The inputs are exactly the features BrainPFN sees
(log-Cholesky FC, with the PCA basis refit on each context). For fixed $(s, r)$ the
posterior mean is $\mu_* = r^2 K_{*}\,(r^2 K + (1 - r^2) I)^{-1} y$. The two baselines
differ only in how they treat $(s, r)$, both using the exact marginal likelihood
$p(y \mid s, r)$ on a grid:
\begin{itemize}
    \item \textbf{GP-BMA} (Bayesian model averaging) places a uniform hyperprior on
    $s \in [1, 3]$ and $r \in [0.2, 0.8]$ -- the bandwidth range of the prior and the
    task-difficulty range of its sampled-$r$ variant, which contains BrainPFN-4.7M's fixed
    $r{=}0.5$ -- and averages the GP predictions over an $11 \times 13$ grid, weighted by
    their posterior probability. It integrates the hyperparameters out, which is the
    computation BrainPFN is trained to amortise, and is the strongest analytic competitor.
    \item \textbf{GP-ML} (type-II maximum likelihood) keeps the single $(s, r)$ that maximises
    the marginal likelihood over a wider grid ($s \in [0.1, 10]$, $r \in [0.05, 0.95]$), the
    standard way GPs are fitted in practice.
\end{itemize}
We also report the ablation variant of BrainPFN-4.7M trained with $r \sim U[0.2, 0.8]$
instead of the fixed $r{=}0.5$ (``sampled $r$'', Sec.~\ref{sec:ablation-study}). All models
share the same folds, 20 context draws per $N$, and test splits; \logauc{} is computed over
$N \in \{5, \dots, 800\}$, and significance is a two-sided paired $t$-test over the 5
fold-level values, reported as $p < 0.05$ or n.s. Real-signal targets are selected as in the rest of the paper, over the models
compared here.

\paragraph{Results.} BrainPFN-4.7M significantly beats both GPs over the whole learning curve
in all three settings, by $+0.010$ to $+0.020$ \logauc{} over GP-BMA
(Tab.~\ref{tab:gp-logauc}); the sampled-$r$ variant is ahead of GP-BMA everywhere but
significantly so only on HCP-A. The gain comes
from the few-shot regime (Tab.~\ref{tab:gp-per-n}): BrainPFN-4.7M is significantly ahead of
GP-BMA at every $N$ from 5 to 100 in all three settings. From $N{=}200$ the two are within
$0.01$ of each other in every setting (GP-BMA is slightly but significantly ahead on HCP-A
for $N \ge 400$), and Ridge is the best model there, so the GP does not
close BrainPFN's large-$N$ gap to Ridge either. GP-ML has the lowest \logauc{} of
all models: with few subjects the marginal likelihood is nearly
flat and its maximiser lands on degenerate corners of the grid, whereas averaging stays near
the centre of the hyperprior.

\begin{table}[h]
 \centering
 \caption{\logauc{} of all models ($N \in \{5, \dots, 800\}$, mean over 5 folds;
 bold: best model) and two-sided paired $t$-test across folds against GP-BMA and GP-ML:
 $\uparrow$/$\downarrow$ marks the row model as significantly better/worse ($p < 0.05$),
 n.s.\ otherwise.}
 \label{tab:gp-logauc}
 \small
 \resizebox{\linewidth}{!}{%
 \begin{tabular}{lccccccccc}
 \toprule
 & \multicolumn{3}{c}{HCP-YA, real-signal (36)} & \multicolumn{3}{c}{HCP-YA, cognition composite} & \multicolumn{3}{c}{HCP-A, real-signal (34)} \\
 \cmidrule(lr){2-4} \cmidrule(lr){5-7} \cmidrule(lr){8-10}
 Model & \logauc{} & vs.\ GP-BMA & vs.\ GP-ML & \logauc{} & vs.\ GP-BMA & vs.\ GP-ML & \logauc{} & vs.\ GP-BMA & vs.\ GP-ML \\
 \midrule
 BrainPFN-4.7M & \textbf{0.137} & $<$0.05\,$\uparrow$ & $<$0.05\,$\uparrow$ & \textbf{0.325} & $<$0.05\,$\uparrow$ & $<$0.05\,$\uparrow$ & \textbf{0.333} & $<$0.05\,$\uparrow$ & $<$0.05\,$\uparrow$ \\
 {\renewcommand{\arraystretch}{0.8}\begin{tabular}[c]{@{}l@{}}BrainPFN-4.7M,\\ sampled $r$\end{tabular}} & 0.129 & n.s. & $<$0.05\,$\uparrow$ & 0.314 & n.s. & $<$0.05\,$\uparrow$ & 0.329 & $<$0.05\,$\uparrow$ & $<$0.05\,$\uparrow$ \\
 GP-BMA & 0.127 & -- & $<$0.05\,$\uparrow$ & 0.305 & -- & $<$0.05\,$\uparrow$ & 0.319 & -- & $<$0.05\,$\uparrow$ \\
 GP-ML & 0.110 & -- & -- & 0.272 & -- & -- & 0.290 & -- & -- \\
 Ridge & 0.126 & n.s. & $<$0.05\,$\uparrow$ & 0.298 & n.s. & $<$0.05\,$\uparrow$ & 0.310 & $<$0.05\,$\downarrow$ & $<$0.05\,$\uparrow$ \\
 \bottomrule
 \end{tabular}%
 }
\end{table}

\begin{table}[h]
 \centering
 \caption{Mean Pearson $r$ at each context size $N$ (mean over targets and 5 folds; bold:
 best model at that $N$). The last row of each block is the two-sided paired $t$-test of
 BrainPFN-4.7M vs.\ GP-BMA across folds ($p < 0.05$ or n.s., in either direction).}
 \label{tab:gp-per-n}
 \small
 \resizebox{\linewidth}{!}{%
 \begin{tabular}{lccccccccc}
 \toprule
 Model & $N{=}5$ & 10 & 20 & 50 & 100 & 200 & 400 & 600 & 800 \\
 \midrule
 \multicolumn{10}{l}{\textit{HCP-YA, real-signal (36 targets)}} \\
 BrainPFN-4.7M & \textbf{0.039} & \textbf{0.059} & \textbf{0.081} & \textbf{0.125} & \textbf{0.155} & 0.182 & 0.221 & 0.238 & 0.250 \\
 {\renewcommand{\arraystretch}{0.8}\begin{tabular}[c]{@{}l@{}}BrainPFN-4.7M,\\ sampled $r$\end{tabular}} & 0.038 & 0.056 & 0.074 & 0.117 & 0.146 & 0.172 & 0.211 & 0.231 & 0.243 \\
 GP-BMA & 0.034 & 0.049 & 0.067 & 0.110 & 0.142 & 0.180 & 0.216 & 0.232 & 0.244 \\
 GP-ML & 0.028 & 0.039 & 0.050 & 0.084 & 0.112 & 0.154 & 0.207 & 0.227 & 0.241 \\
 Ridge & 0.029 & 0.043 & 0.060 & 0.102 & 0.136 & \textbf{0.184} & \textbf{0.226} & \textbf{0.245} & \textbf{0.260} \\
 \quad $p$ (4.7M vs.\ GP-BMA) & $<$0.05 & $<$0.05 & $<$0.05 & $<$0.05 & $<$0.05 & n.s. & $<$0.05 & n.s. & n.s. \\
 \midrule
 \multicolumn{10}{l}{\textit{HCP-YA, cognition composite}} \\
 BrainPFN-4.7M & \textbf{0.121} & \textbf{0.179} & \textbf{0.201} & \textbf{0.319} & \textbf{0.376} & 0.424 & 0.475 & 0.501 & 0.511 \\
 {\renewcommand{\arraystretch}{0.8}\begin{tabular}[c]{@{}l@{}}BrainPFN-4.7M,\\ sampled $r$\end{tabular}} & 0.120 & 0.173 & 0.188 & 0.299 & 0.364 & 0.409 & 0.467 & 0.497 & 0.509 \\
 GP-BMA & 0.096 & 0.136 & 0.168 & 0.281 & 0.355 & 0.427 & 0.483 & 0.504 & 0.515 \\
 GP-ML & 0.076 & 0.106 & 0.131 & 0.225 & 0.276 & 0.422 & 0.479 & 0.499 & 0.509 \\
 Ridge & 0.083 & 0.131 & 0.141 & 0.254 & 0.340 & \textbf{0.435} & \textbf{0.500} & \textbf{0.533} & \textbf{0.552} \\
 \quad $p$ (4.7M vs.\ GP-BMA) & $<$0.05 & $<$0.05 & $<$0.05 & $<$0.05 & $<$0.05 & n.s. & n.s. & n.s. & n.s. \\
 \midrule
 \multicolumn{10}{l}{\textit{HCP-A, real-signal (34 targets)}} \\
 BrainPFN-4.7M & 0.154 & \textbf{0.215} & \textbf{0.278} & \textbf{0.337} & \textbf{0.375} & 0.396 & 0.430 & 0.449 & 0.460 \\
 {\renewcommand{\arraystretch}{0.8}\begin{tabular}[c]{@{}l@{}}BrainPFN-4.7M,\\ sampled $r$\end{tabular}} & \textbf{0.155} & 0.214 & 0.277 & 0.332 & 0.369 & 0.391 & 0.422 & 0.440 & 0.450 \\
 GP-BMA & 0.136 & 0.187 & 0.247 & 0.314 & 0.364 & 0.399 & 0.435 & 0.455 & 0.468 \\
 GP-ML & 0.115 & 0.154 & 0.202 & 0.263 & 0.313 & 0.393 & 0.433 & 0.453 & 0.465 \\
 Ridge & 0.112 & 0.159 & 0.222 & 0.297 & 0.355 & \textbf{0.403} & \textbf{0.447} & \textbf{0.472} & \textbf{0.488} \\
 \quad $p$ (4.7M vs.\ GP-BMA) & $<$0.05 & $<$0.05 & $<$0.05 & $<$0.05 & $<$0.05 & n.s. & $<$0.05 & $<$0.05 & $<$0.05 \\
 \bottomrule
 \end{tabular}%
 }
\end{table}

\clearpage

\subsection{Complete Ablation Study Results}
\label{sec:appendix-ablations}

Section~\ref{sec:ablation-study} reports the ablation study (kernel choice, model size, and
fixed vs. sampled $r$, all against the flagship BrainPFN-4.7M) on HCP-Aging only. Here we give
the same comparison, first summarized across all four datasets, then broken down by score
domain for each dataset individually (HCP-Aging's own by-domain breakdown is
Figure~\ref{fig:ablation-study-summary}).

\begin{figure}[h]
    \centering
    \includegraphics[width=0.9\linewidth]{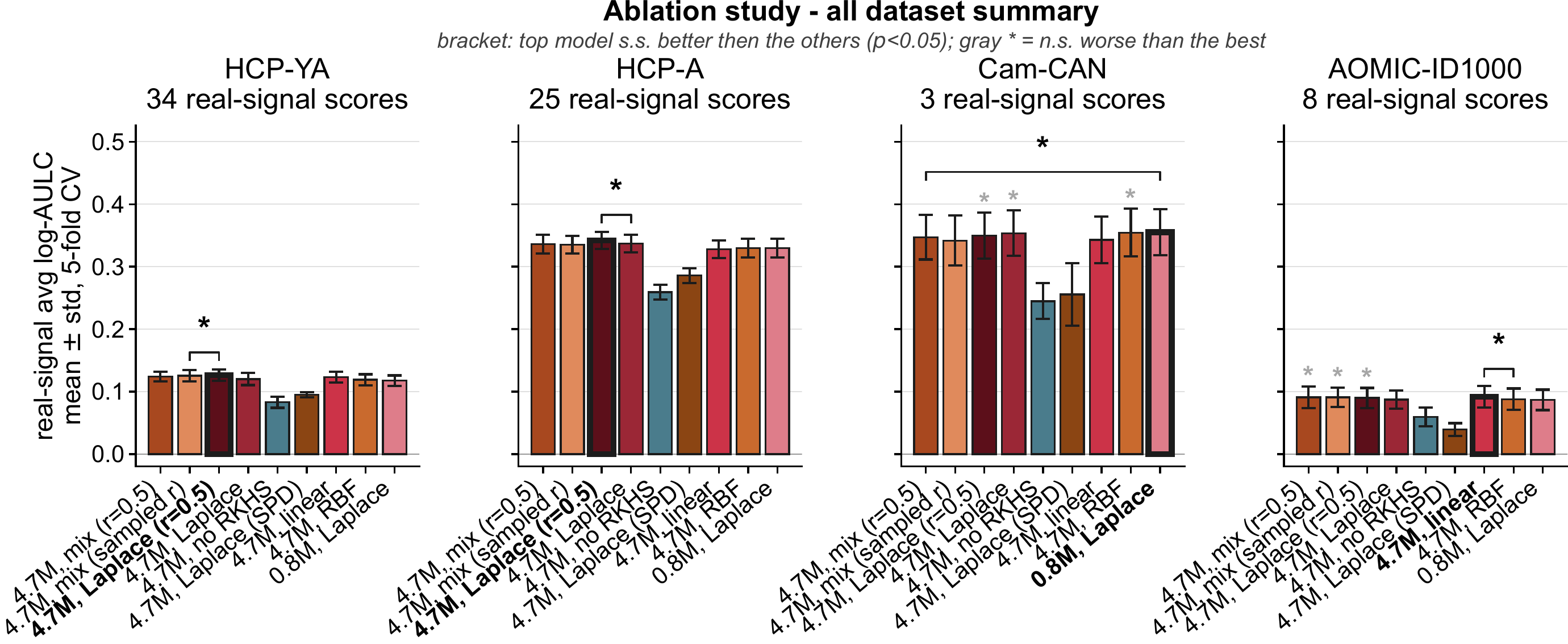}
    \caption{Ablations vs. the flagship (BrainPFN-4.7M, Laplace kernel, $r{=}0.5$), summarized
    per dataset. Unlike the by-domain panels below, each dataset's bar here averages only over
    that dataset's real-signal scores (the union, across the five plotted checkpoints, of each
    model's own scores with detectable signal vs. Ridge), so these numbers do not match the
    "ALL" panel of the corresponding by-domain figure, which averages unfiltered over every
    baseline-covered score.}
    \label{fig:ablation-alldatasets-summary}
\end{figure}

\begin{figure}
    \centering
    \includegraphics[width=0.9\linewidth]{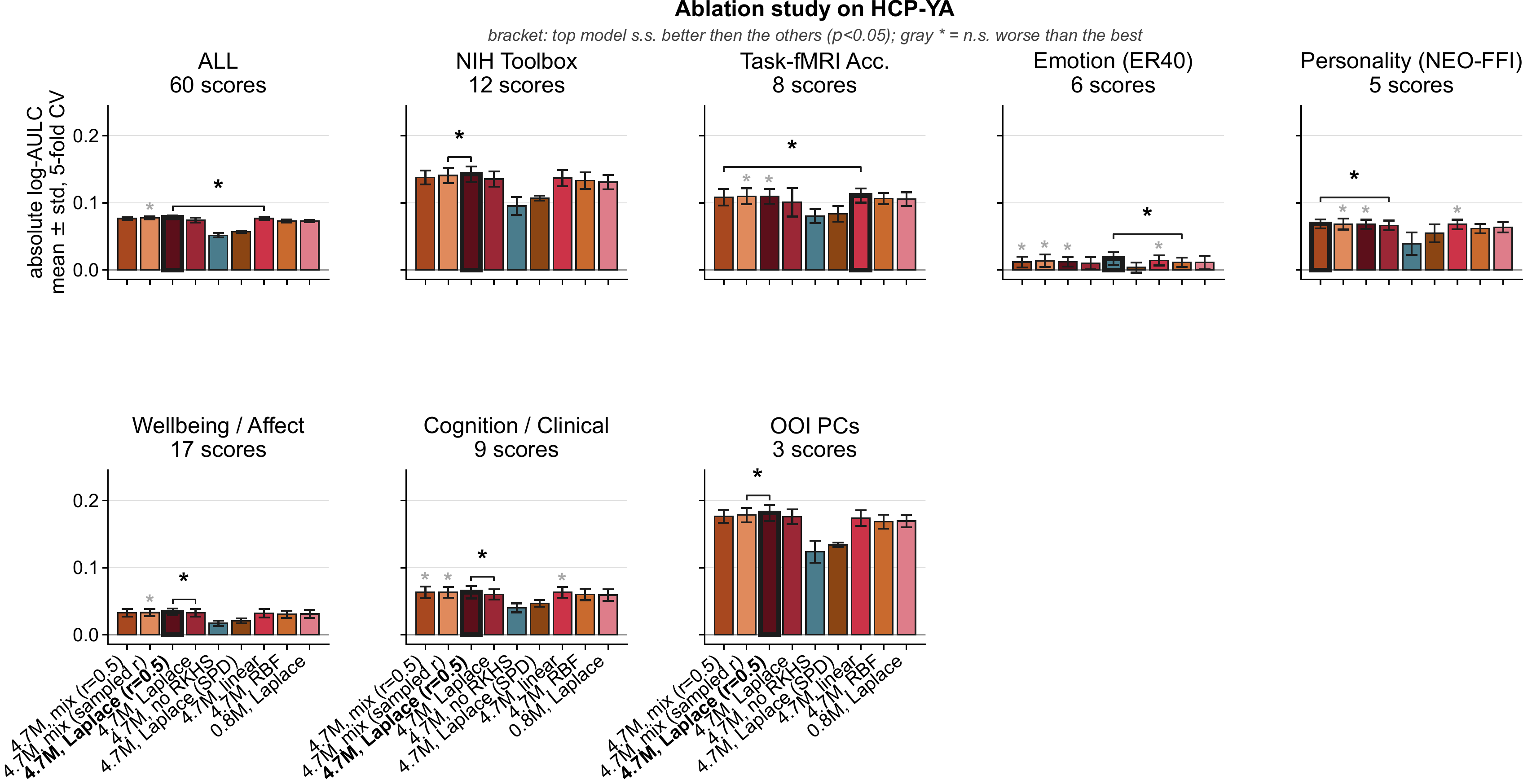}
    \caption{Ablation study on HCP-YA, by score domain.}
    \label{fig:ablation-study-hcp}
\end{figure}

\begin{figure}
    \centering
    \includegraphics[width=0.9\linewidth]{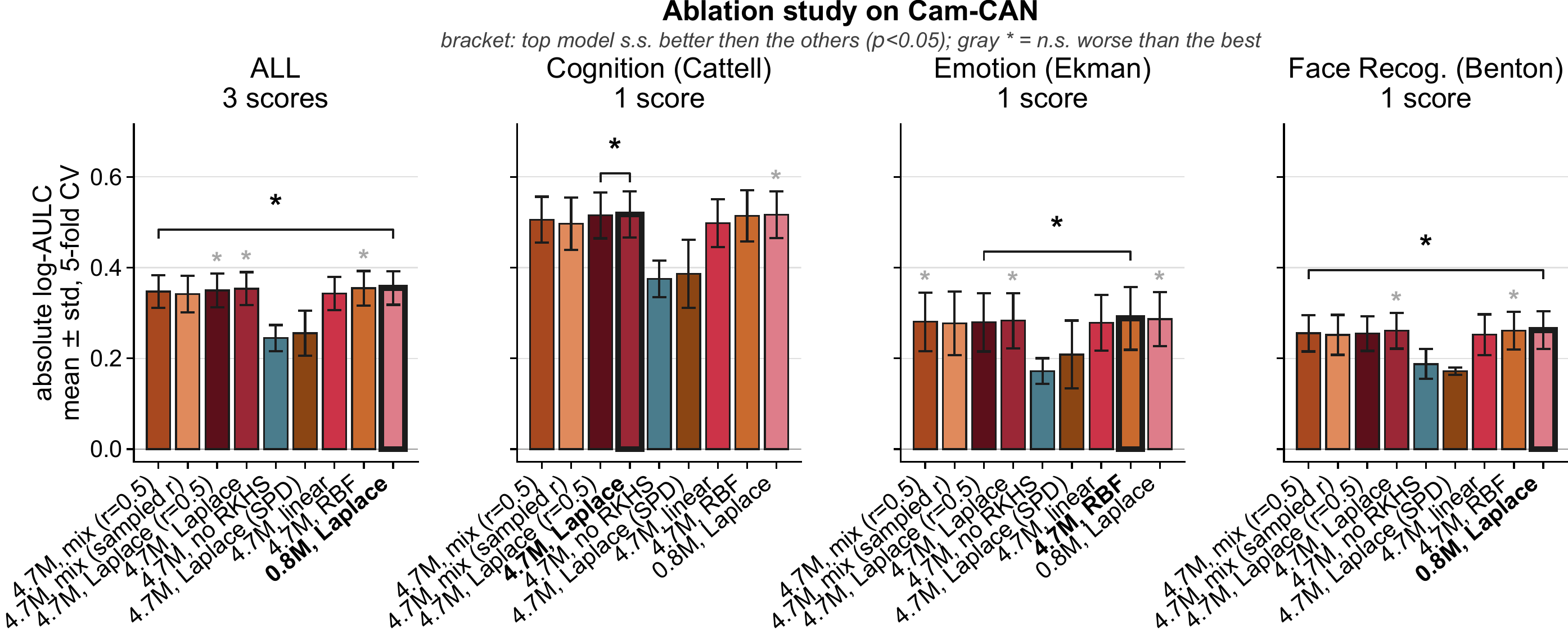}
    \caption{Ablation study on Cam-CAN, by score domain.}
    \label{fig:ablation-study-camcan}
\end{figure}

\begin{figure}
    \centering
    \includegraphics[width=0.9\linewidth]{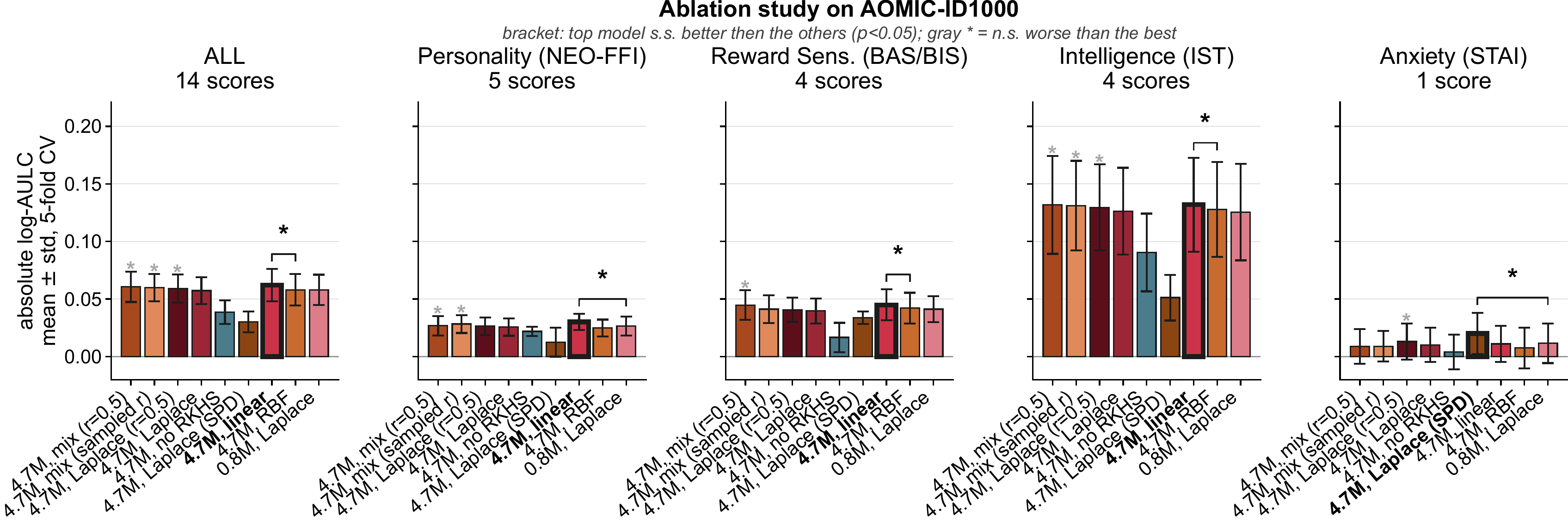}
    \caption{Ablation study on AOMIC-ID1000, by score domain.}
    \label{fig:ablation-study-aomic-id1000}
\end{figure}

\clearpage

\subsection{Crossover between BrainPFN and Ridge across context sizes}
\label{sec:appendix-crossover}

\begin{figure}[h!]
 \centering
 \includegraphics[width=\linewidth]{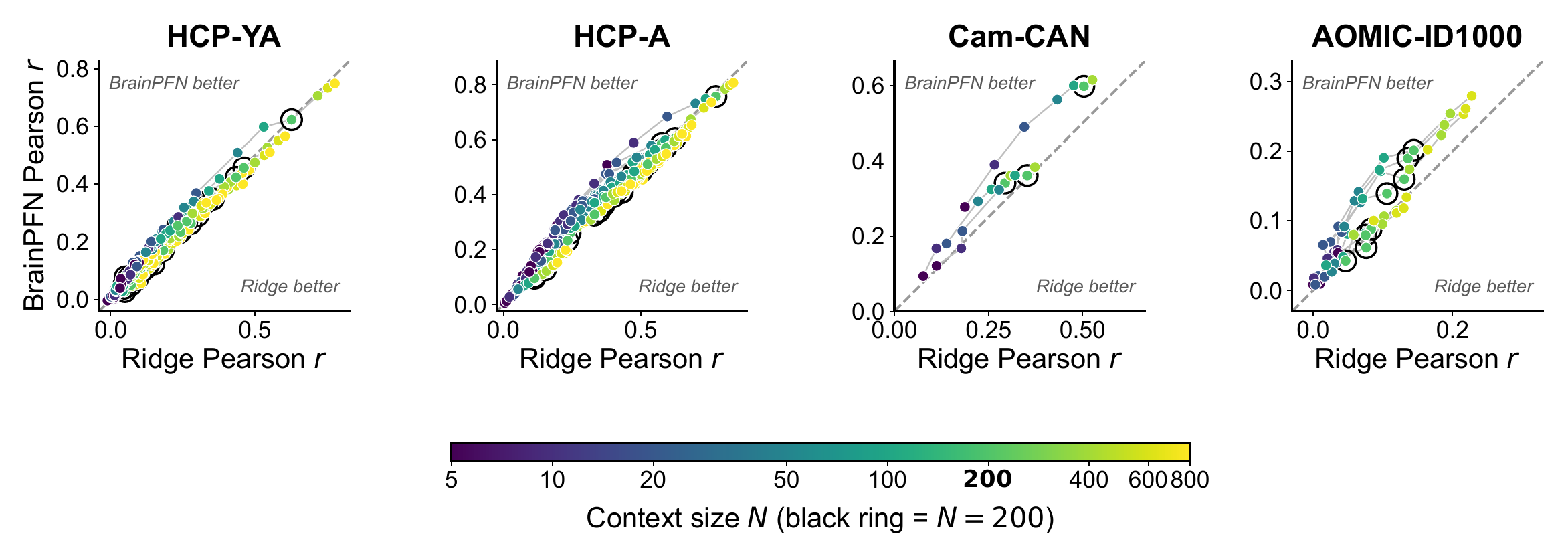}
 \caption{Ridge Pearson $r$ versus BrainPFN Pearson $r$ for every real-signal target, one point
 per (target, context size $N$), coloured by $N$ (light = small, dark = large); points for the
 same target at consecutive $N$ are joined by a thin grey line. Points above the dashed
 diagonal are context sizes at which BrainPFN wins; $N{=}200$ points are marked with a black
 ring. On HCP and HCP-Aging, trajectories cross the diagonal right around $N{=}200$; on Cam-CAN
 and AOMIC-ID1000, they never do within the tested range.}
 \label{fig:crossover_parity}
\end{figure}

We define each target's \emph{crossover} $N$ as the smallest context size beyond which Ridge exceeds BrainPFN for every larger $N$ tested: the practical sample size below which BrainPFN is the better choice. Most HCP-YA and HCP-A targets cross over near $N{=}150$--$200$; no target on Cam-CAN or AOMIC-ID1000 crosses over within their respective pool size. %
Table~\ref{tab:crossover} reports this crossover for
representative targets spanning cognition, physical, and personality measures; a working-memory
accuracy score on HCP crosses earliest, at $N{=}50$, while a matrix-reasoning score never
crosses within the tested range. Figure~\ref{fig:crossover_parity} shows the same crossover
directly for every real-signal target rather than this representative handful.

\begin{table}[h]
 \centering
 \caption{Context-size crossover for representative targets: the smallest $N$ at which
 Ridge's Pearson $r$ overtakes BrainPFN's and remains higher for every larger $N$ tested.
 Most HCP targets cross between $N{=}50$ and $N{=}200$; no target on Cam-CAN or AOMIC-ID1000
 crosses within the tested range.}
 \label{tab:crossover}
 \small
 \begin{tabular}{llc}
 \toprule
 Dataset & Target & Crossover $N$ \\
 \midrule
 HCP & \texttt{WM\_Task\_Acc} (working-memory accuracy) & 50 \\
 HCP & \texttt{pc2} (cognition composite) & 200 \\
 HCP & \texttt{PicVocab\_AgeAdj} & 200 \\
 HCP & \texttt{Strength\_Unadj} (grip strength) & 200 \\
 HCP & \texttt{PMAT24\_A\_CR} (matrix reasoning) & never ($N \le 800$) \\
 Cam-CAN & \texttt{Cattell} (fluid intelligence) & never ($N \le 400$) \\
 Cam-CAN & \texttt{BentonFaces} (face recognition) & never ($N \le 400$) \\
 AOMIC-ID1000 & \texttt{IST\_intelligence\_total} & never ($N \le 600$) \\
 \bottomrule
 \end{tabular}
\end{table}

\subsection{Movie-watching vs. resting-state functional connectivity on HCP}
\label{sec:appendix-movie-vs-rest-hcp}

\begin{figure}[h]
    \centering
    \includegraphics[width=0.9\linewidth]{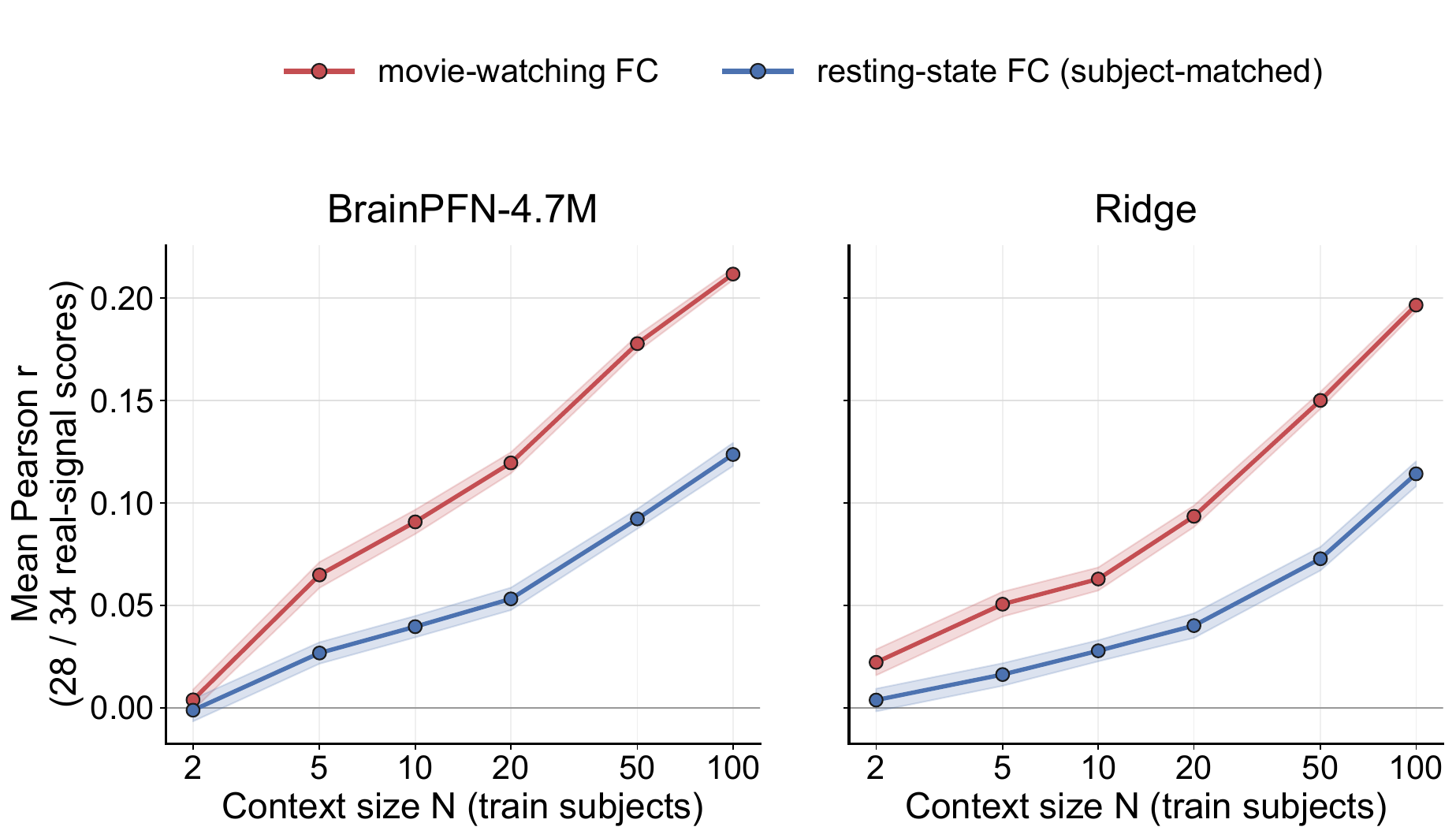}
    \caption{HCP: mean Pearson $r$ vs. context size $N$ for movie-watching FC and
    subject-matched resting-state FC, restricted to each FC input's own real-signal scores
    (28 for movie-watching, 34 for resting-state), 5-fold CV, 20 repeats per fold.}
    \label{fig:movie-vs-rest}
\end{figure}

Figure~\ref{fig:movie-vs-rest} compares FC computed from the HCP 7T movie-watching scans
against FC from each subject's own resting-state scan (subject-matched to the movie cohort),
for BrainPFN-4.7M and Ridge alike. Of the four HCP 7T movie-watching runs
(\texttt{MOVIE1--4\_7T}), we use \texttt{MOVIE1\_7T\_AP}, which covers the most subjects
(176) among the four. Movie-watching FC gives a consistently higher signal than
resting-state FC for both models at every context size, and BrainPFN's margin over Ridge is
preserved across both FC inputs -- the gain from movie-derived FC and the gain from BrainPFN
over Ridge are additive rather than substituting for one another.

\clearpage

\section{Full List of Results}
\label{sec:appendix-full-results}

\subsection{Per-score, all-model comparison, all 5-fold benchmarks}
\label{sec:appendix-all-score-model-comparison}

The per-score breakdown above (Sec.~\ref{sec:appendix-perscore}) pairs each model with its
own Ridge baseline, one model per figure. This section instead gives every score its own
panel with all 7 models plotted together (same style as
Figure~\ref{fig:headline-results}'s curve panels: one line per model, colored and
sized as in the legend, BrainPFN variants shown bold, plus the shared canonical-Ridge
dashed reference curve, "*" marking $N$ where a model is significantly ahead of its own
Ridge -- one-sided paired $t$-test on the 5 fold-level means, Holm-corrected across that
panel's own $N$-grid), split per dataset and paginated the same way the per-model perscore
figures are (panels ordered by peak Pearson $r$ across every model; \texttt{COG} tags the
dataset's designated cognition target, \texttt{SIG} tags a score with real signal for
BrainPFN-4.7M vs. Ridge). HCP-YA (60 scores) is split across 7 pages, HCP-A across 3, and
AOMIC-ID1000 across 2; Cam-CAN (3 scores) fits on one page.

\begin{figure}[h]
    \centering
    \includegraphics[width=0.84\linewidth]{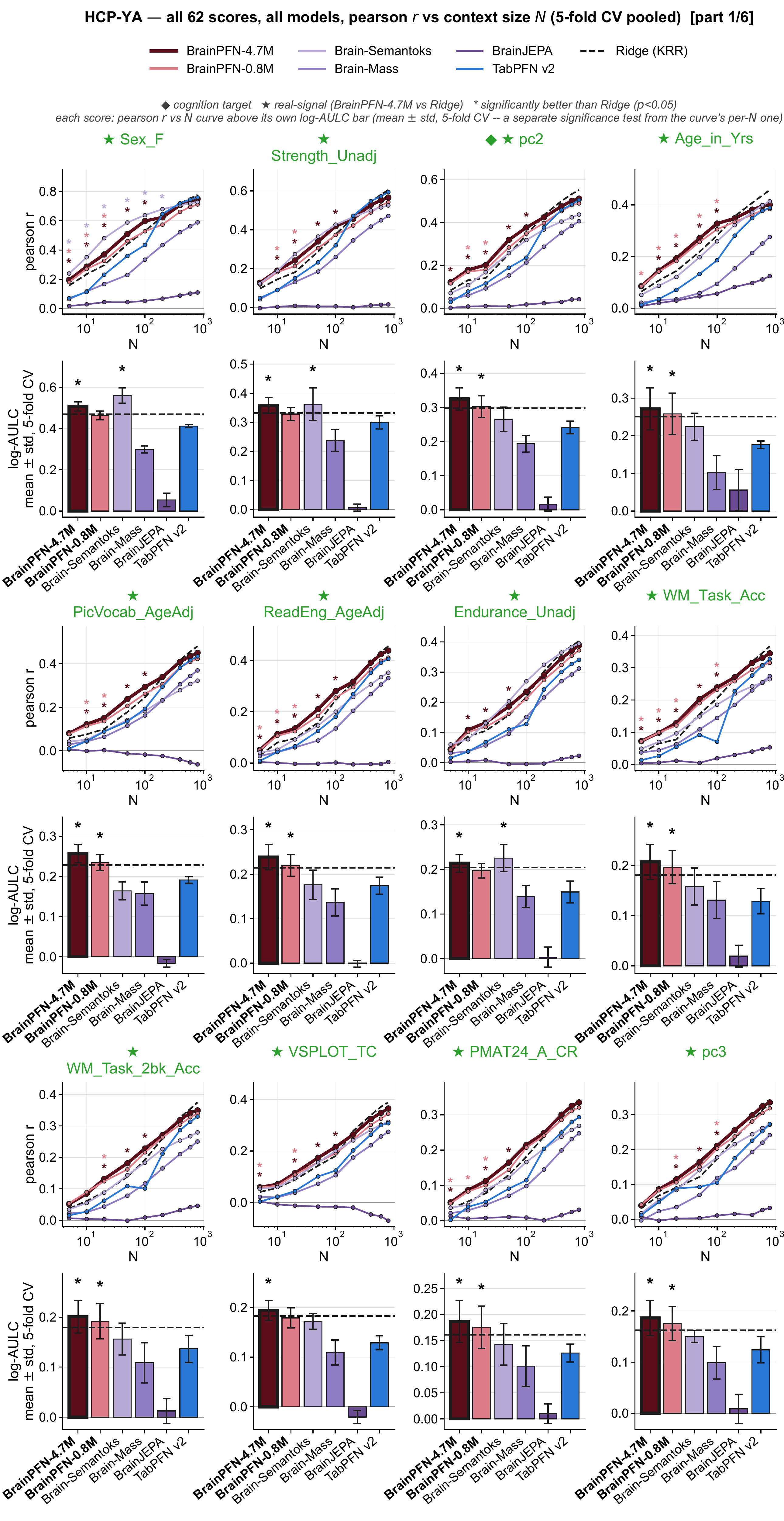}
    \caption{All 7 models vs. shared Ridge reference, all scores, HCP-YA, 5-fold CV (1/6).}
    \label{fig:all-scores-hcp-1}
\end{figure}

\begin{figure}
    \centering
    \includegraphics[width=0.84\linewidth]{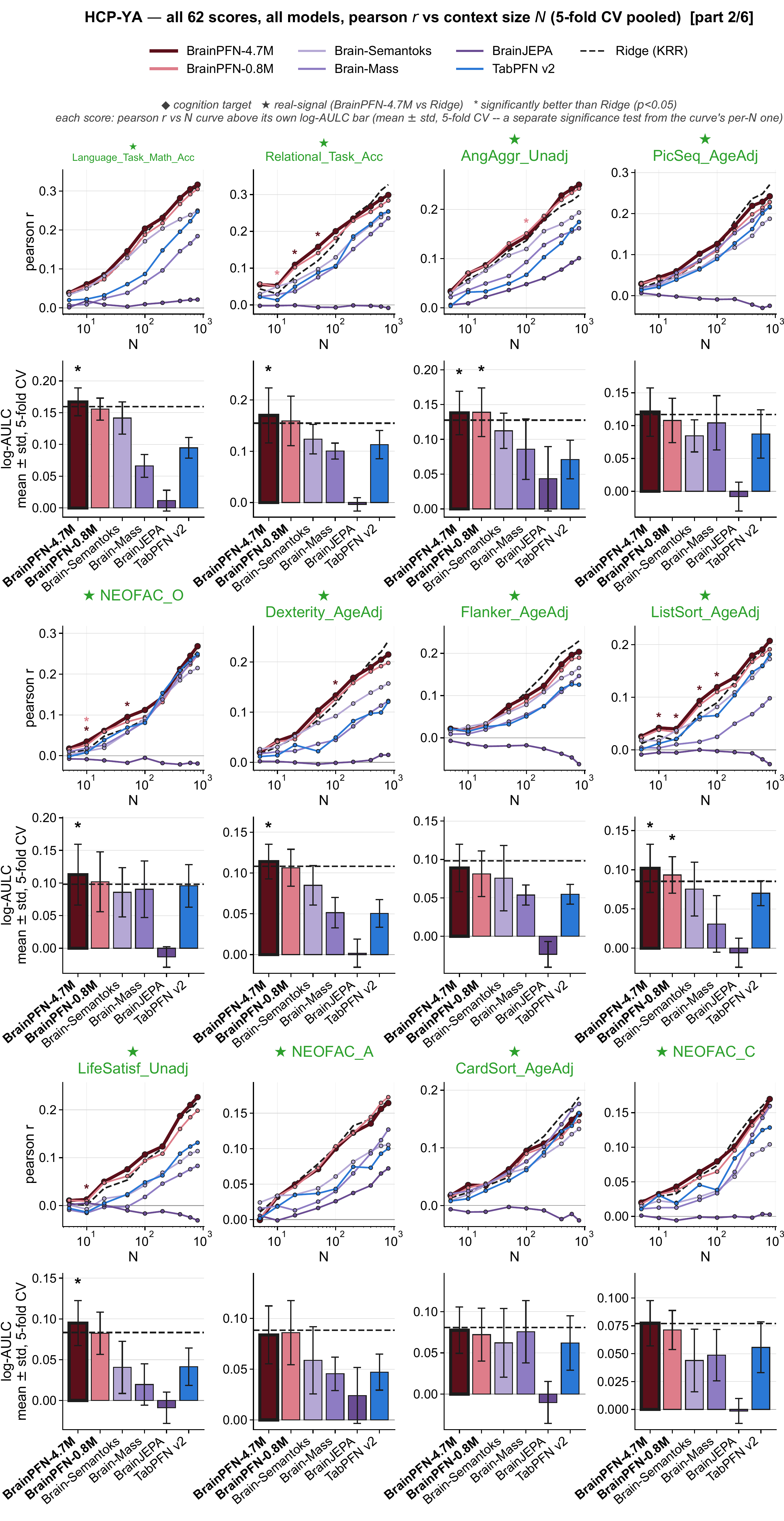}
    \caption{All 7 models vs. shared Ridge reference, all scores, HCP-YA, 5-fold CV (2/6).}
    \label{fig:all-scores-hcp-2}
\end{figure}

\begin{figure}
    \centering
    \includegraphics[width=0.84\linewidth]{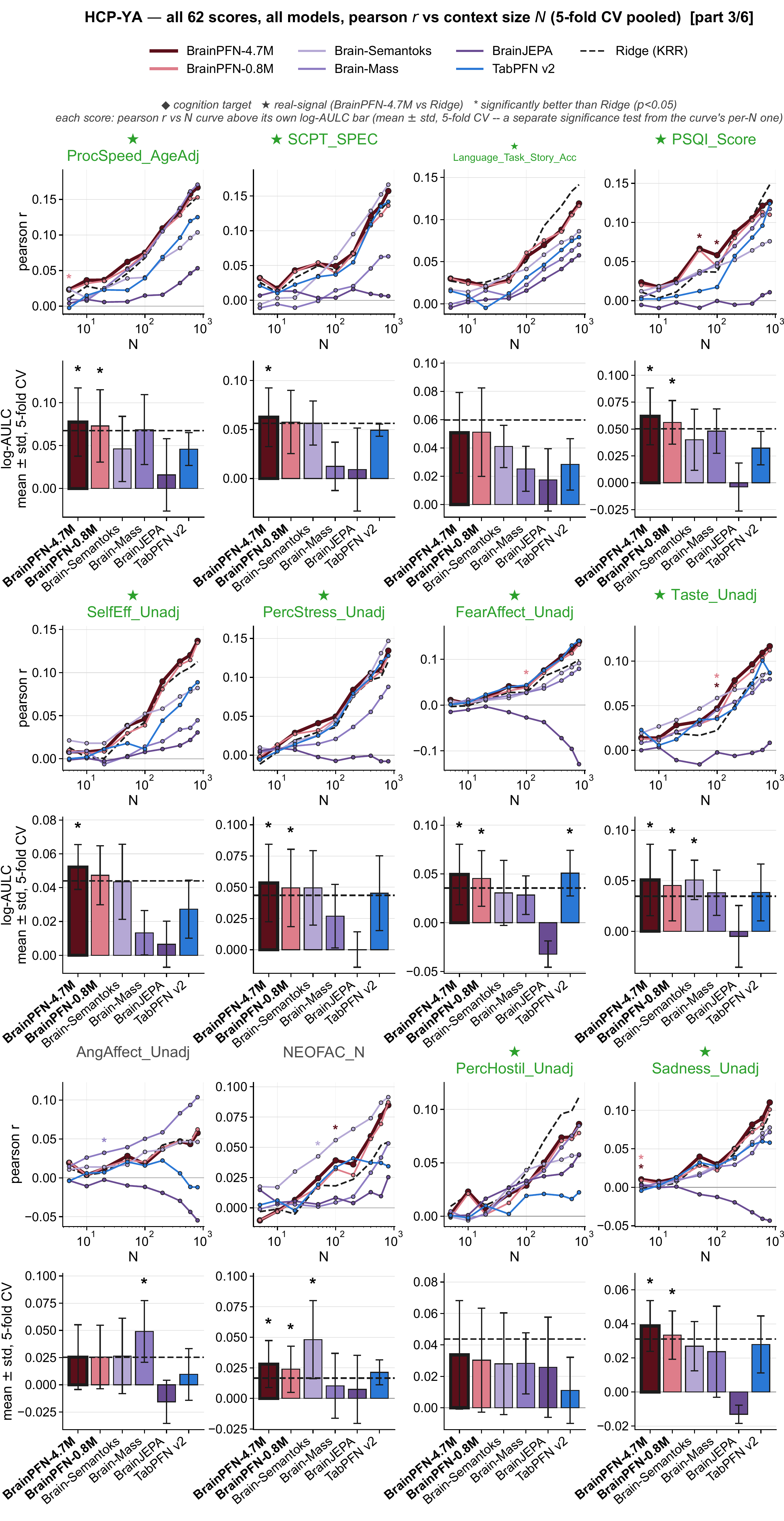}
    \caption{All 7 models vs. shared Ridge reference, all scores, HCP-YA, 5-fold CV (3/6).}
    \label{fig:all-scores-hcp-3}
\end{figure}

\begin{figure}
    \centering
    \includegraphics[width=0.84\linewidth]{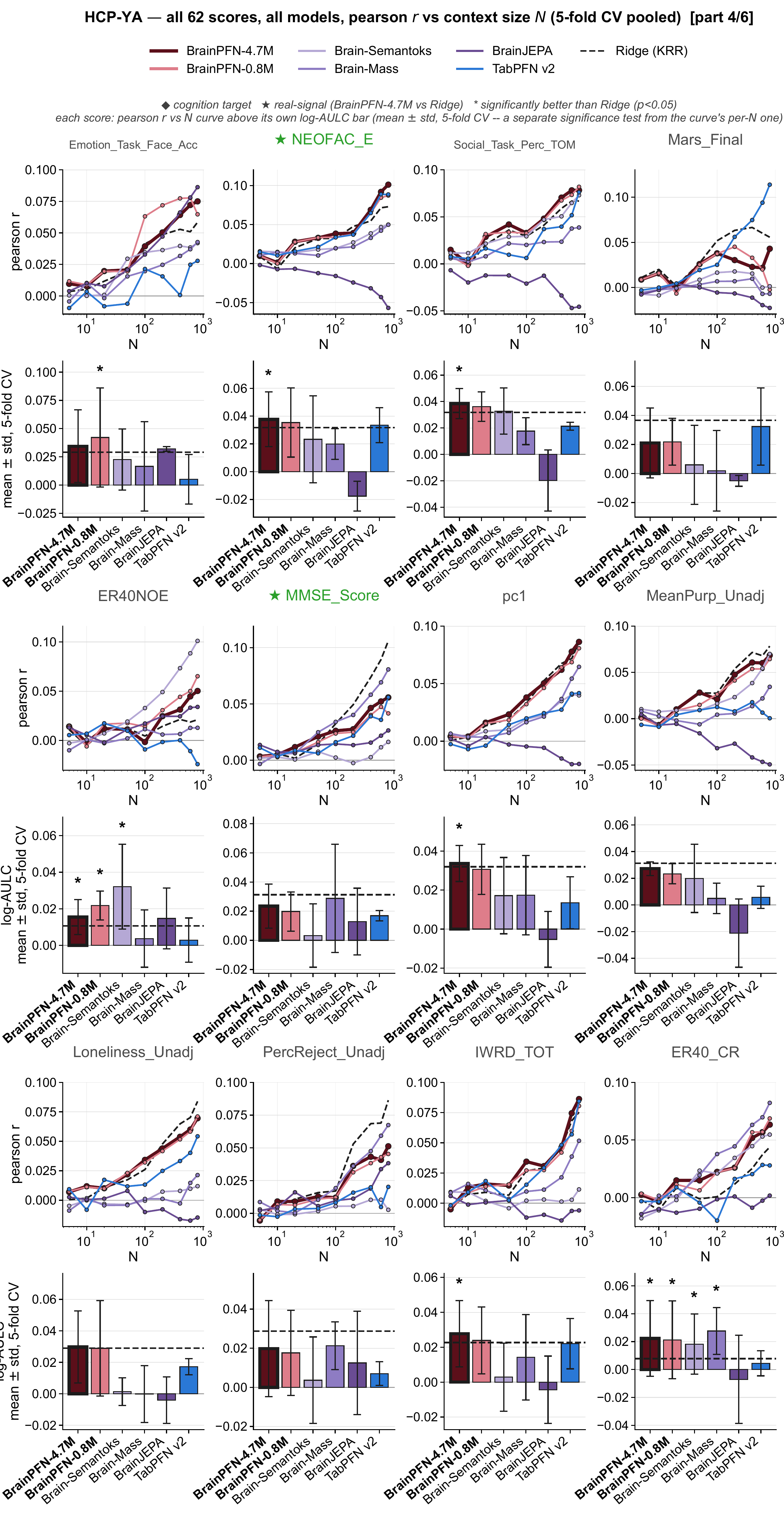}
    \caption{All 7 models vs. shared Ridge reference, all scores, HCP-YA, 5-fold CV (4/6).}
    \label{fig:all-scores-hcp-4}
\end{figure}

\begin{figure}
    \centering
    \includegraphics[width=0.84\linewidth]{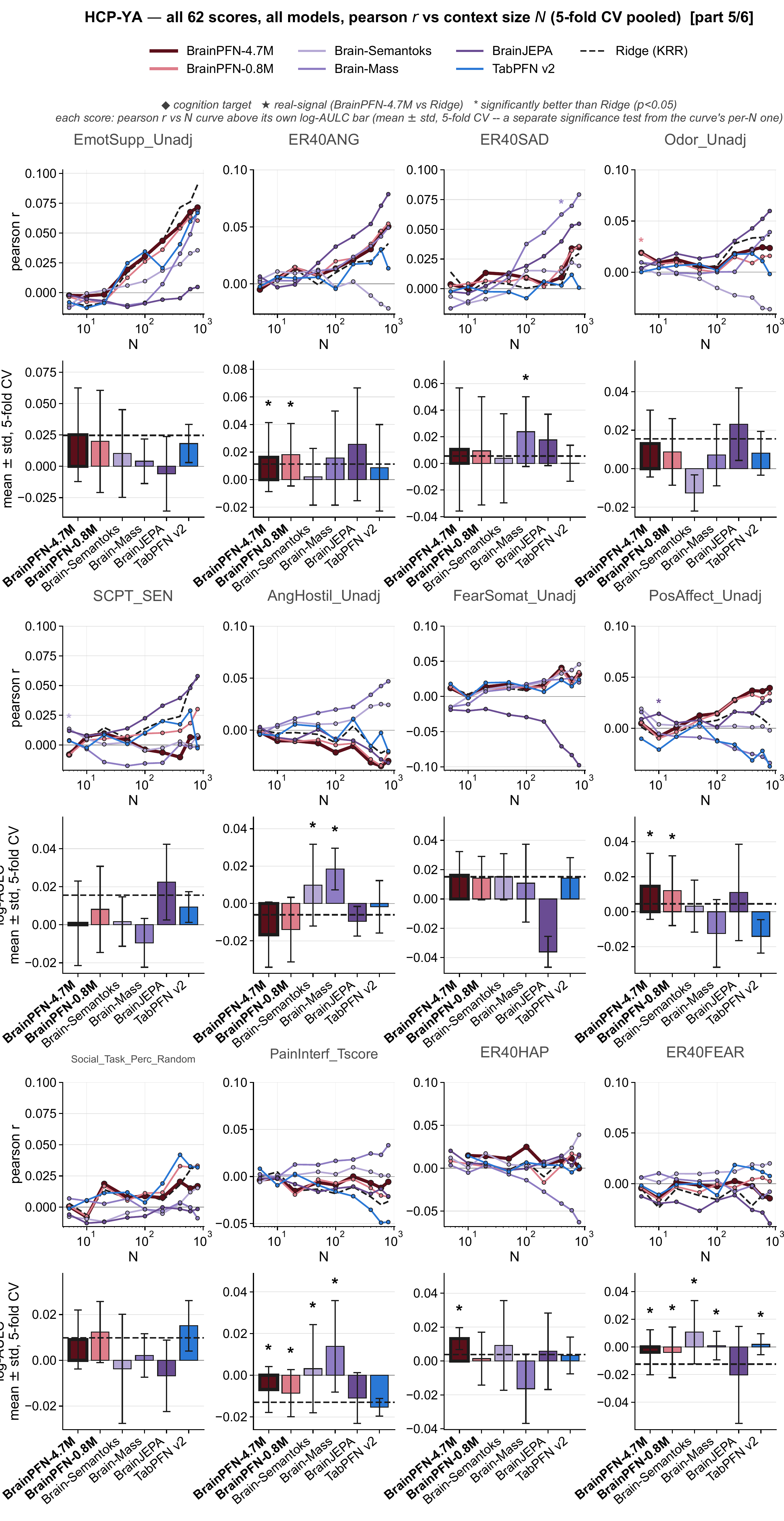}
    \caption{All 7 models vs. shared Ridge reference, all scores, HCP-YA, 5-fold CV (5/6).}
    \label{fig:all-scores-hcp-5}
\end{figure}

\begin{figure}
    \centering
    \includegraphics[width=0.84\linewidth]{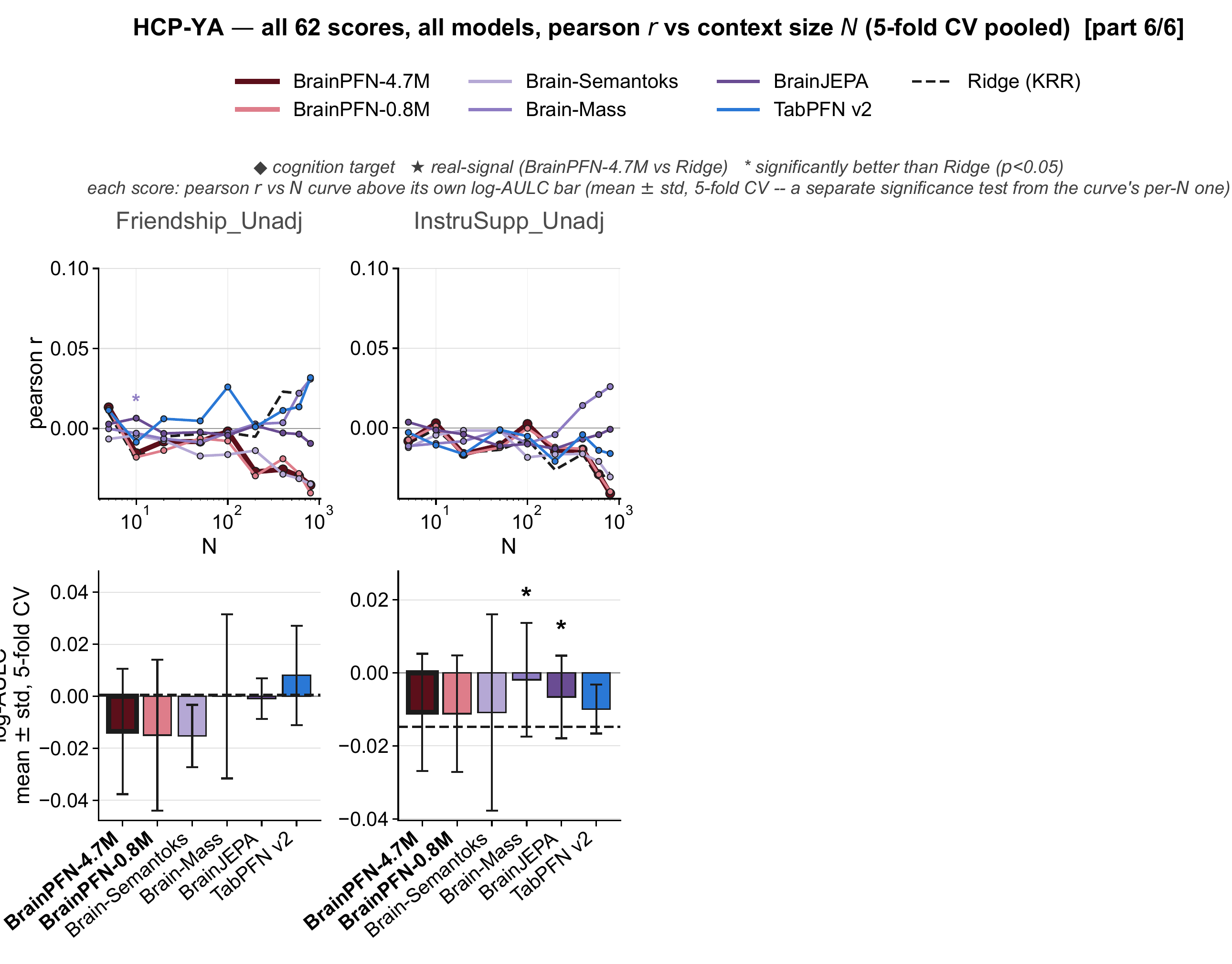}
    \caption{All 7 models vs. shared Ridge reference, all scores, HCP-YA, 5-fold CV (6/6).}
    \label{fig:all-scores-hcp-6}
\end{figure}

\begin{figure}
    \centering
    \includegraphics[width=0.84\linewidth]{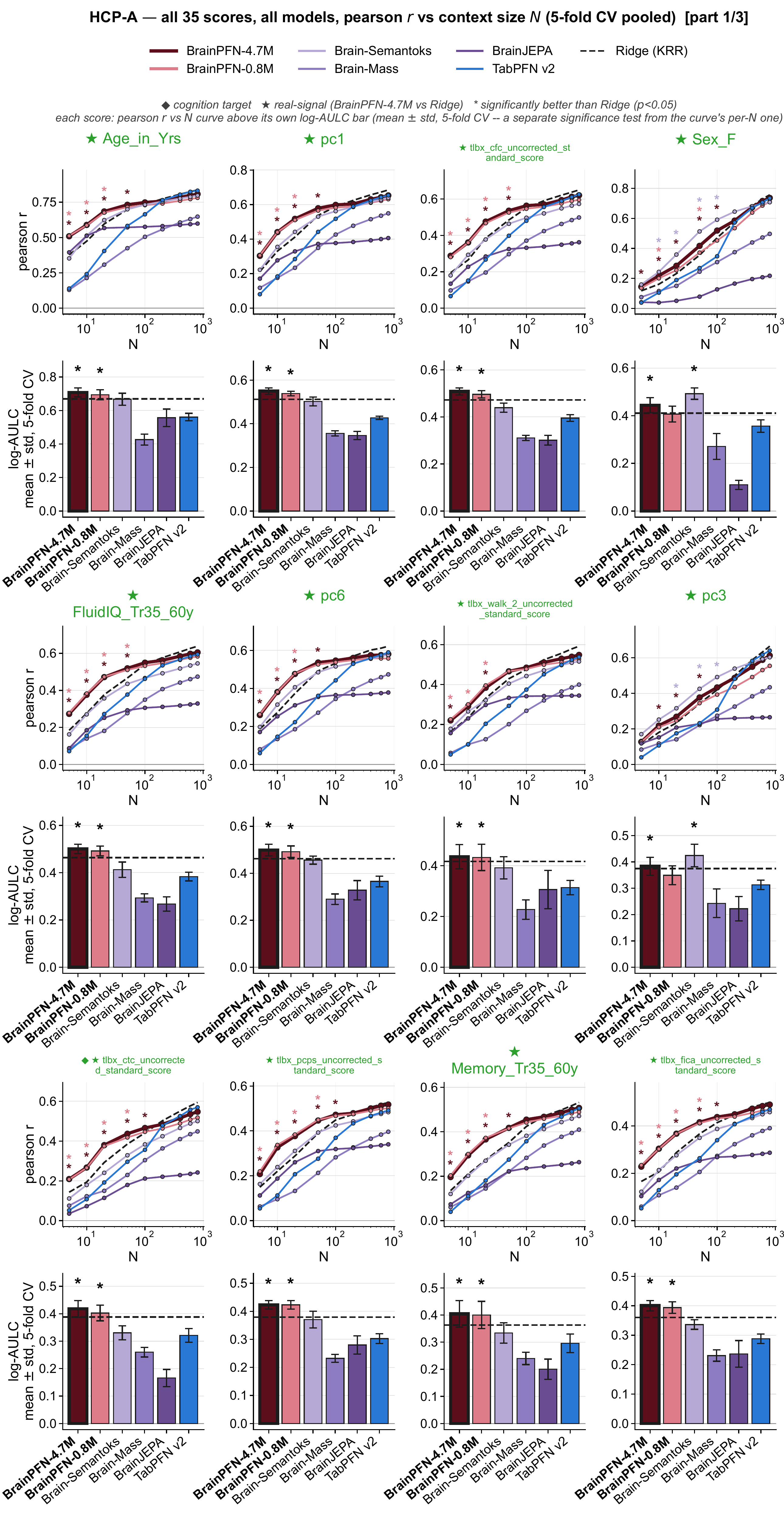}
    \caption{All 7 models vs. shared Ridge reference, all scores, HCP-A, 5-fold CV (1/3).}
    \label{fig:all-scores-hcp-aging-1}
\end{figure}

\begin{figure}
    \centering
    \includegraphics[width=0.84\linewidth]{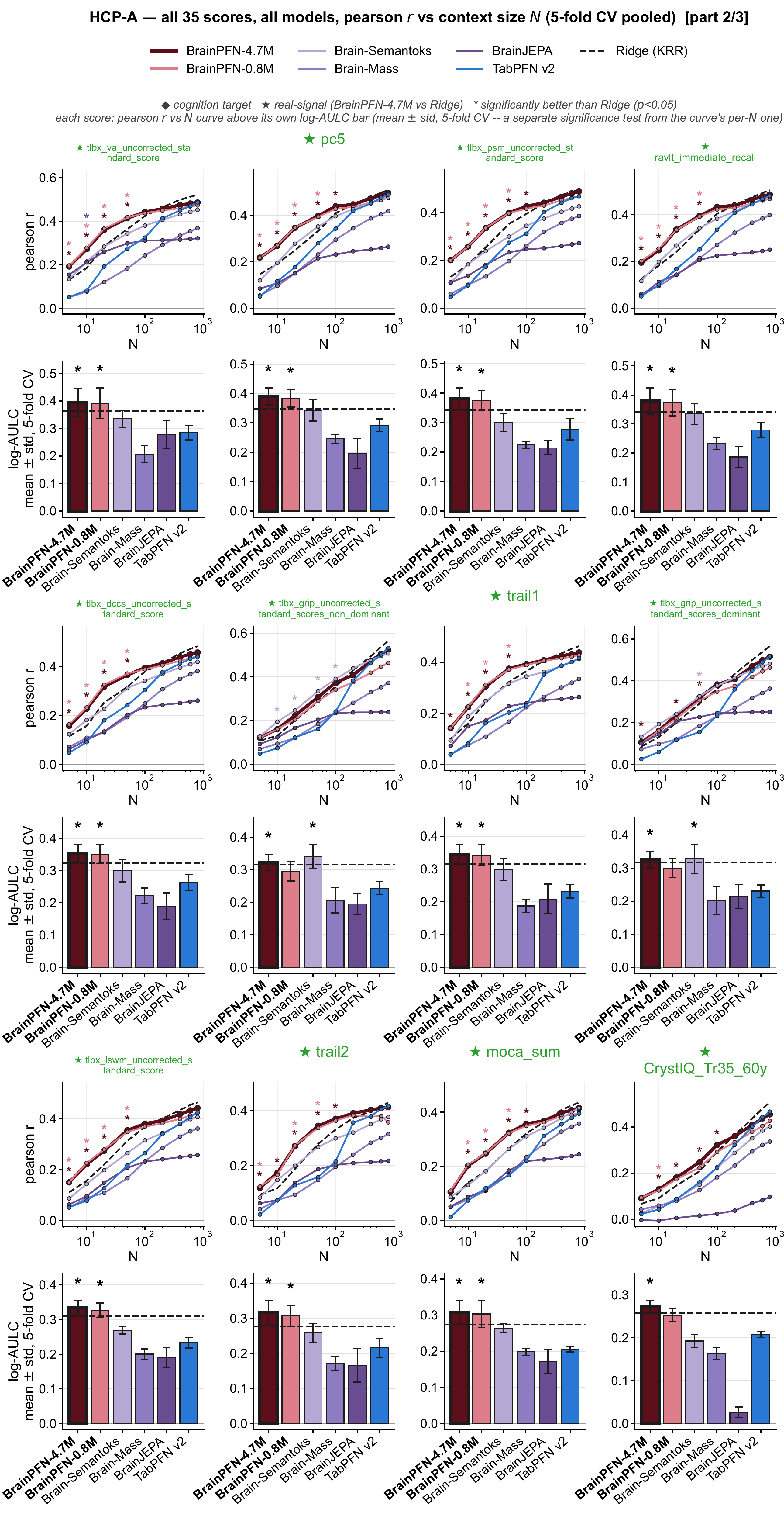}
    \caption{All 7 models vs. shared Ridge reference, all scores, HCP-A, 5-fold CV (2/3).}
    \label{fig:all-scores-hcp-aging-2}
\end{figure}

\begin{figure}
    \centering
    \includegraphics[width=0.84\linewidth]{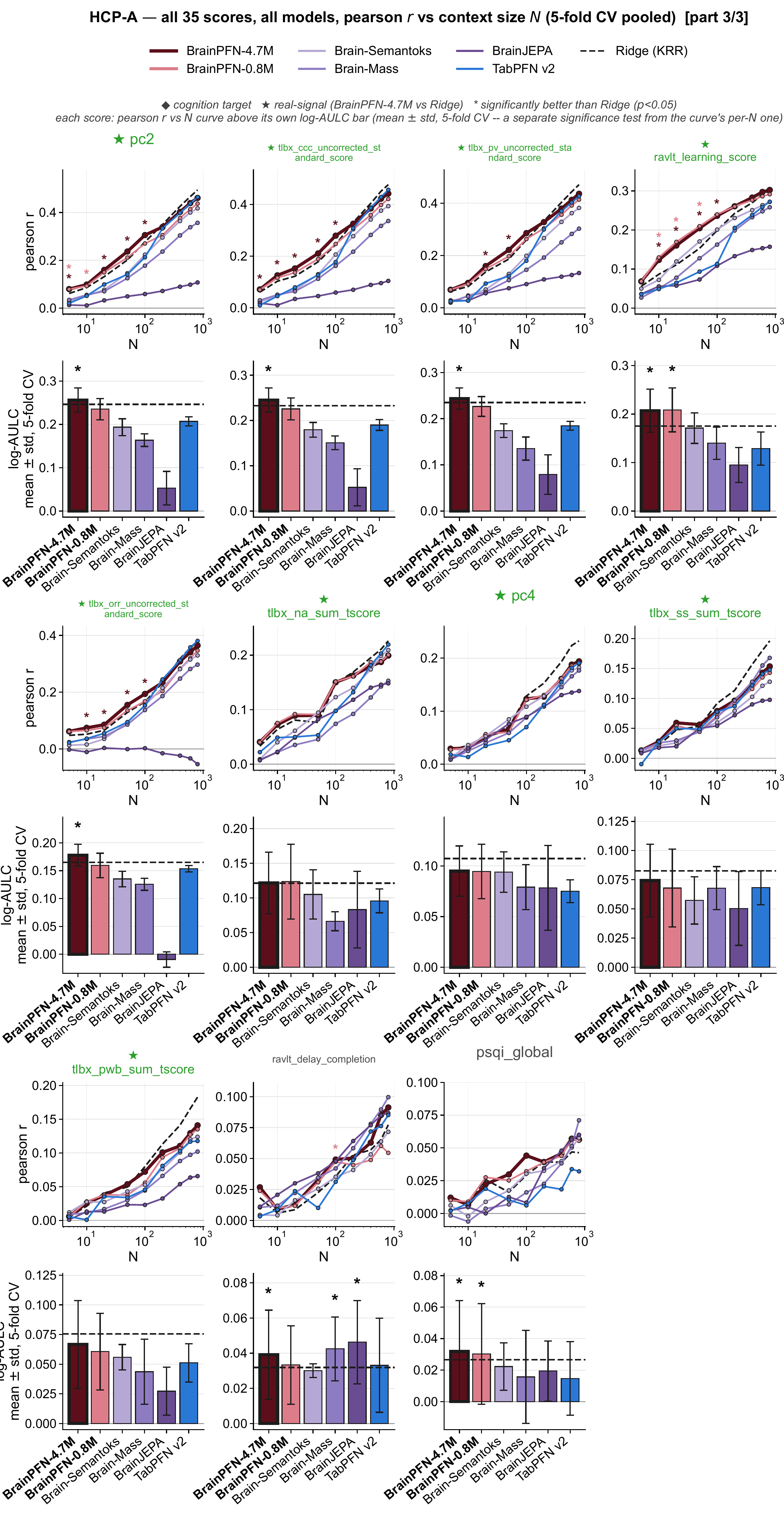}
    \caption{All 7 models vs. shared Ridge reference, all scores, HCP-A, 5-fold CV (3/3).}
    \label{fig:all-scores-hcp-aging-3}
\end{figure}

\begin{figure}
    \centering
    \includegraphics[width=0.84\linewidth]{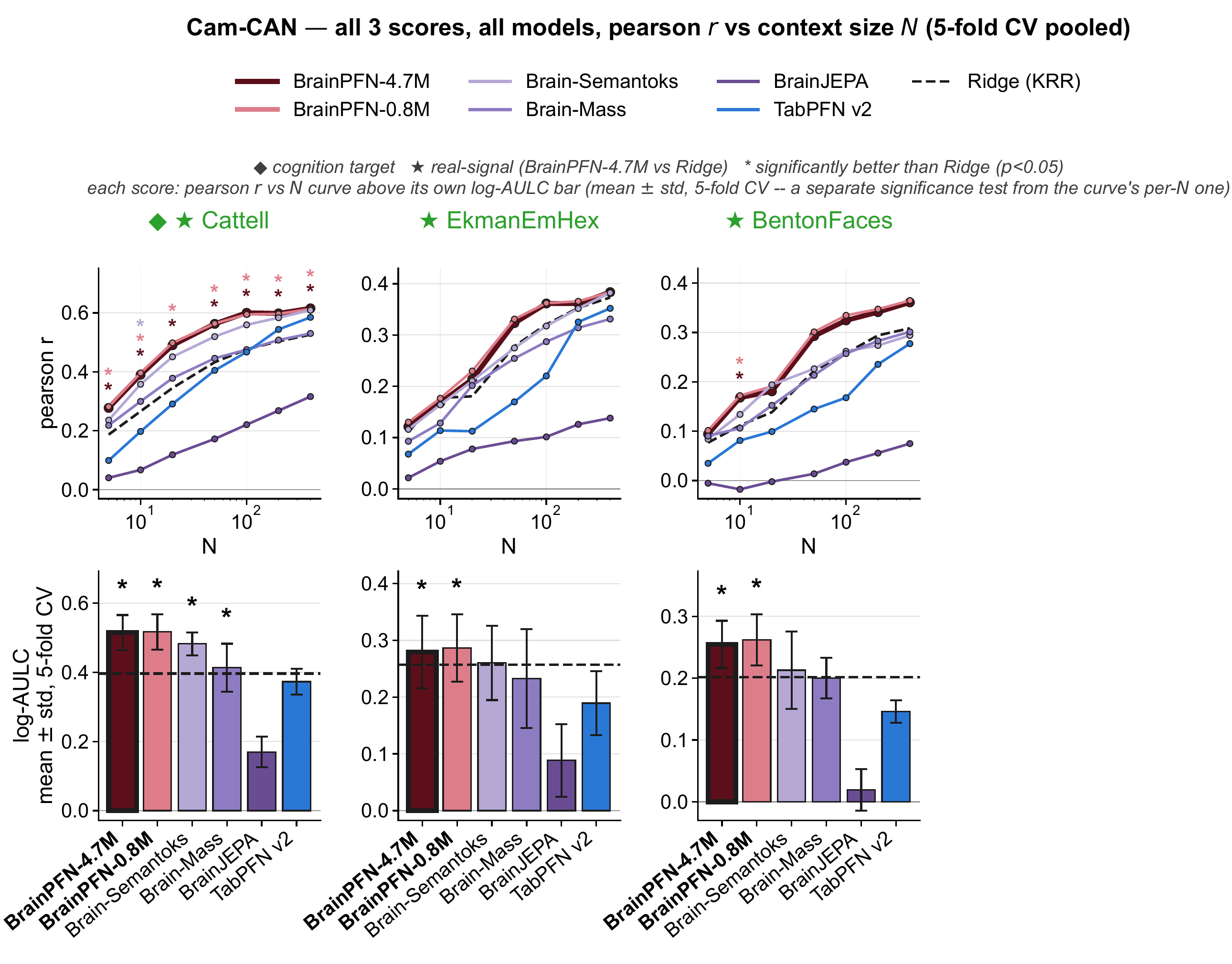}
    \caption{All 7 models vs. shared Ridge reference, all scores, Cam-CAN, 5-fold CV.}
    \label{fig:all-scores-camcan}
\end{figure}

\begin{figure}
    \centering
    \includegraphics[width=0.84\linewidth]{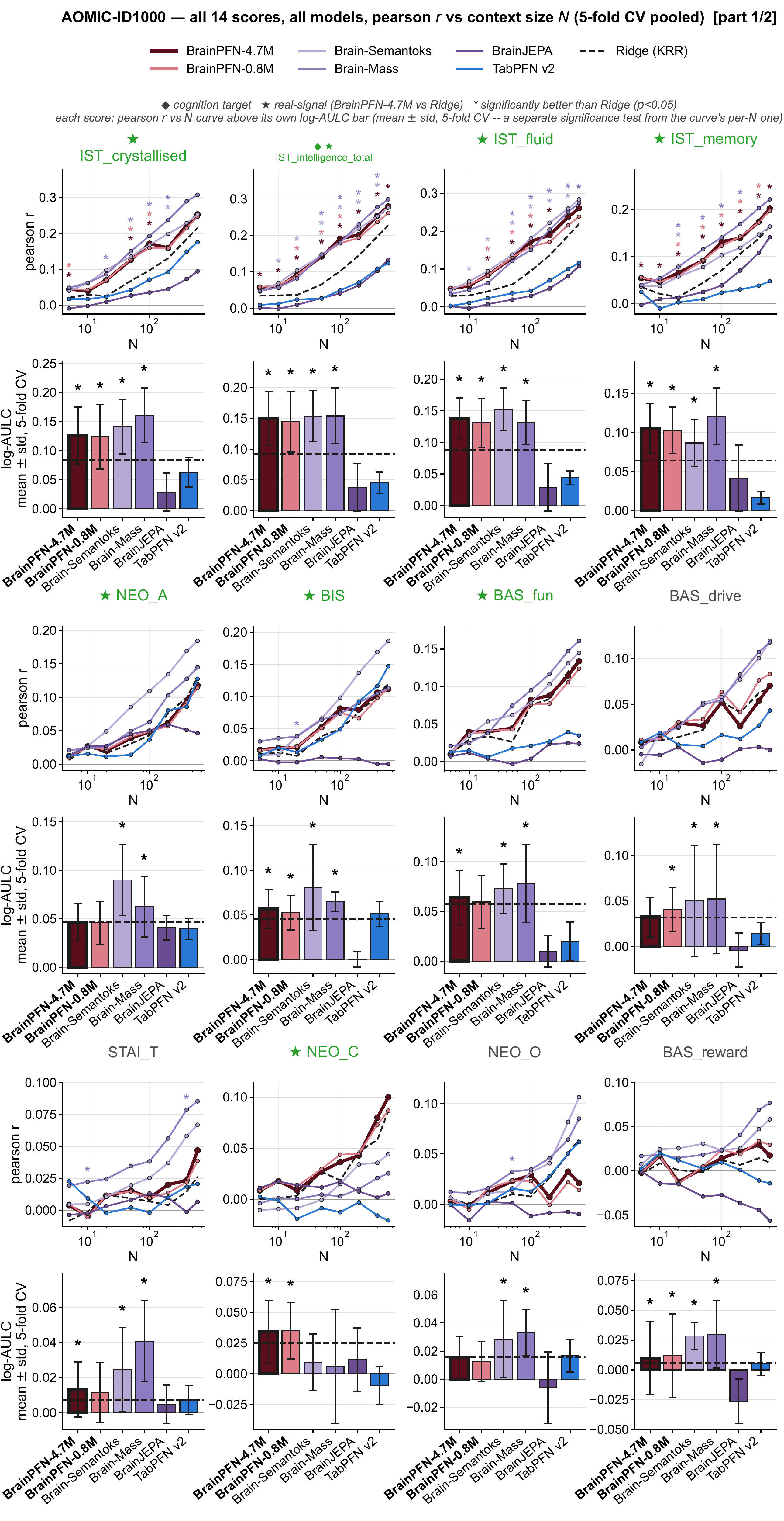}
    \caption{All 7 models vs. shared Ridge reference, all scores, AOMIC-ID1000, 5-fold CV (1/2).}
    \label{fig:all-scores-aomic-id1000-1}
\end{figure}

\begin{figure}
    \centering
    \includegraphics[width=0.84\linewidth]{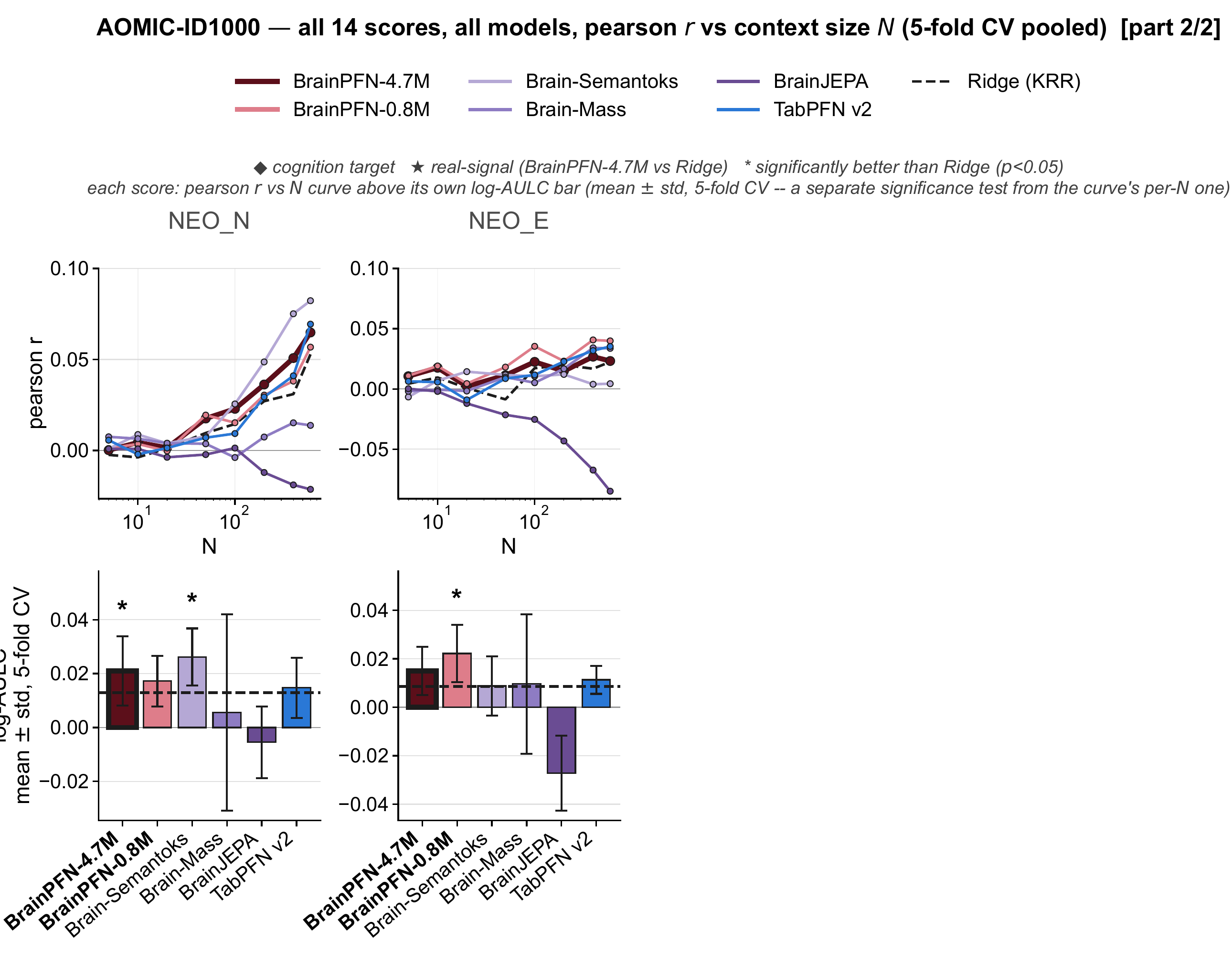}
    \caption{All 7 models vs. shared Ridge reference, all scores, AOMIC-ID1000, 5-fold CV (2/2).}
    \label{fig:all-scores-aomic-id1000-2}
\end{figure}

\clearpage

\subsection{Per-score learning curves, all 5-fold benchmarks}
\label{sec:appendix-perscore}

For completeness, this section reports the full per-score breakdown (all $\sim$60 behavioral
scores, Pearson $r$ vs. context size $N$, pooled across the 5-fold CV) underlying the
aggregate results of Figure~\ref{fig:headline-results}, for every model and every
dataset that has a complete 5-fold run -- i.e. every (model, dataset) pair that contributes
to \texttt{5fold\_model\_comparison\_avg\_curve.pdf}: BrainPFN-4.7M, BrainPFN-0.8M,
Brain-Semantoks, Brain-Mass, BrainJEPA and TabPFN v2, on HCP-YA, HCP-A, Cam-CAN,
and AOMIC-ID1000. HCP-YA (60 scores) is split across 3
pages and HCP-A across 2; every other panel fits on a single page.

\subsubsection{BrainPFN-4.7M}

\begin{figure}[h]
    \centering
    \includegraphics[width=\linewidth]{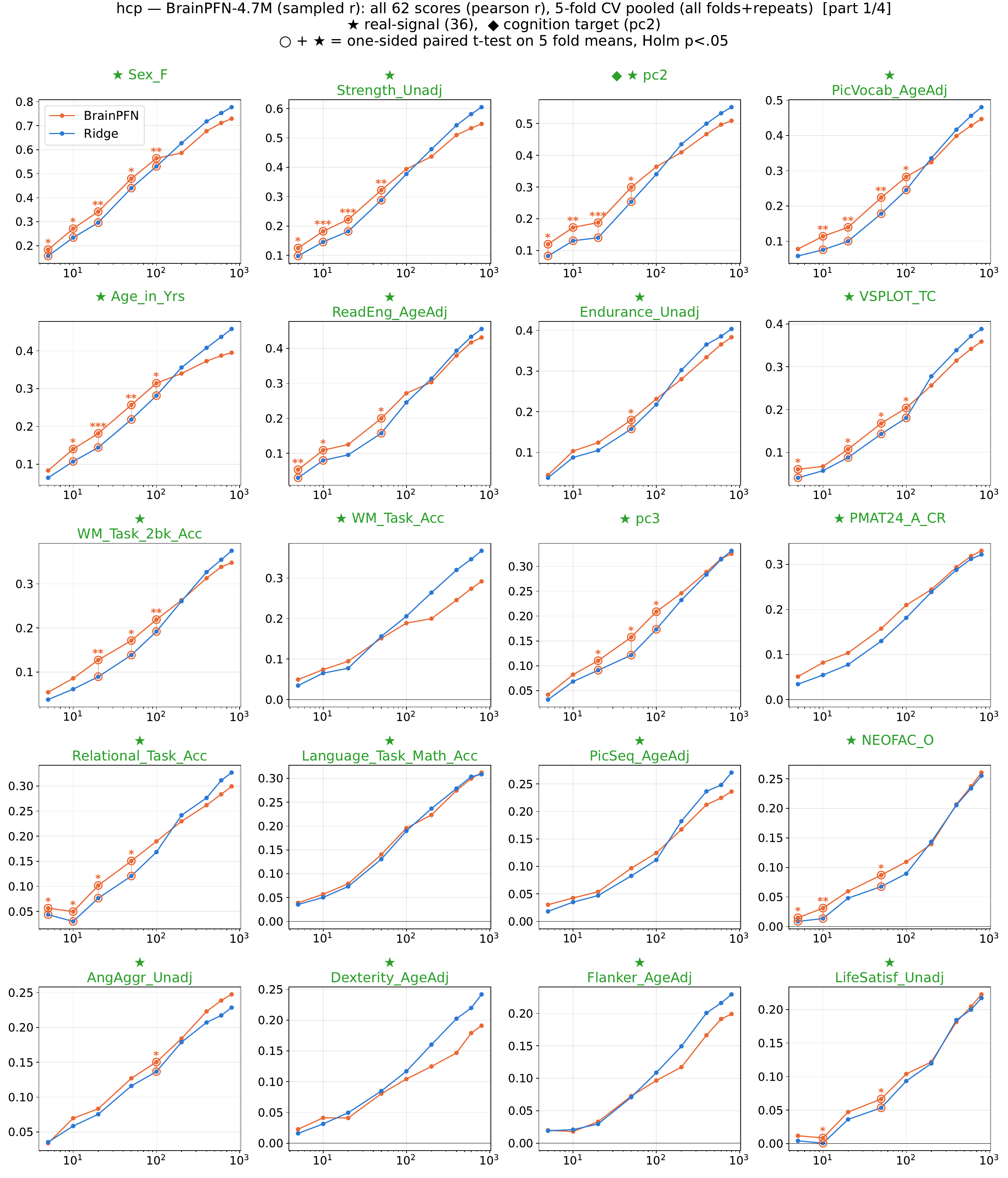}
    \caption{BrainPFN-4.7M, all scores, HCP-YA, 5-fold CV (1/3).}
    \label{fig:perscore-brainpfn-hcp-1}
\end{figure}

\begin{figure}
    \centering
    \includegraphics[width=\linewidth]{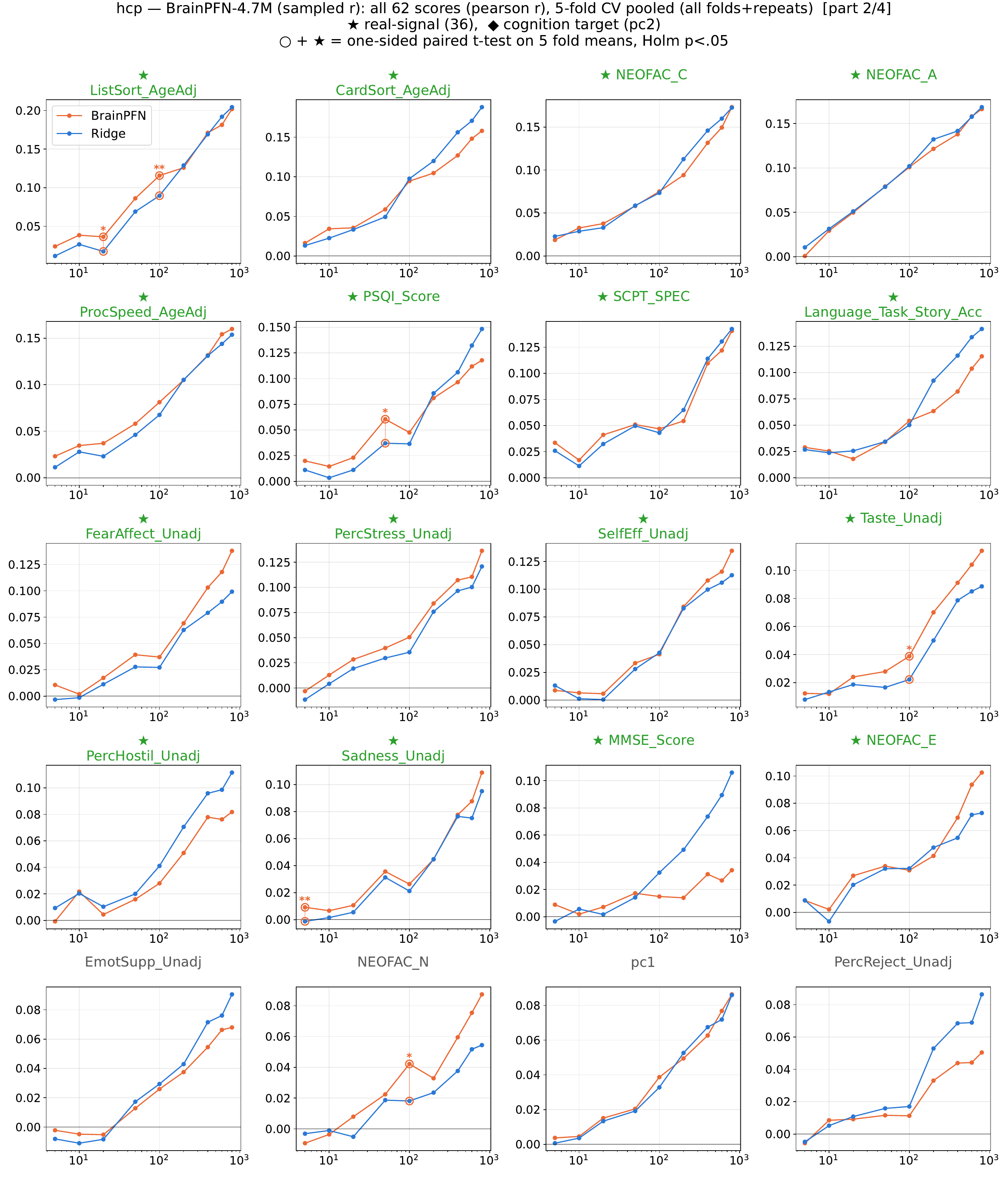}
    \caption{BrainPFN-4.7M, all scores, HCP-YA, 5-fold CV (2/3).}
    \label{fig:perscore-brainpfn-hcp-2}
\end{figure}

\begin{figure}
    \centering
    \includegraphics[width=\linewidth]{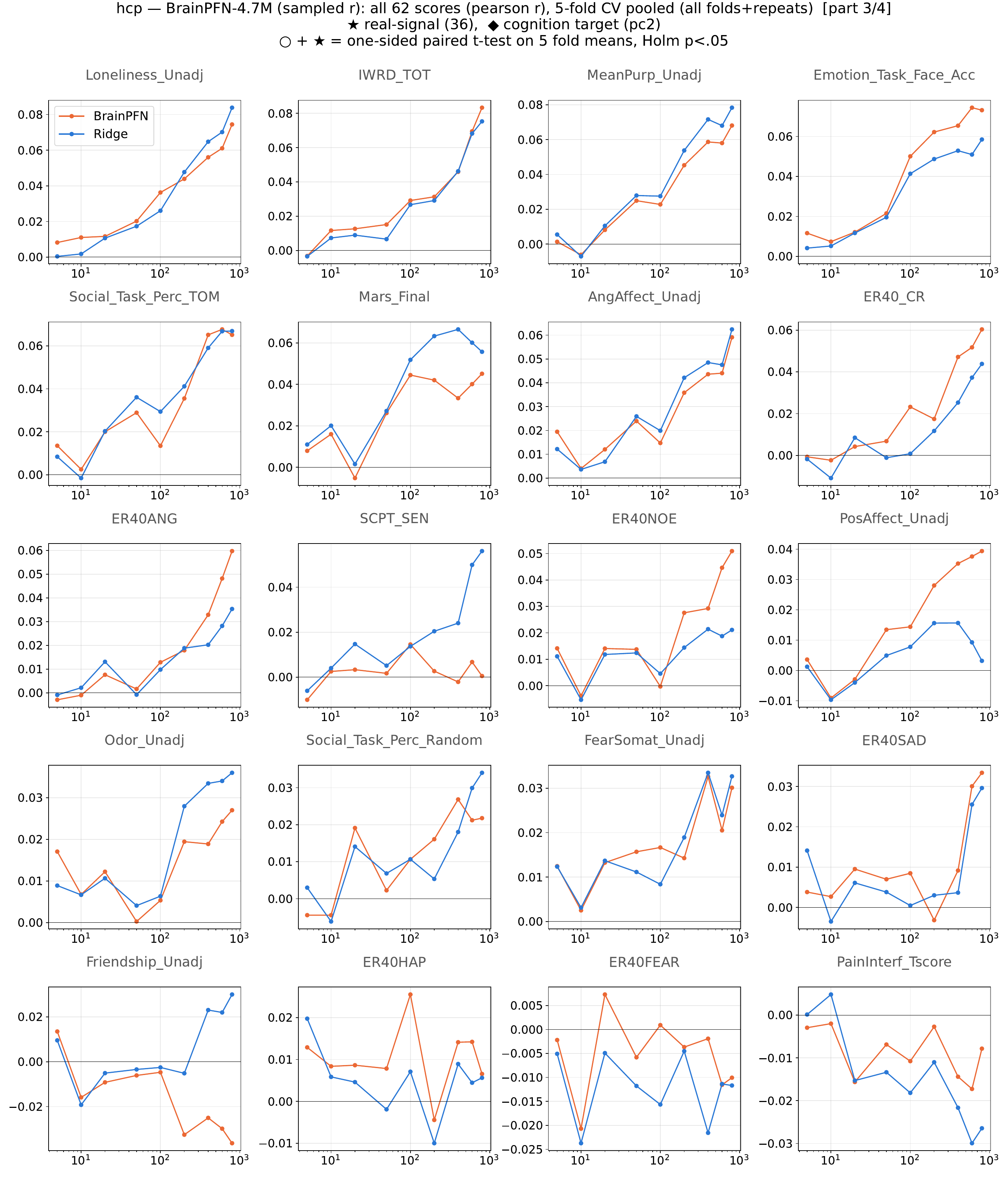}
    \caption{BrainPFN-4.7M, all scores, HCP-YA, 5-fold CV (3/3).}
    \label{fig:perscore-brainpfn-hcp-3}
\end{figure}

\begin{figure}
    \centering
    \includegraphics[width=\linewidth]{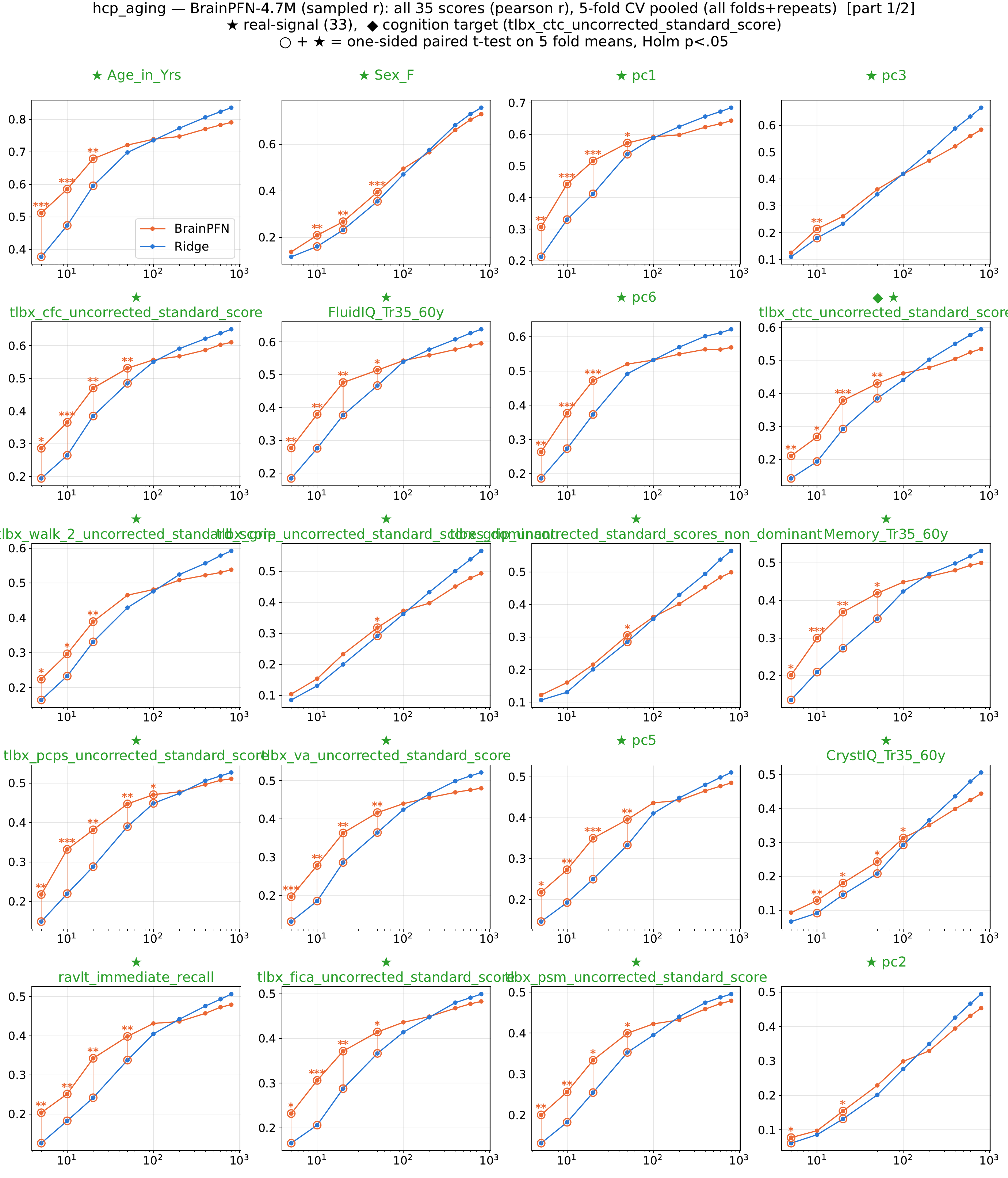}
    \caption{BrainPFN-4.7M, all scores, HCP-A, 5-fold CV (1/2).}
    \label{fig:perscore-brainpfn-hcp-aging-1}
\end{figure}

\begin{figure}
    \centering
    \includegraphics[width=\linewidth]{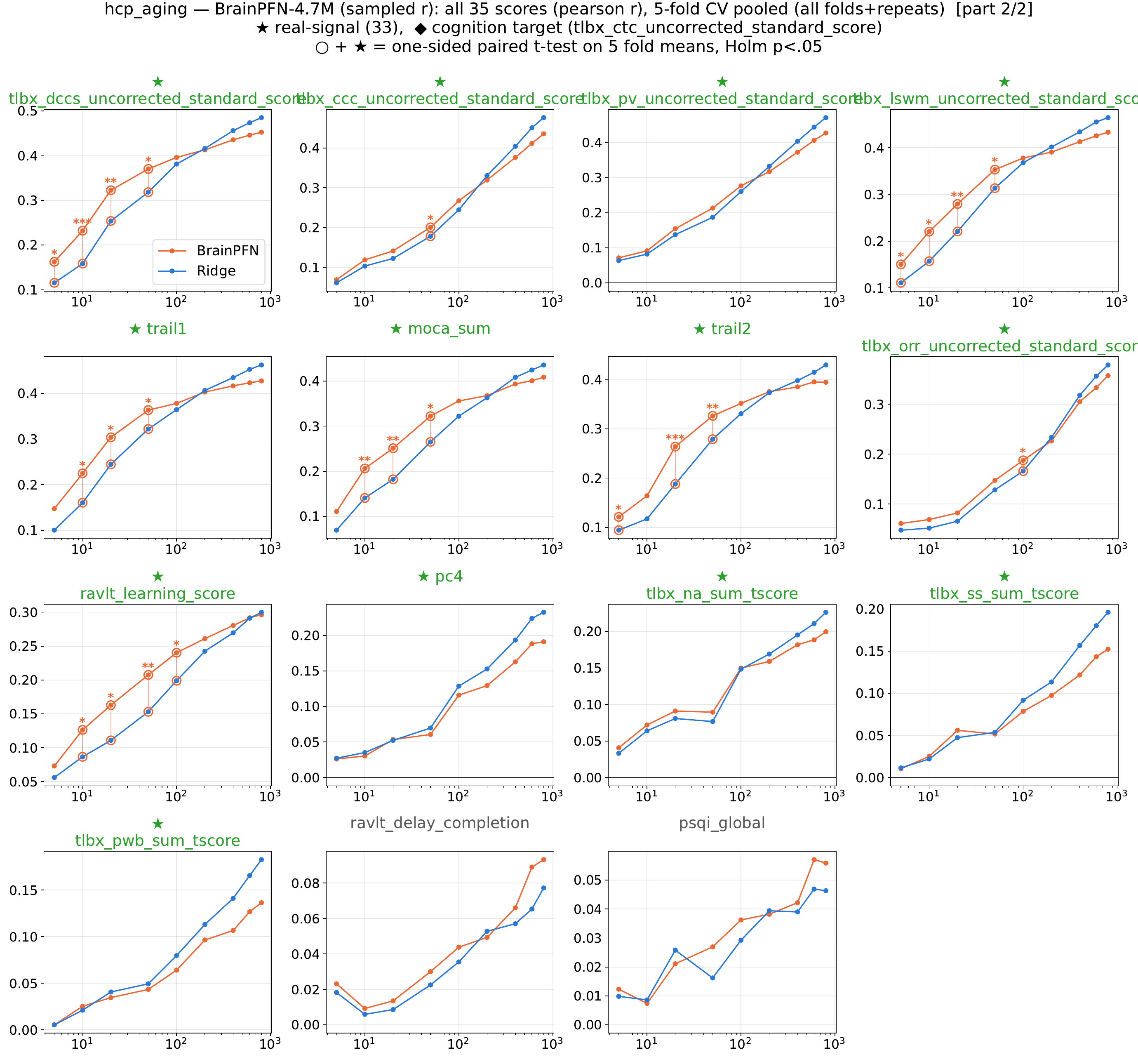}
    \caption{BrainPFN-4.7M, all scores, HCP-A, 5-fold CV (2/2).}
    \label{fig:perscore-brainpfn-hcp-aging-2}
\end{figure}

\begin{figure}
    \centering
    \includegraphics[width=\linewidth]{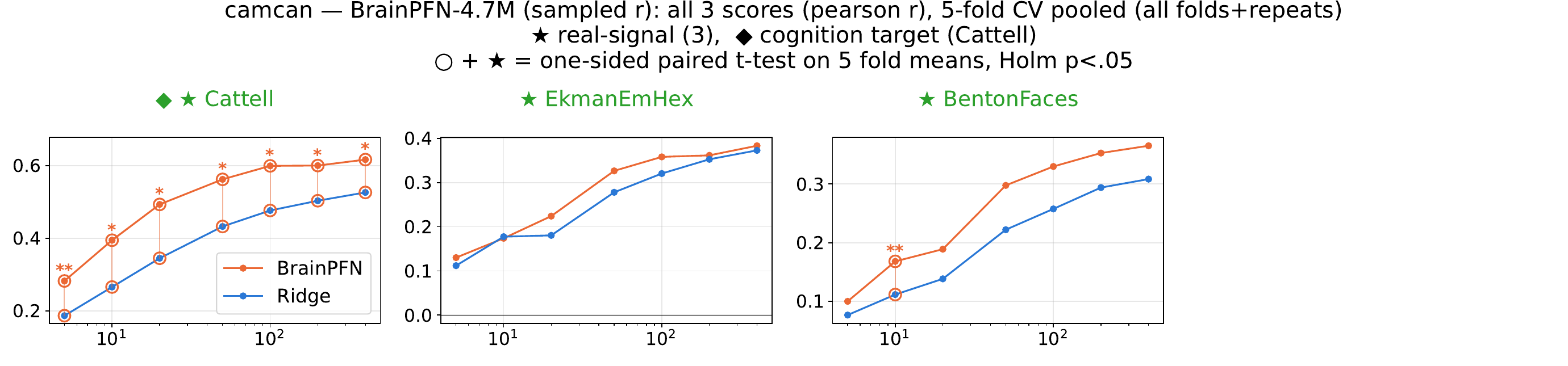}
    \caption{BrainPFN-4.7M, all scores, Cam-CAN, 5-fold CV.}
    \label{fig:perscore-brainpfn-camcan}
\end{figure}

\begin{figure}
    \centering
    \includegraphics[width=\linewidth]{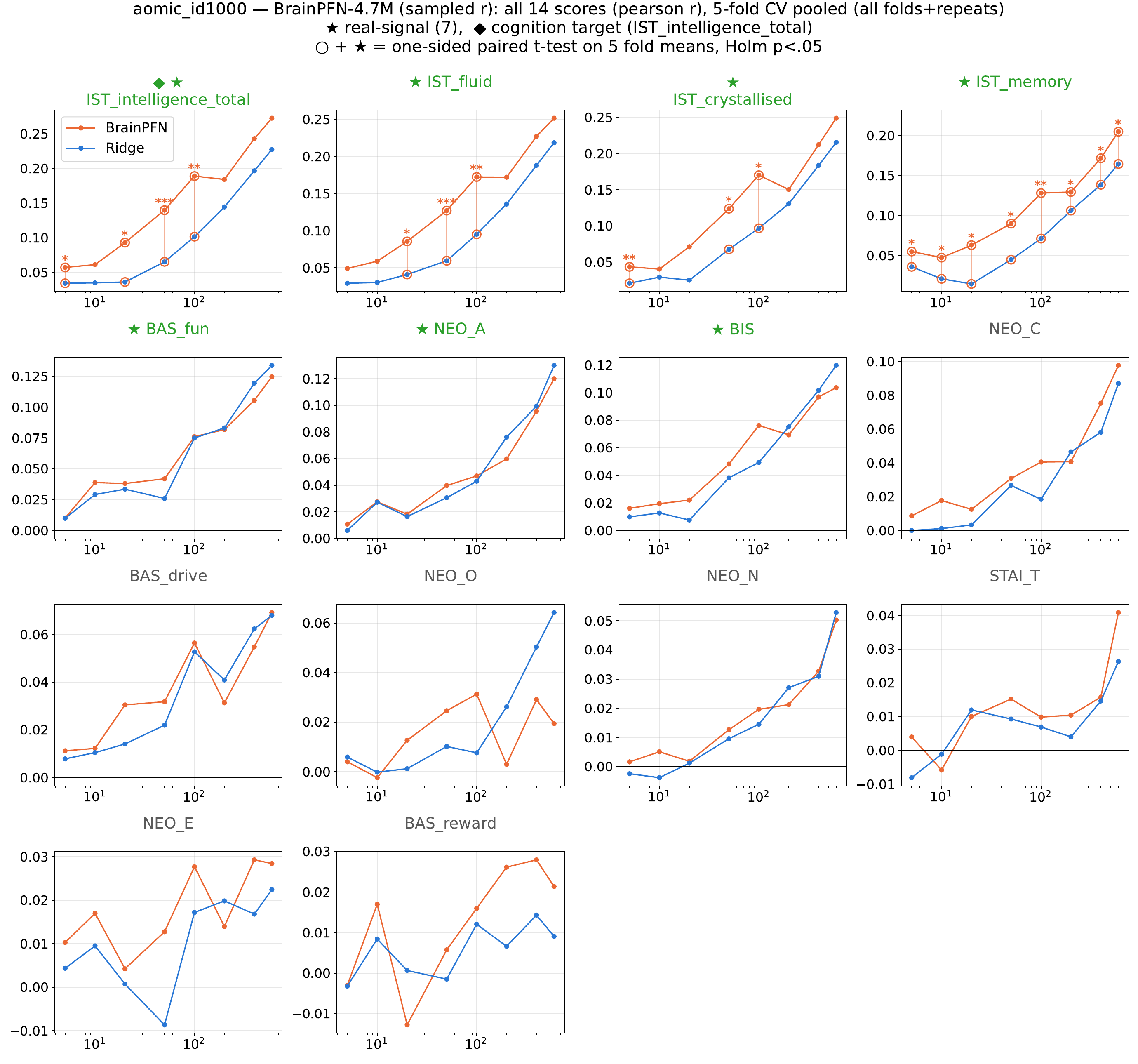}
    \caption{BrainPFN-4.7M, all scores, AOMIC-ID1000, 5-fold CV.}
    \label{fig:perscore-brainpfn-aomic-id1000}
\end{figure}

\clearpage

\subsubsection{BrainPFN-0.8M}

\begin{figure}[h]
    \centering
    \includegraphics[width=\linewidth]{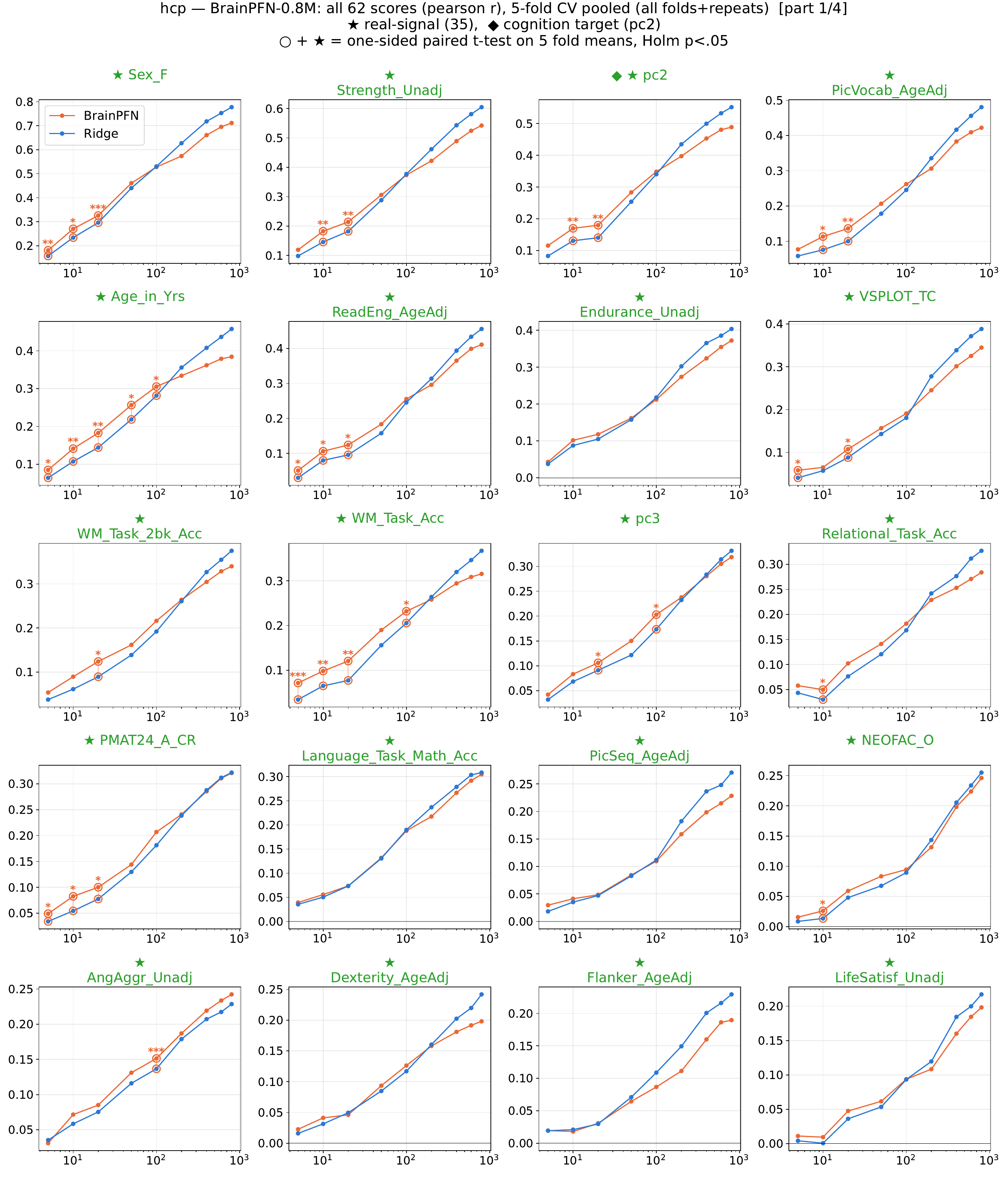}
    \caption{BrainPFN-0.8M, all scores, HCP-YA, 5-fold CV (1/3).}
    \label{fig:perscore-brainpfn08-hcp-1}
\end{figure}

\begin{figure}
    \centering
    \includegraphics[width=\linewidth]{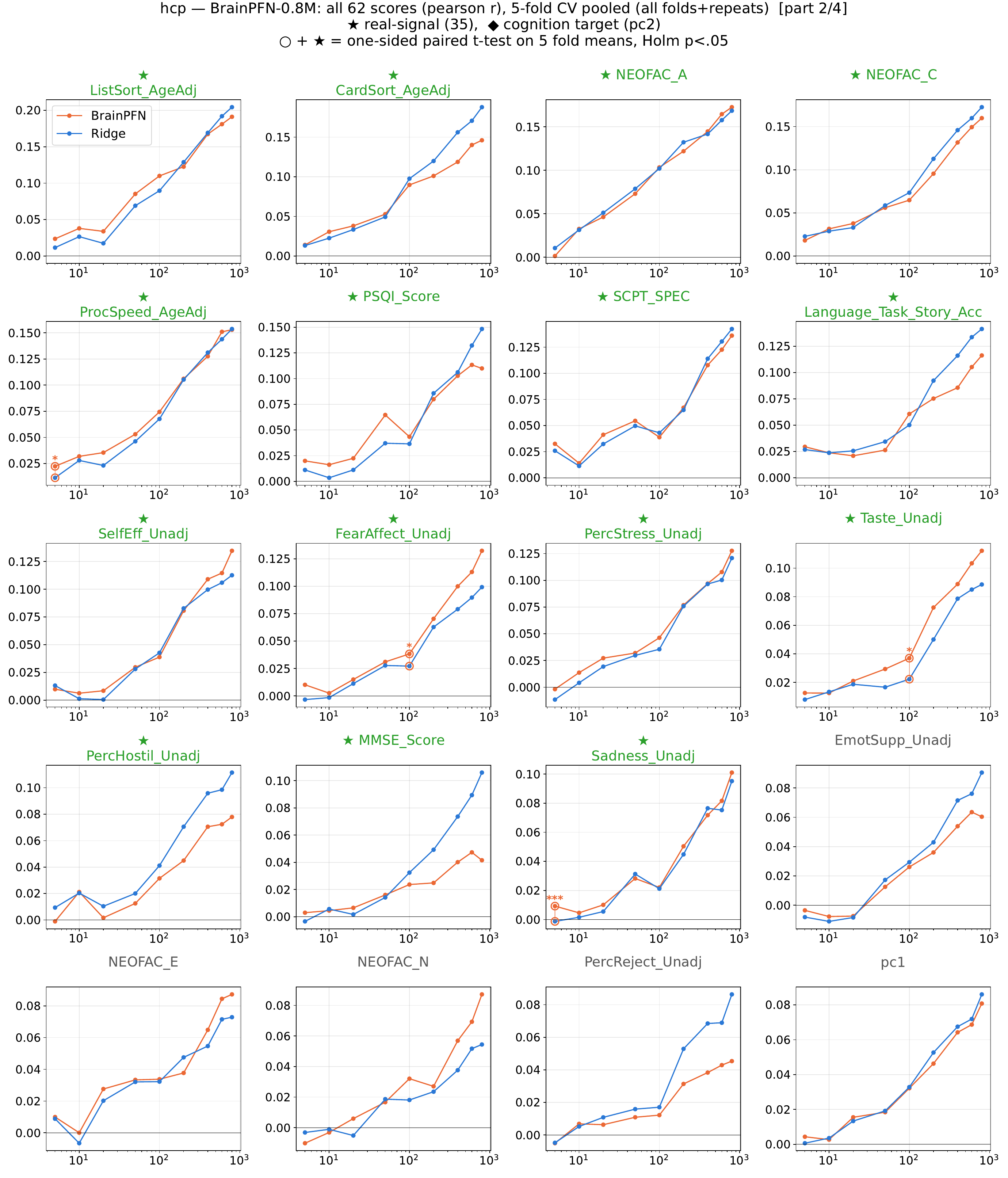}
    \caption{BrainPFN-0.8M, all scores, HCP-YA, 5-fold CV (2/3).}
    \label{fig:perscore-brainpfn08-hcp-2}
\end{figure}

\begin{figure}
    \centering
    \includegraphics[width=\linewidth]{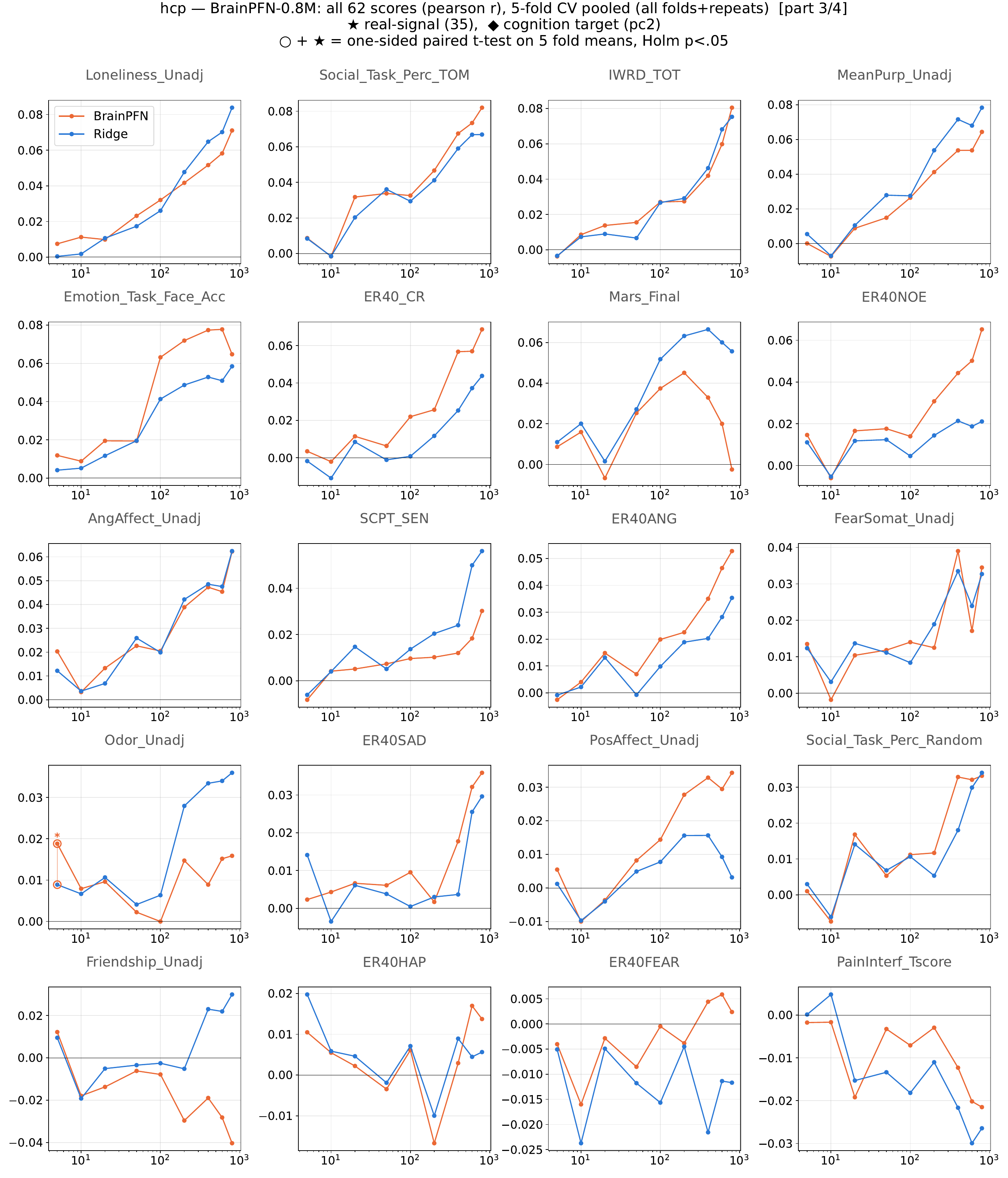}
    \caption{BrainPFN-0.8M, all scores, HCP-YA, 5-fold CV (3/3).}
    \label{fig:perscore-brainpfn08-hcp-3}
\end{figure}

\begin{figure}
    \centering
    \includegraphics[width=\linewidth]{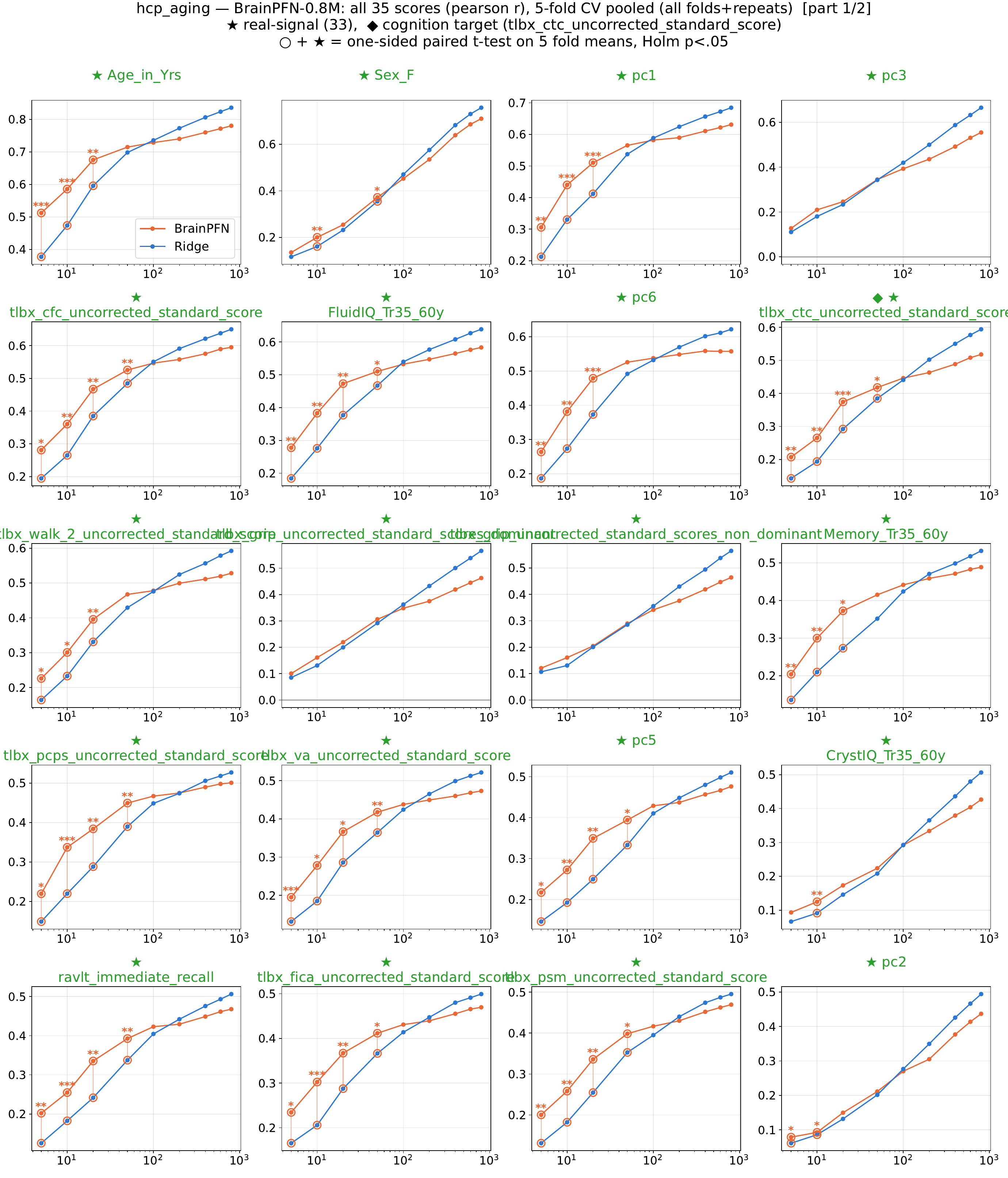}
    \caption{BrainPFN-0.8M, all scores, HCP-A, 5-fold CV (1/2).}
    \label{fig:perscore-brainpfn08-hcp-aging-1}
\end{figure}

\begin{figure}
    \centering
    \includegraphics[width=\linewidth]{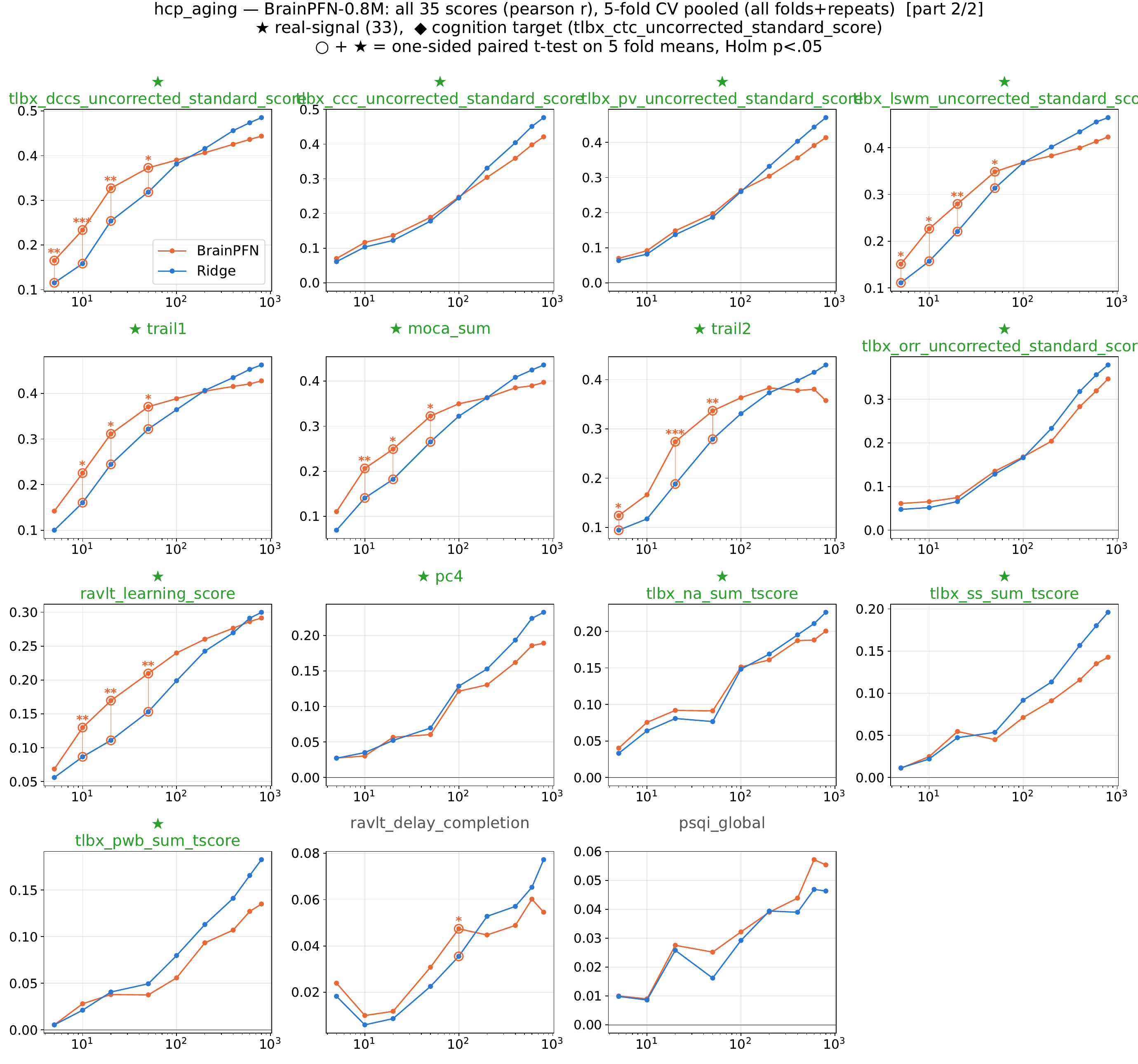}
    \caption{BrainPFN-0.8M, all scores, HCP-A, 5-fold CV (2/2).}
    \label{fig:perscore-brainpfn08-hcp-aging-2}
\end{figure}

\begin{figure}
    \centering
    \includegraphics[width=\linewidth]{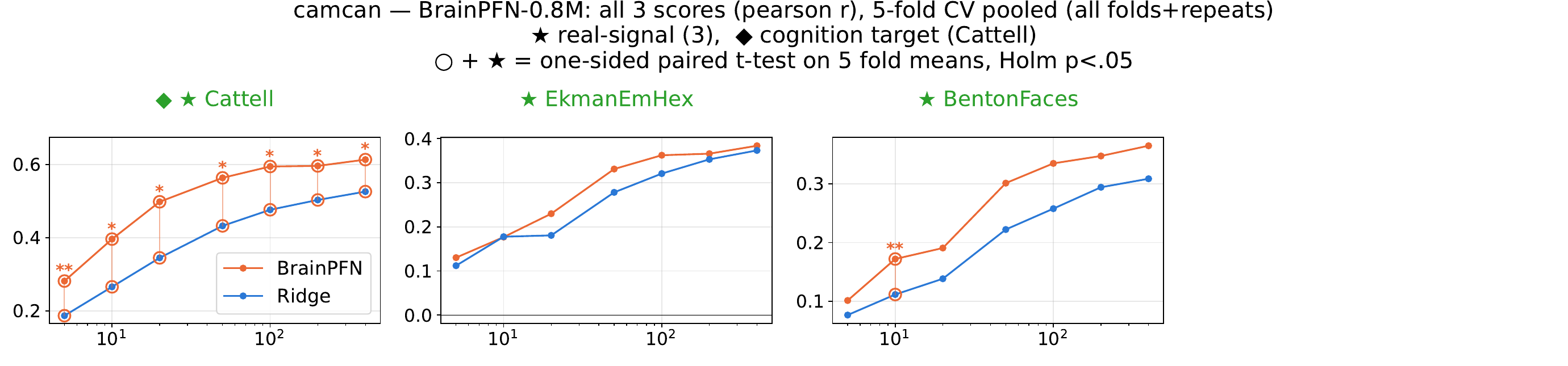}
    \caption{BrainPFN-0.8M, all scores, Cam-CAN, 5-fold CV.}
    \label{fig:perscore-brainpfn08-camcan}
\end{figure}

\begin{figure}
    \centering
    \includegraphics[width=\linewidth]{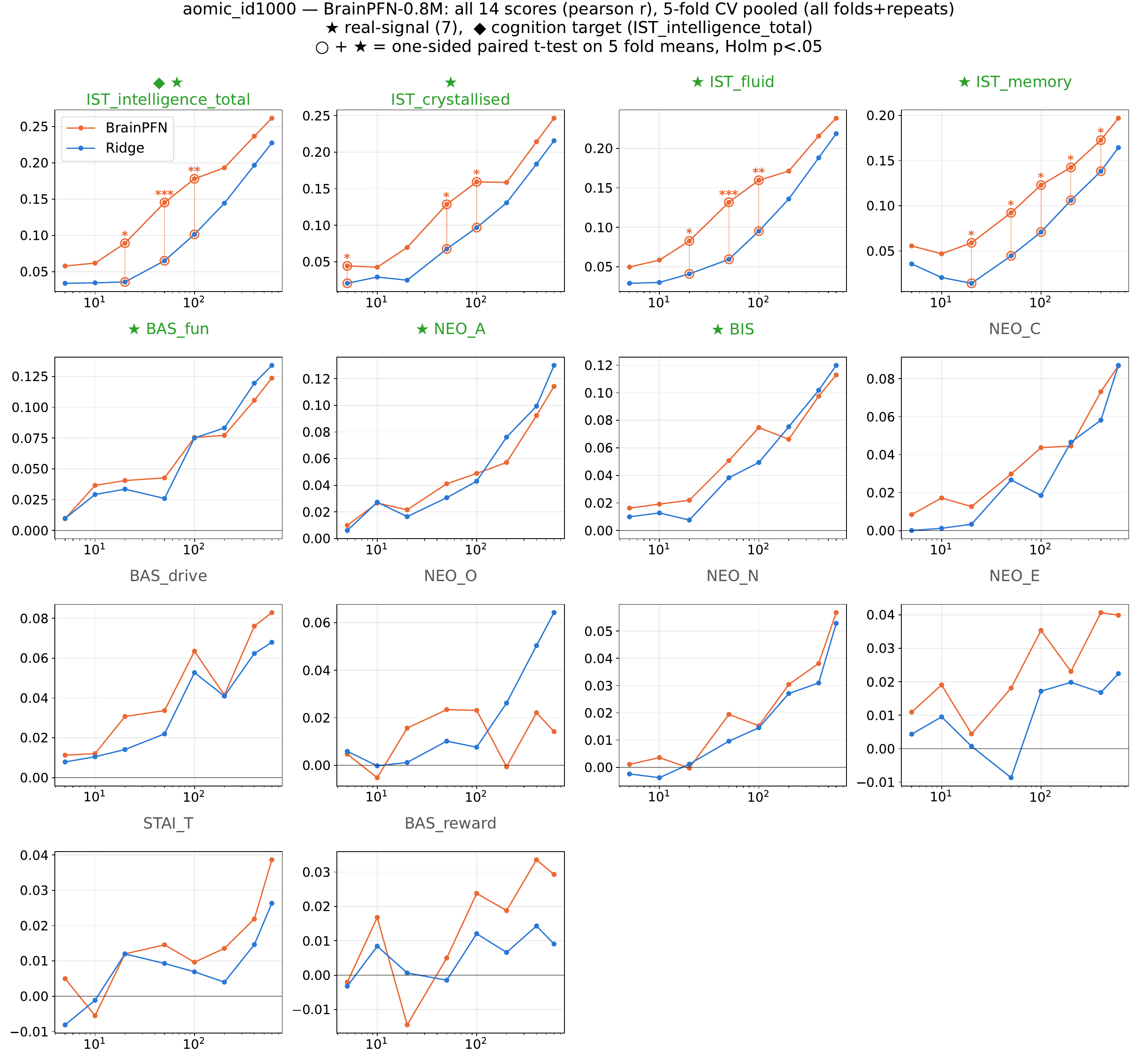}
    \caption{BrainPFN-0.8M, all scores, AOMIC-ID1000, 5-fold CV.}
    \label{fig:perscore-brainpfn08-aomic-id1000}
\end{figure}

\clearpage

\subsubsection{Brain-Semantoks}

\begin{figure}[h]
    \centering
    \includegraphics[width=\linewidth]{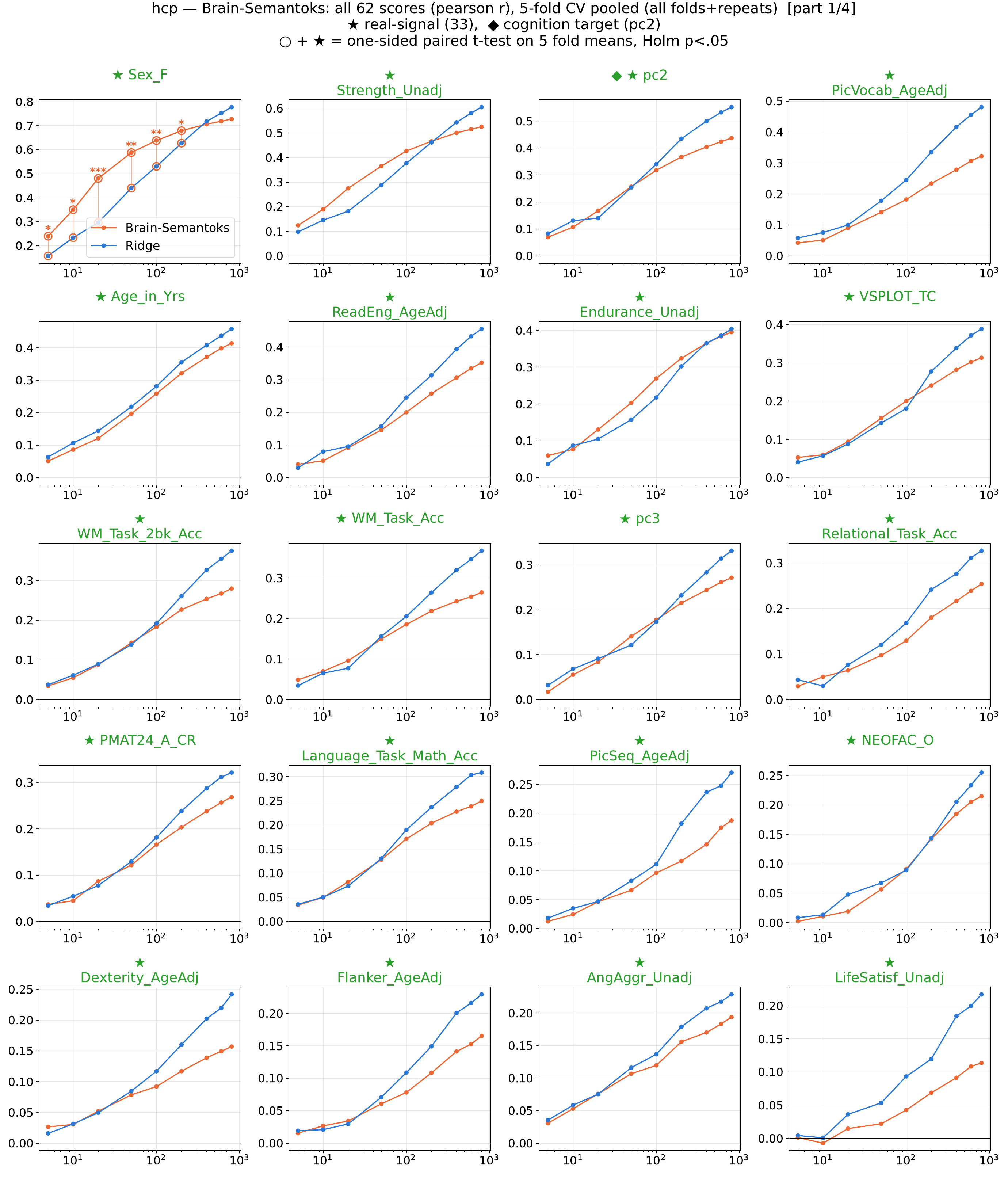}
    \caption{Brain-Semantoks, all scores, HCP-YA, 5-fold CV (1/3).}
    \label{fig:perscore-brainsemantoks-hcp-1}
\end{figure}

\begin{figure}
    \centering
    \includegraphics[width=\linewidth]{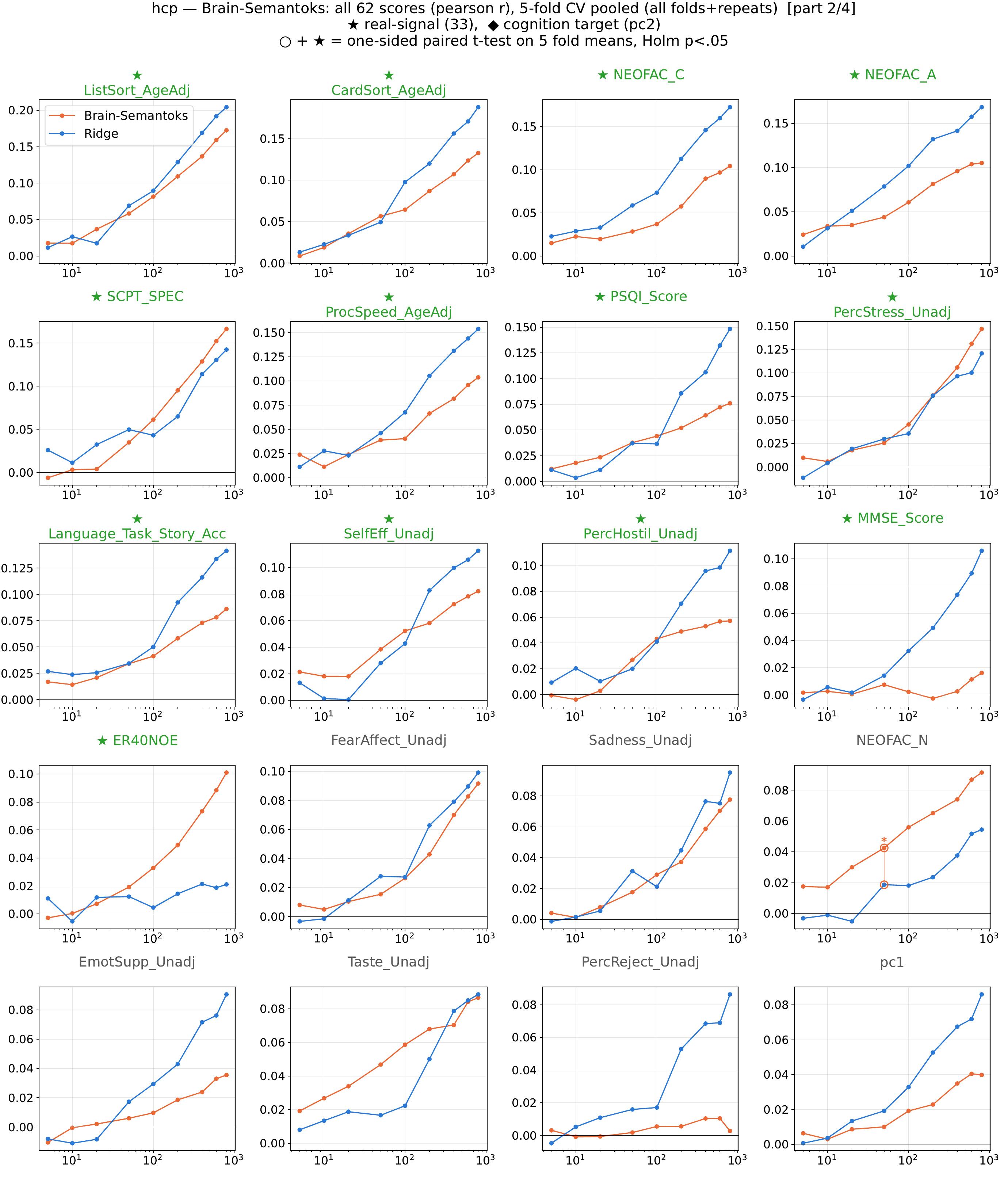}
    \caption{Brain-Semantoks, all scores, HCP-YA, 5-fold CV (2/3).}
    \label{fig:perscore-brainsemantoks-hcp-2}
\end{figure}

\begin{figure}
    \centering
    \includegraphics[width=\linewidth]{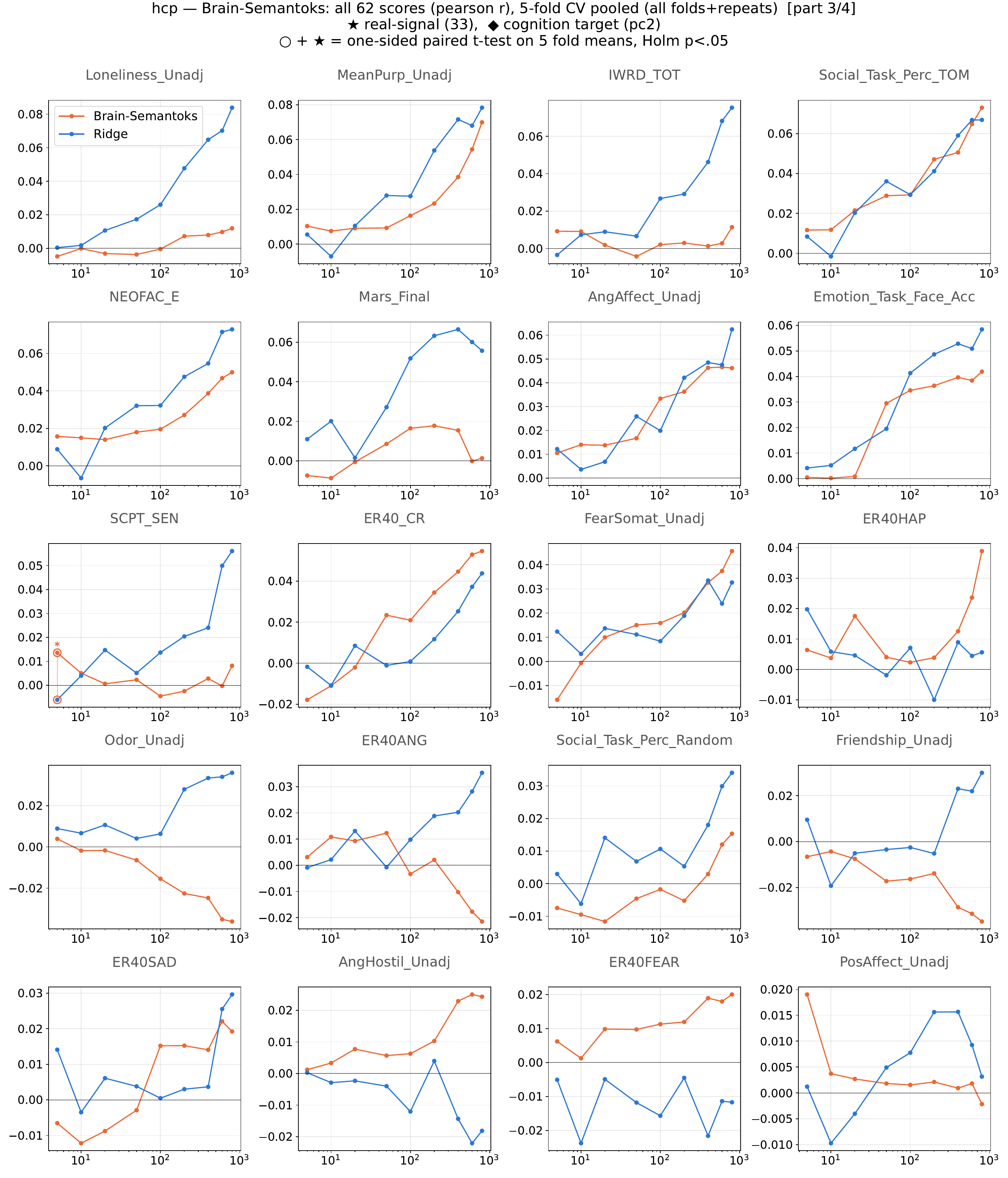}
    \caption{Brain-Semantoks, all scores, HCP-YA, 5-fold CV (3/3).}
    \label{fig:perscore-brainsemantoks-hcp-3}
\end{figure}

\begin{figure}
    \centering
    \includegraphics[width=\linewidth]{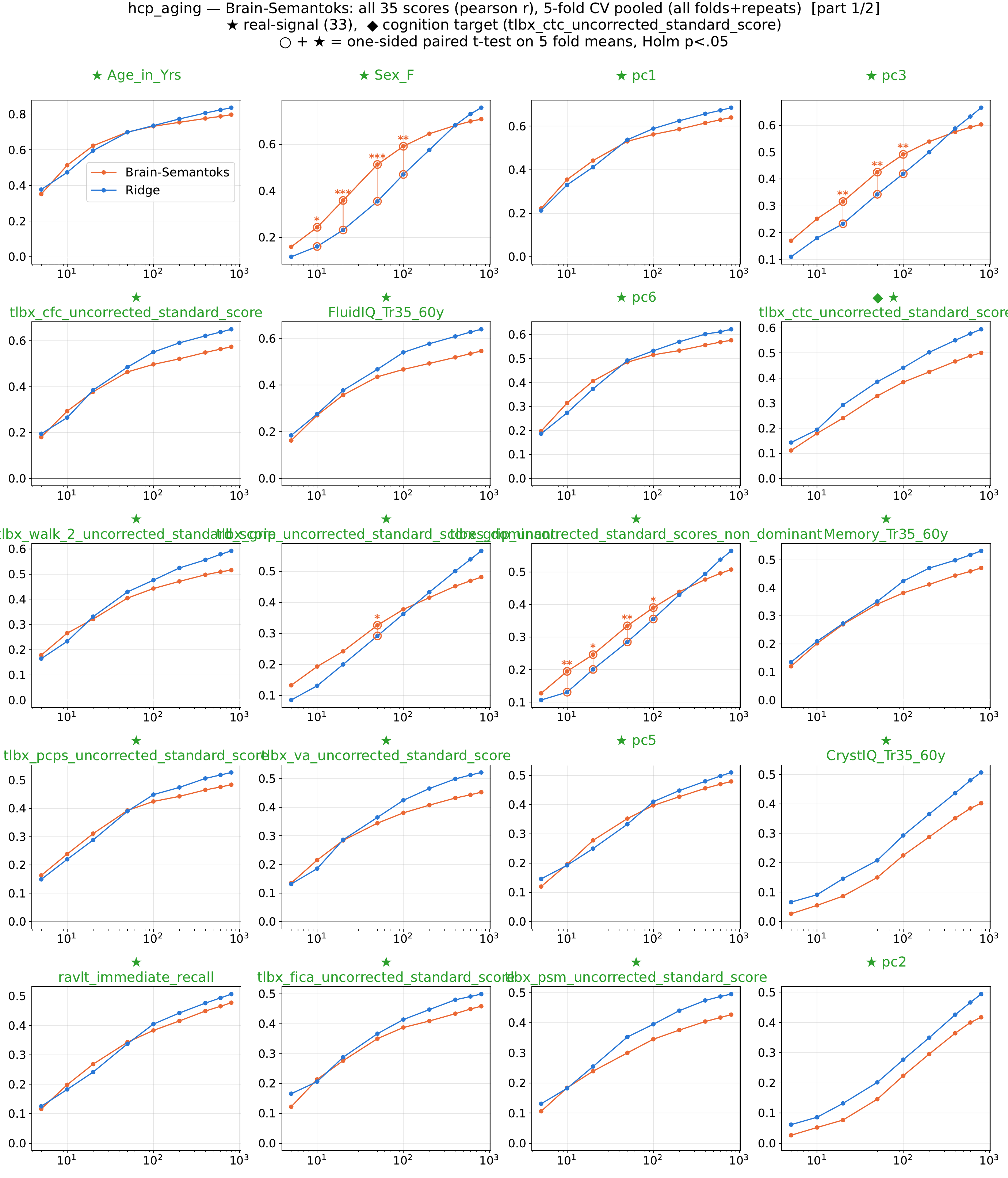}
    \caption{Brain-Semantoks, all scores, HCP-A, 5-fold CV (1/2).}
    \label{fig:perscore-brainsemantoks-hcp-aging-1}
\end{figure}

\begin{figure}
    \centering
    \includegraphics[width=\linewidth]{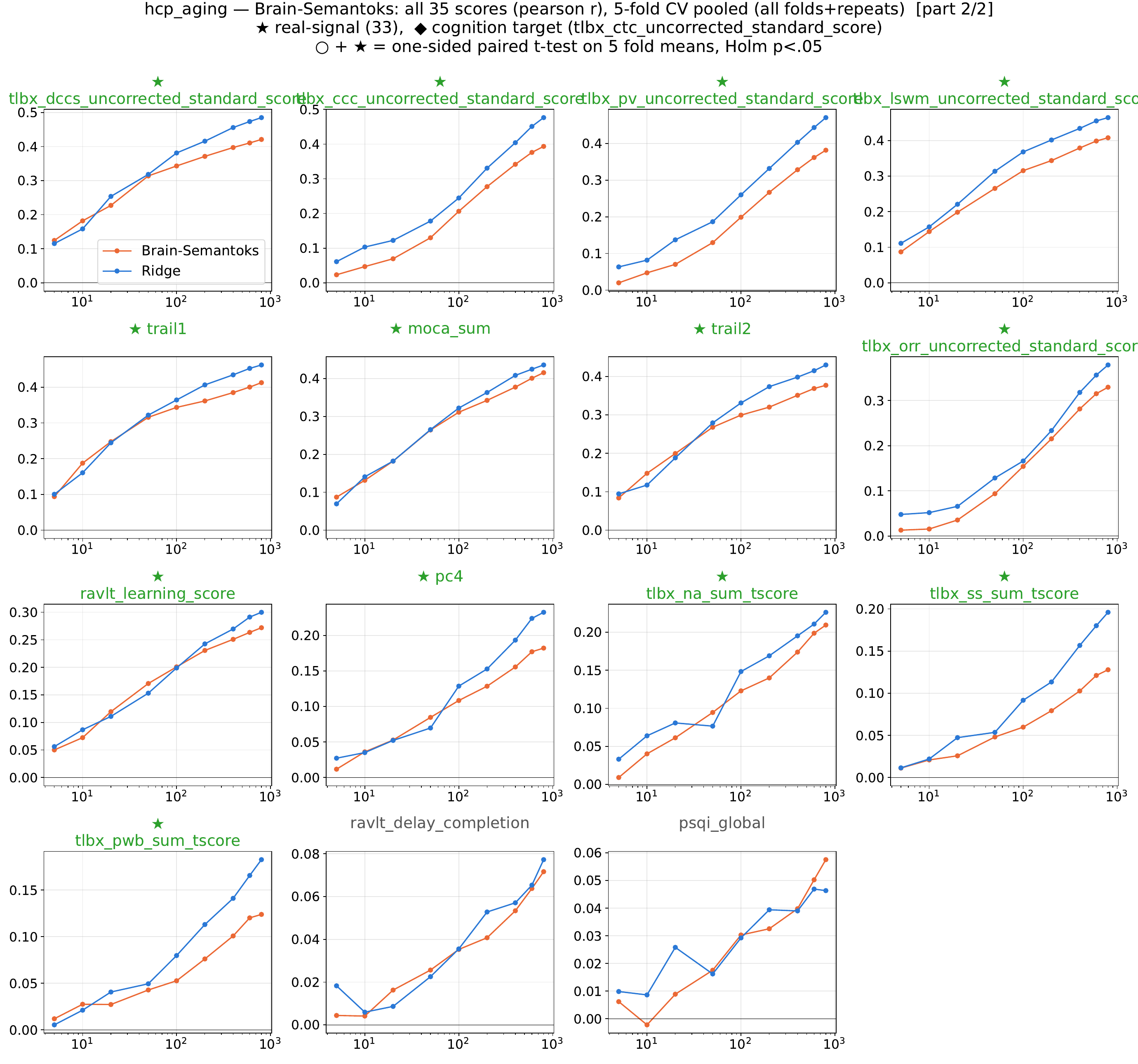}
    \caption{Brain-Semantoks, all scores, HCP-A, 5-fold CV (2/2).}
    \label{fig:perscore-brainsemantoks-hcp-aging-2}
\end{figure}

\begin{figure}
    \centering
    \includegraphics[width=\linewidth]{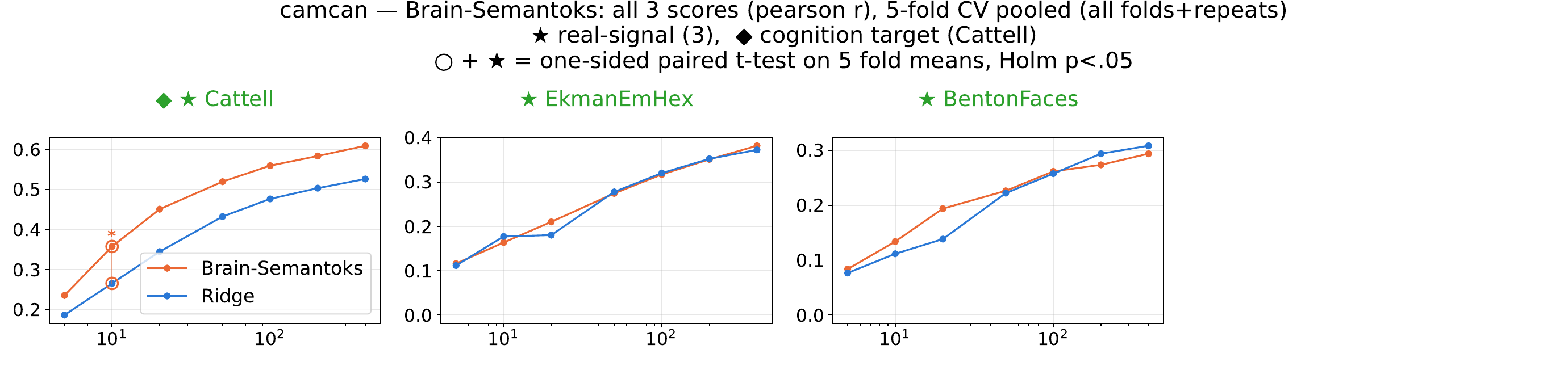}
    \caption{Brain-Semantoks, all scores, Cam-CAN, 5-fold CV.}
    \label{fig:perscore-brainsemantoks-camcan}
\end{figure}

\begin{figure}
    \centering
    \includegraphics[width=\linewidth]{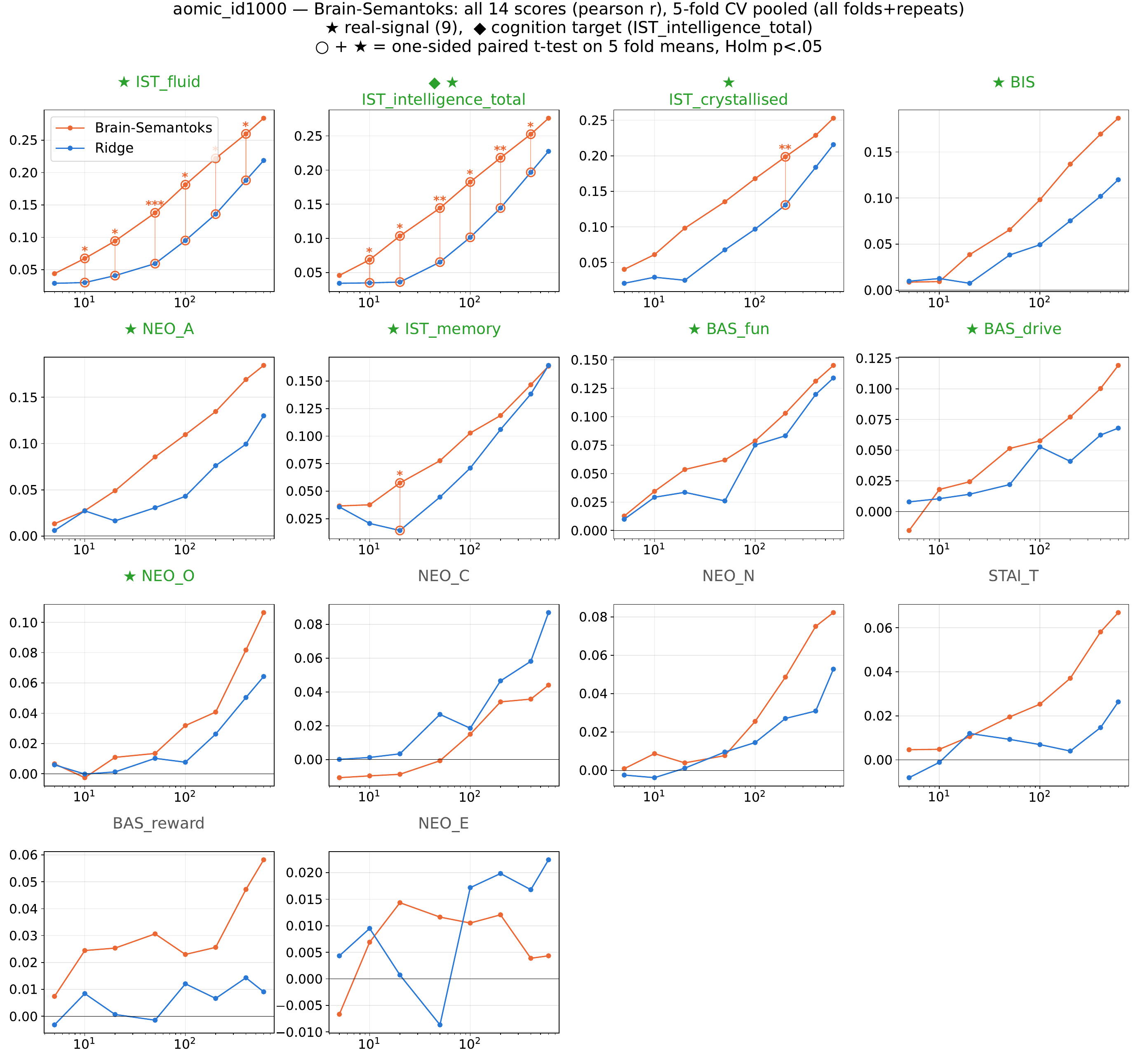}
    \caption{Brain-Semantoks, all scores, AOMIC-ID1000, 5-fold CV.}
    \label{fig:perscore-brainsemantoks-aomic-id1000}
\end{figure}

\clearpage

\subsubsection{Brain-Mass}

\begin{figure}[h]
    \centering
    \includegraphics[width=\linewidth]{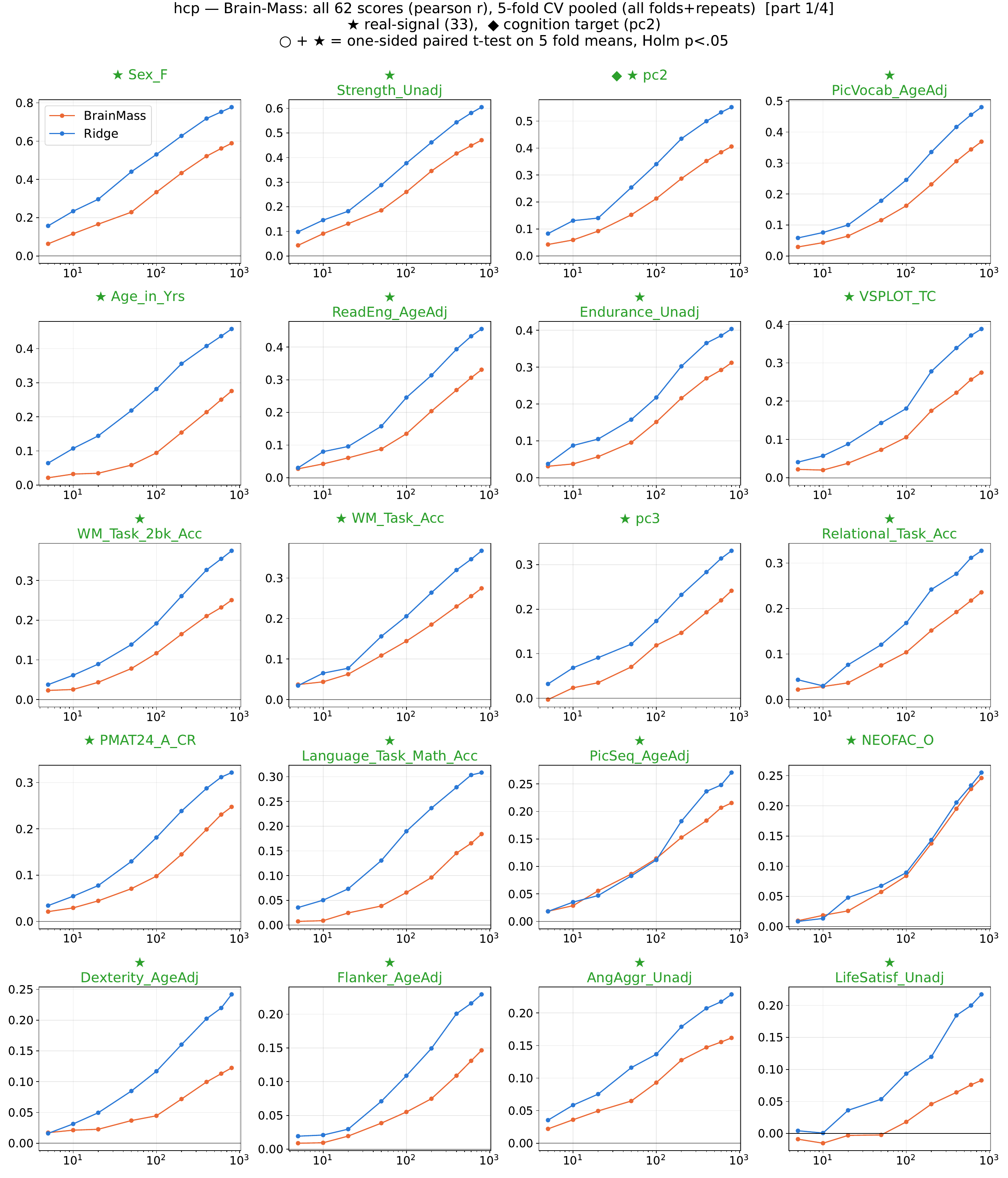}
    \caption{Brain-Mass, all scores, HCP-YA, 5-fold CV (1/3).}
    \label{fig:perscore-brainmass-hcp-1}
\end{figure}

\begin{figure}
    \centering
    \includegraphics[width=\linewidth]{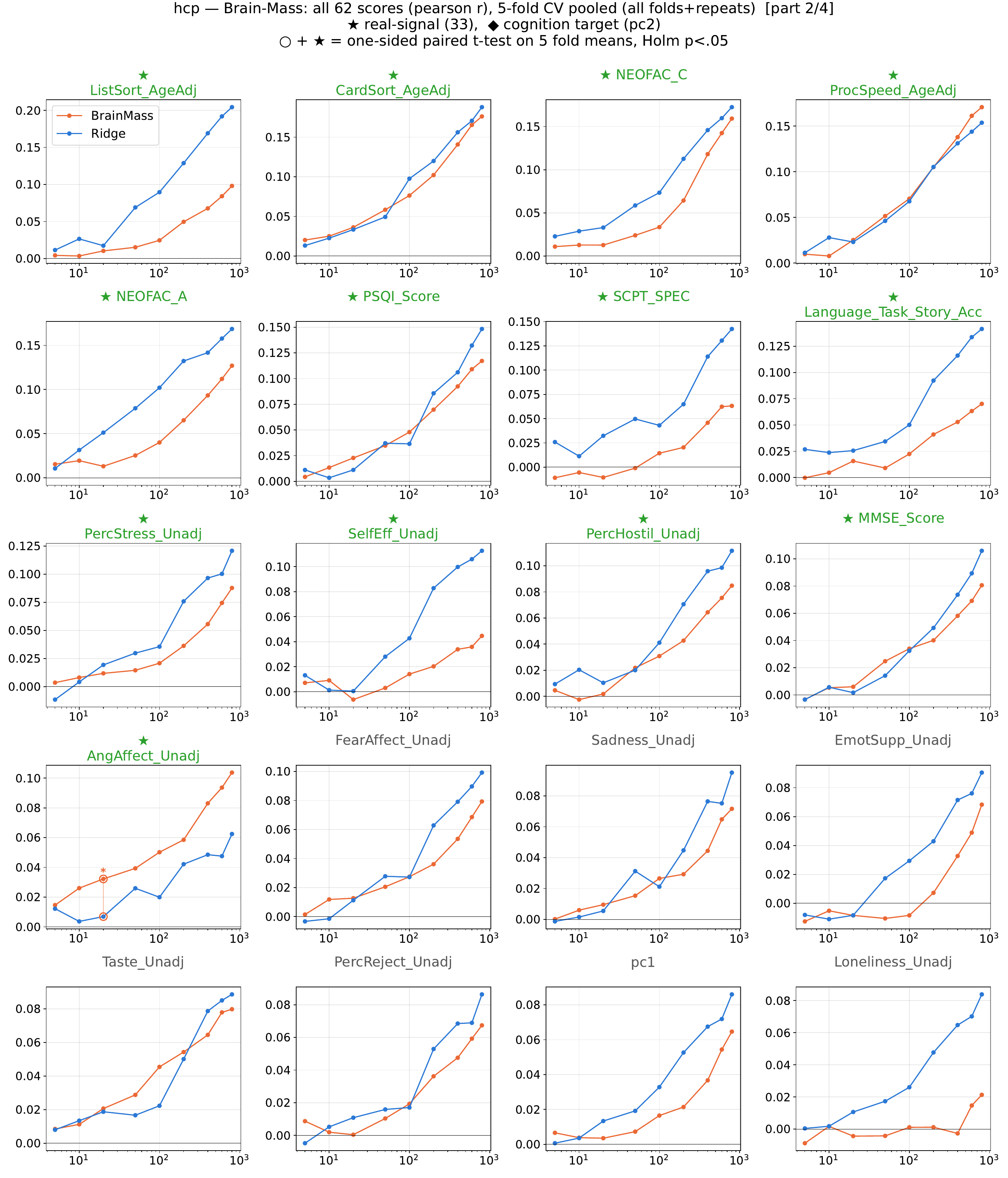}
    \caption{Brain-Mass, all scores, HCP-YA, 5-fold CV (2/3).}
    \label{fig:perscore-brainmass-hcp-2}
\end{figure}

\begin{figure}
    \centering
    \includegraphics[width=\linewidth]{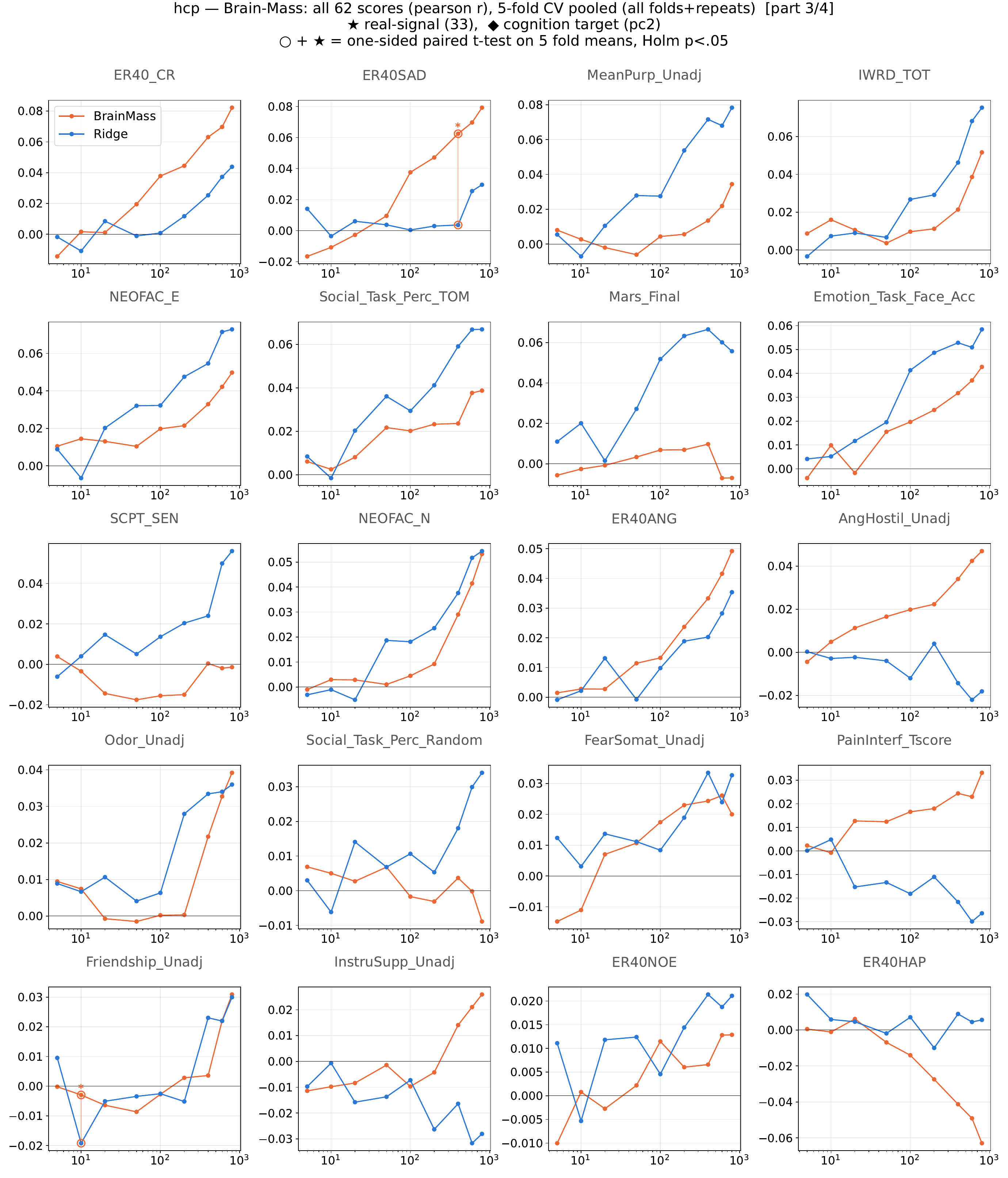}
    \caption{Brain-Mass, all scores, HCP-YA, 5-fold CV (3/3).}
    \label{fig:perscore-brainmass-hcp-3}
\end{figure}

\begin{figure}
    \centering
    \includegraphics[width=\linewidth]{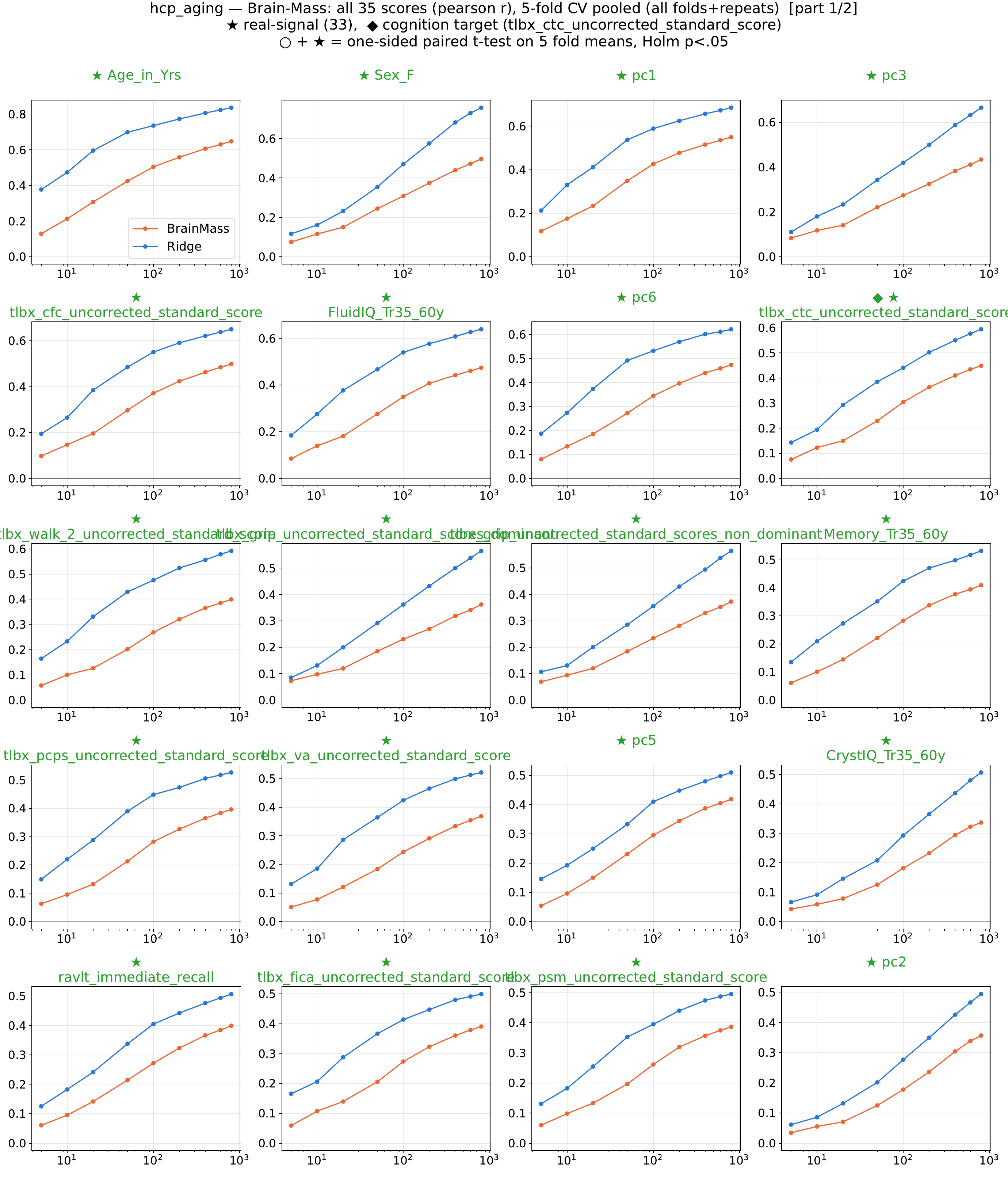}
    \caption{Brain-Mass, all scores, HCP-A, 5-fold CV (1/2).}
    \label{fig:perscore-brainmass-hcp-aging-1}
\end{figure}

\begin{figure}
    \centering
    \includegraphics[width=\linewidth]{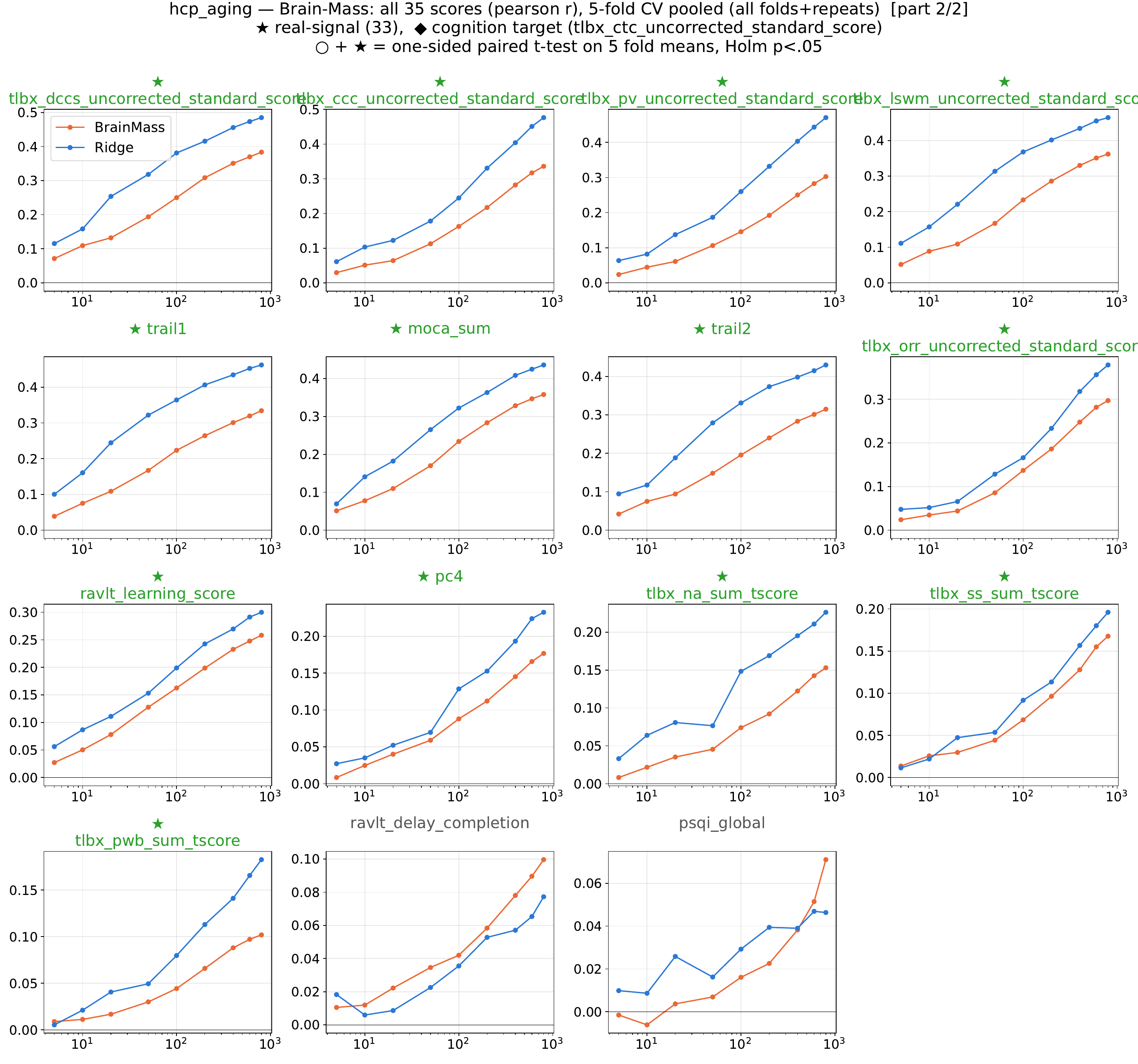}
    \caption{Brain-Mass, all scores, HCP-A, 5-fold CV (2/2).}
    \label{fig:perscore-brainmass-hcp-aging-2}
\end{figure}

\begin{figure}
    \centering
    \includegraphics[width=\linewidth]{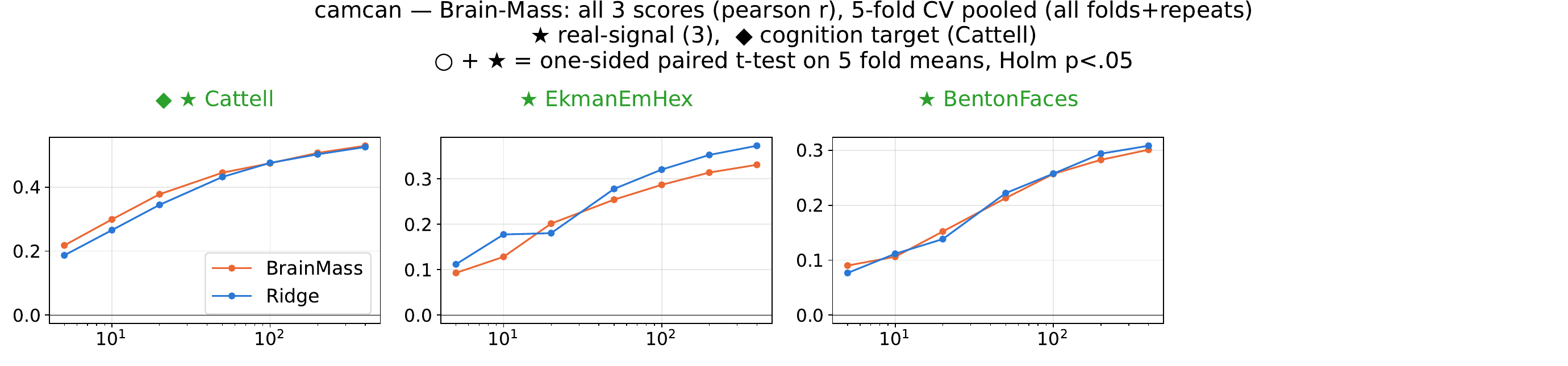}
    \caption{Brain-Mass, all scores, Cam-CAN, 5-fold CV.}
    \label{fig:perscore-brainmass-camcan}
\end{figure}

\begin{figure}
    \centering
    \includegraphics[width=\linewidth]{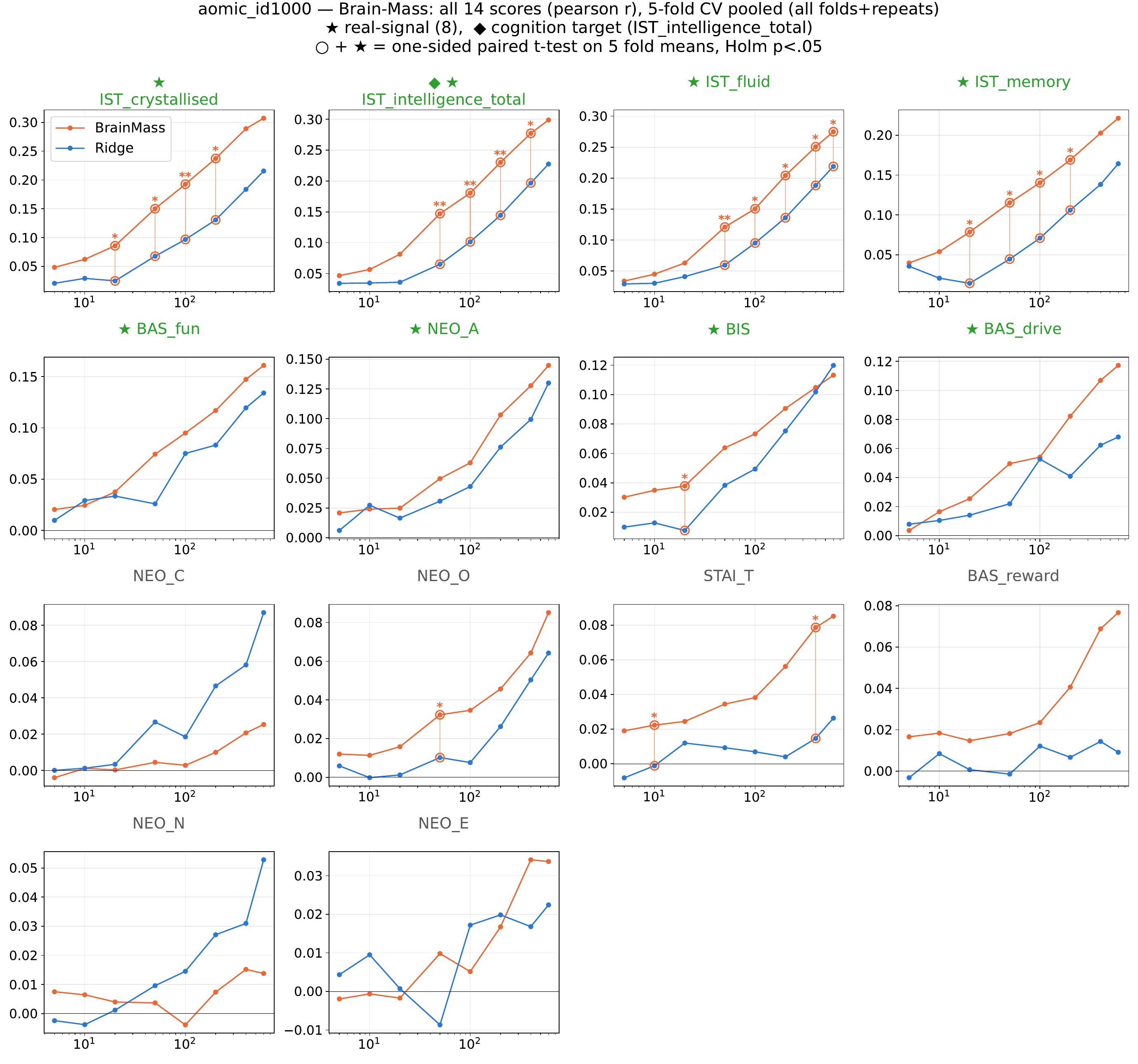}
    \caption{Brain-Mass, all scores, AOMIC-ID1000, 5-fold CV.}
    \label{fig:perscore-brainmass-aomic-id1000}
\end{figure}

\clearpage

\subsubsection{BrainJEPA}

\begin{figure}[h]
    \centering
    \includegraphics[width=\linewidth]{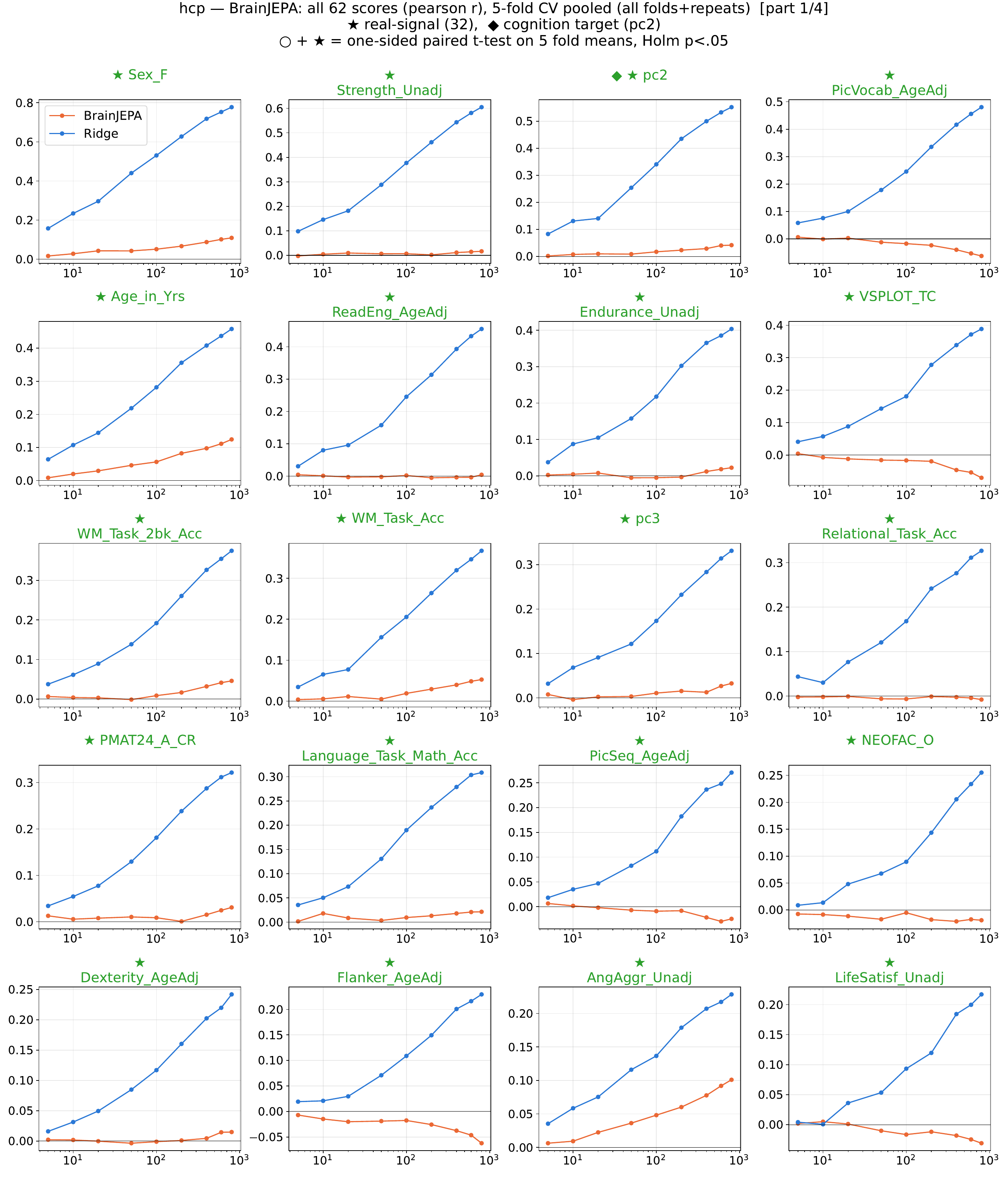}
    \caption{BrainJEPA, all scores, HCP-YA, 5-fold CV (1/3).}
    \label{fig:perscore-brainjepa-hcp-1}
\end{figure}

\begin{figure}
    \centering
    \includegraphics[width=\linewidth]{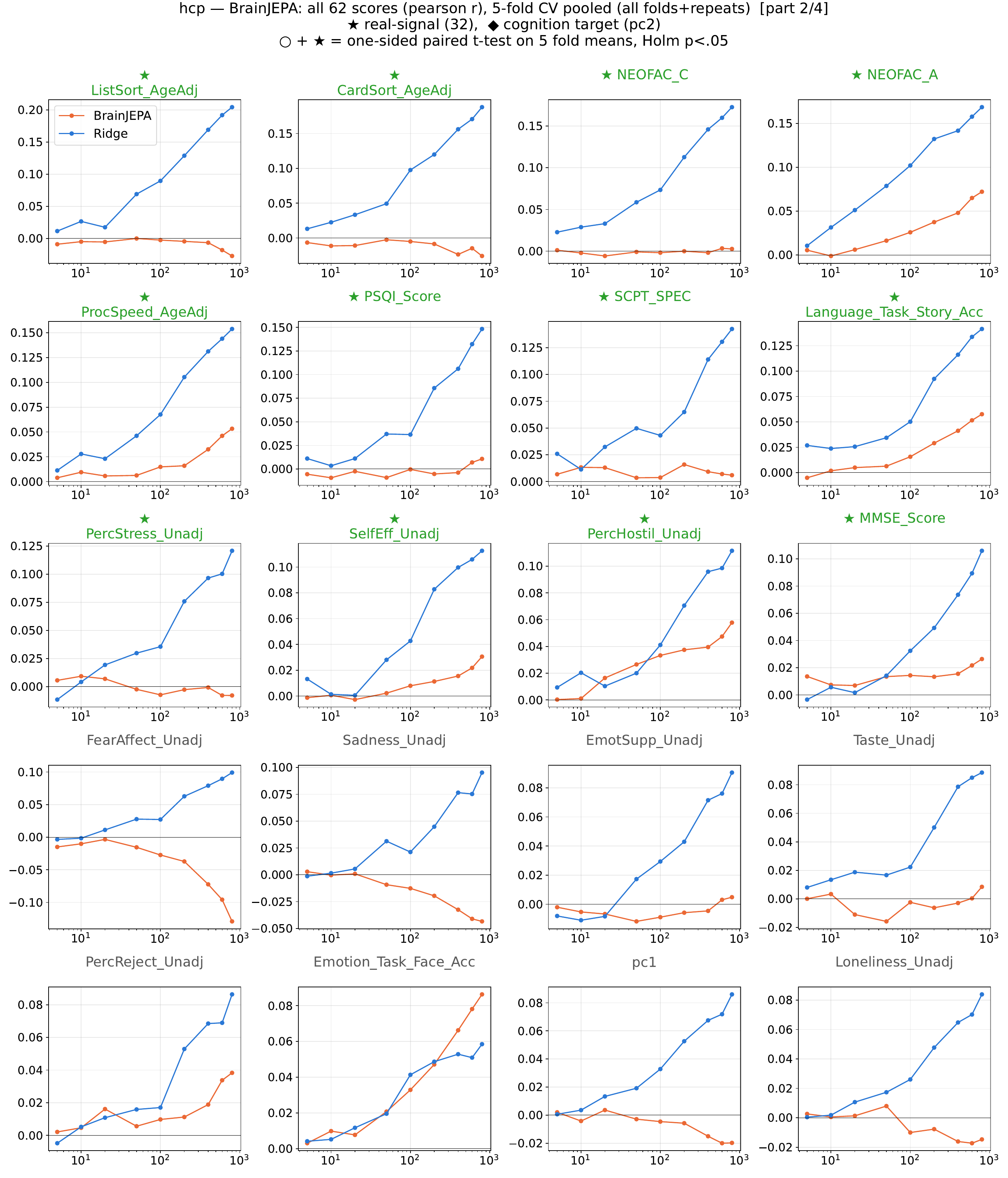}
    \caption{BrainJEPA, all scores, HCP-YA, 5-fold CV (2/3).}
    \label{fig:perscore-brainjepa-hcp-2}
\end{figure}

\begin{figure}
    \centering
    \includegraphics[width=\linewidth]{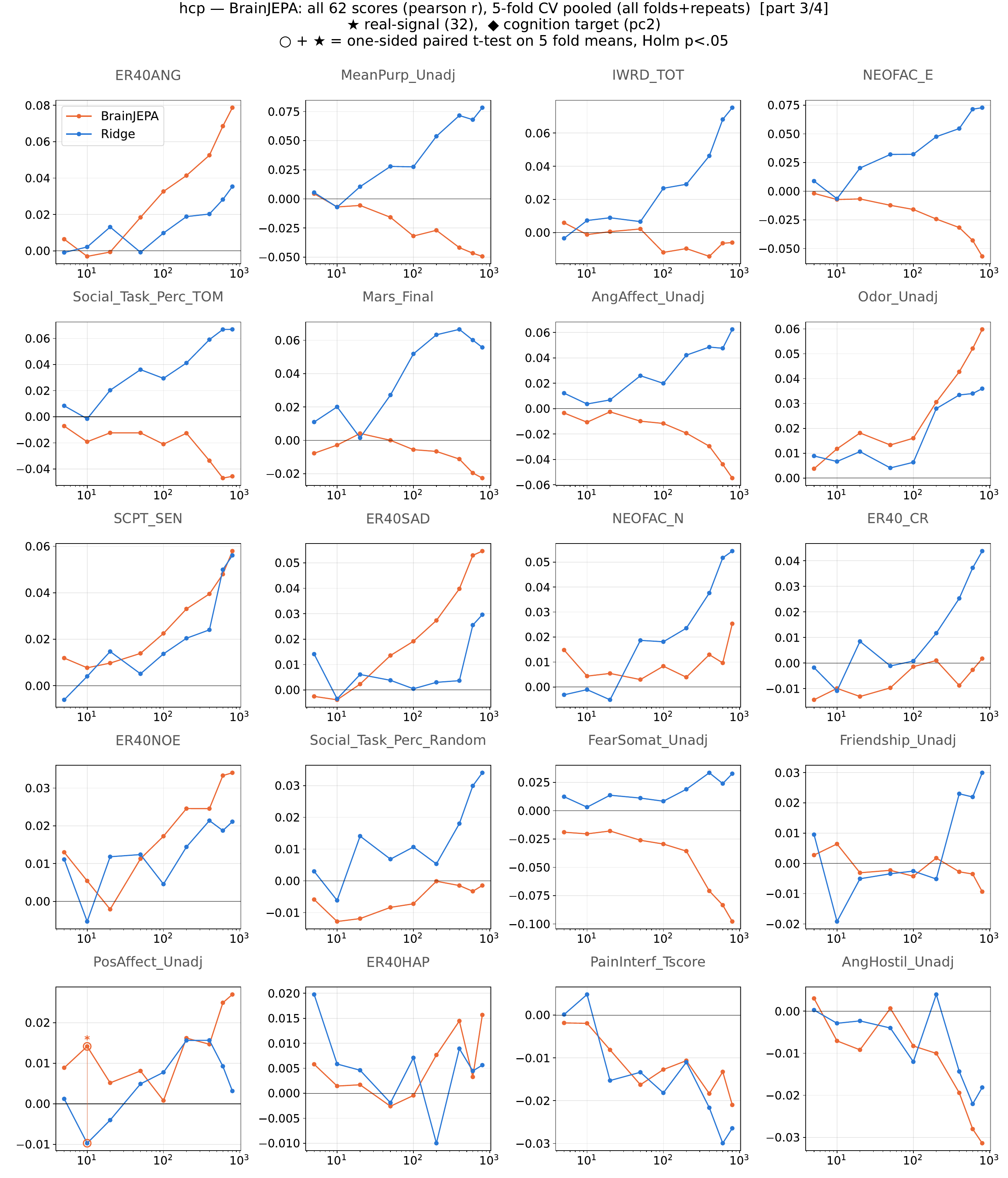}
    \caption{BrainJEPA, all scores, HCP-YA, 5-fold CV (3/3).}
    \label{fig:perscore-brainjepa-hcp-3}
\end{figure}

\begin{figure}
    \centering
    \includegraphics[width=\linewidth]{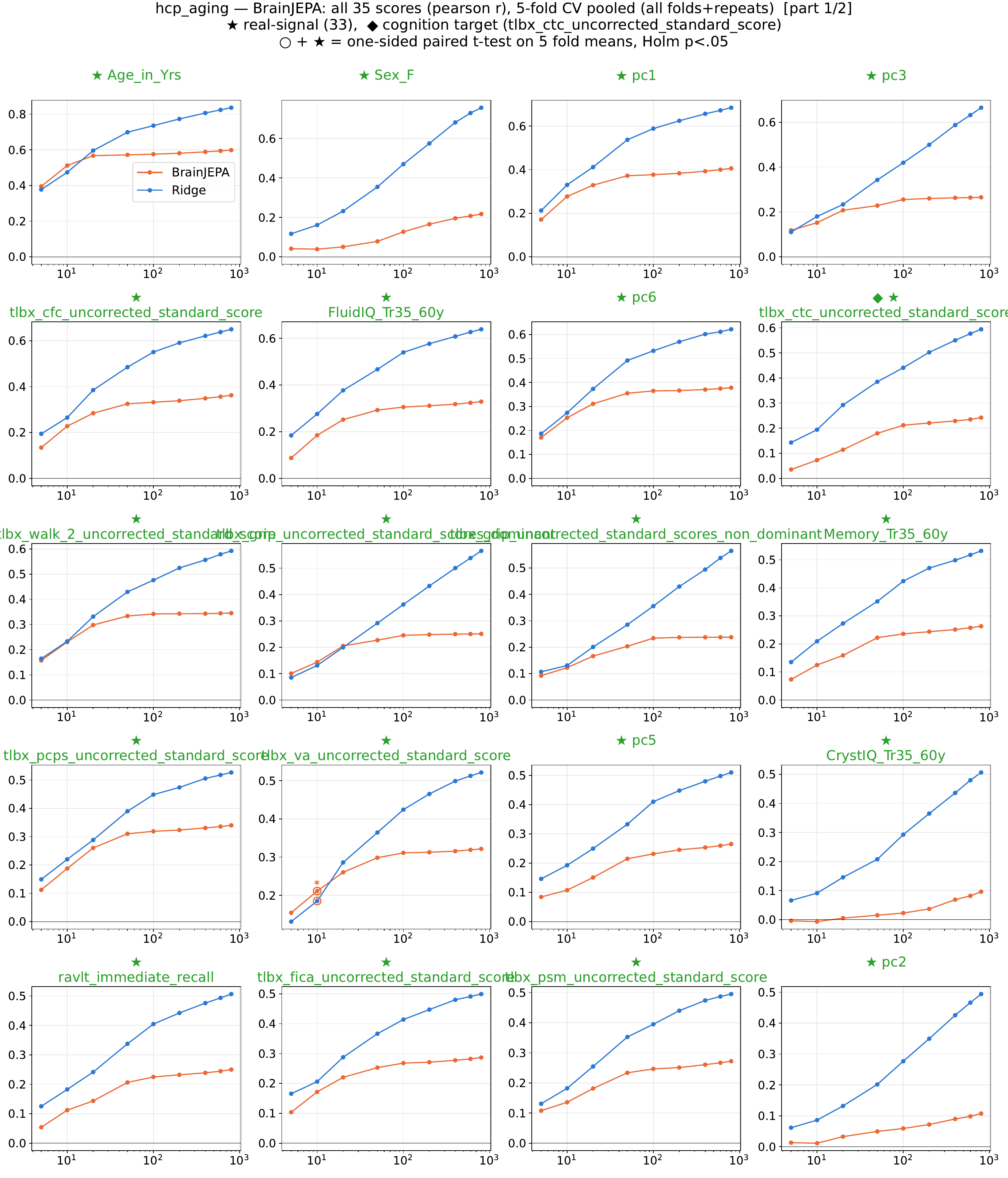}
    \caption{BrainJEPA, all scores, HCP-A, 5-fold CV (1/2).}
    \label{fig:perscore-brainjepa-hcp-aging-1}
\end{figure}

\begin{figure}
    \centering
    \includegraphics[width=\linewidth]{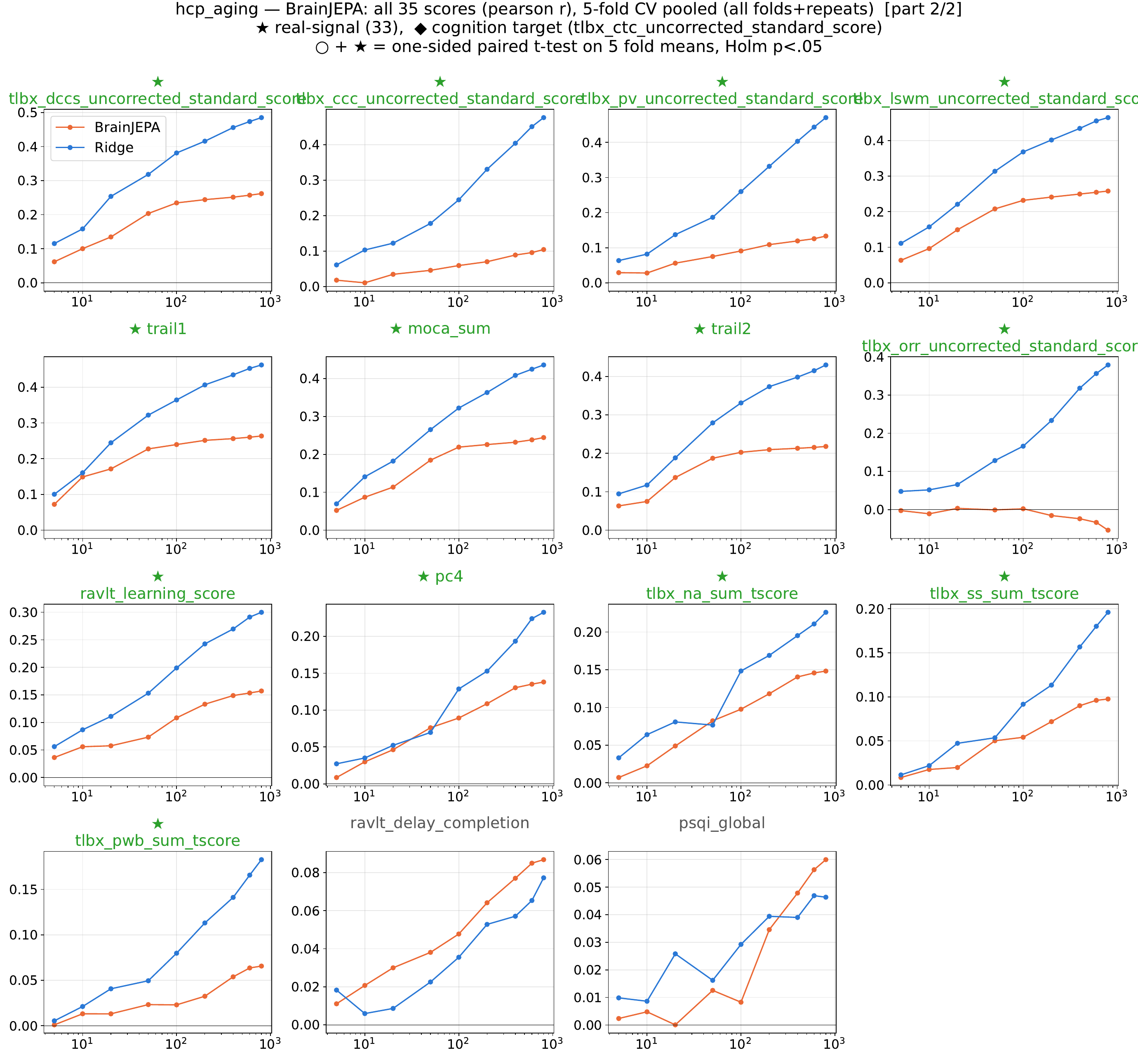}
    \caption{BrainJEPA, all scores, HCP-A, 5-fold CV (2/2).}
    \label{fig:perscore-brainjepa-hcp-aging-2}
\end{figure}

\begin{figure}
    \centering
    \includegraphics[width=\linewidth]{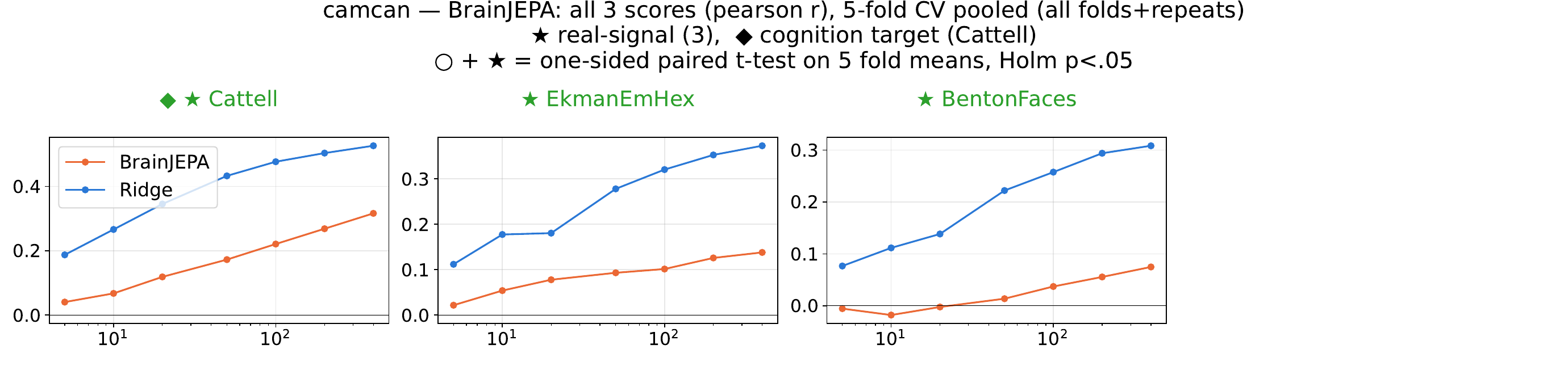}
    \caption{BrainJEPA, all scores, Cam-CAN, 5-fold CV.}
    \label{fig:perscore-brainjepa-camcan}
\end{figure}

\begin{figure}
    \centering
    \includegraphics[width=\linewidth]{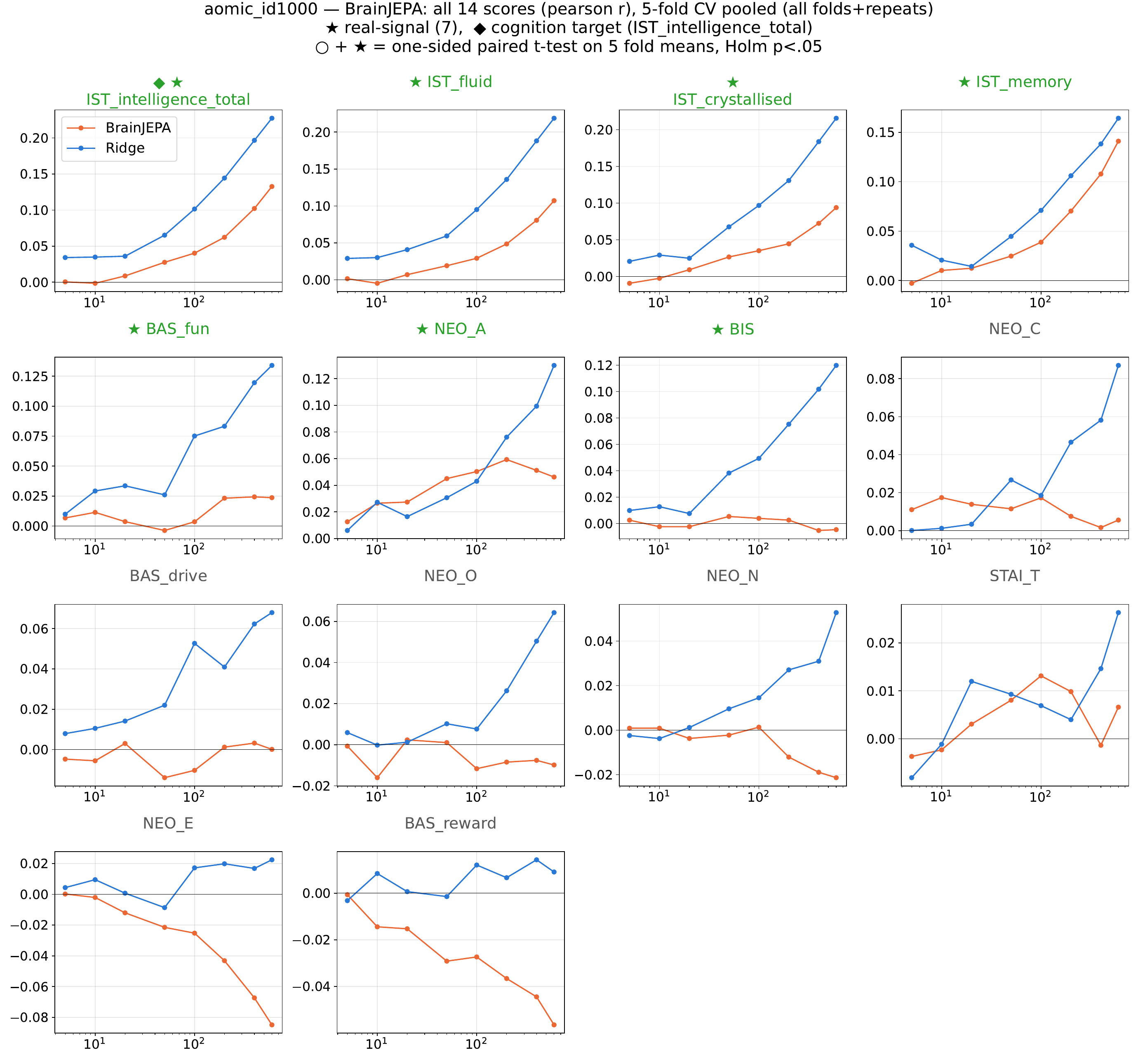}
    \caption{BrainJEPA, all scores, AOMIC-ID1000, 5-fold CV.}
    \label{fig:perscore-brainjepa-aomic-id1000}
\end{figure}

\clearpage

\subsubsection{TabPFN v2}

\begin{figure}[h]
    \centering
    \includegraphics[width=\linewidth]{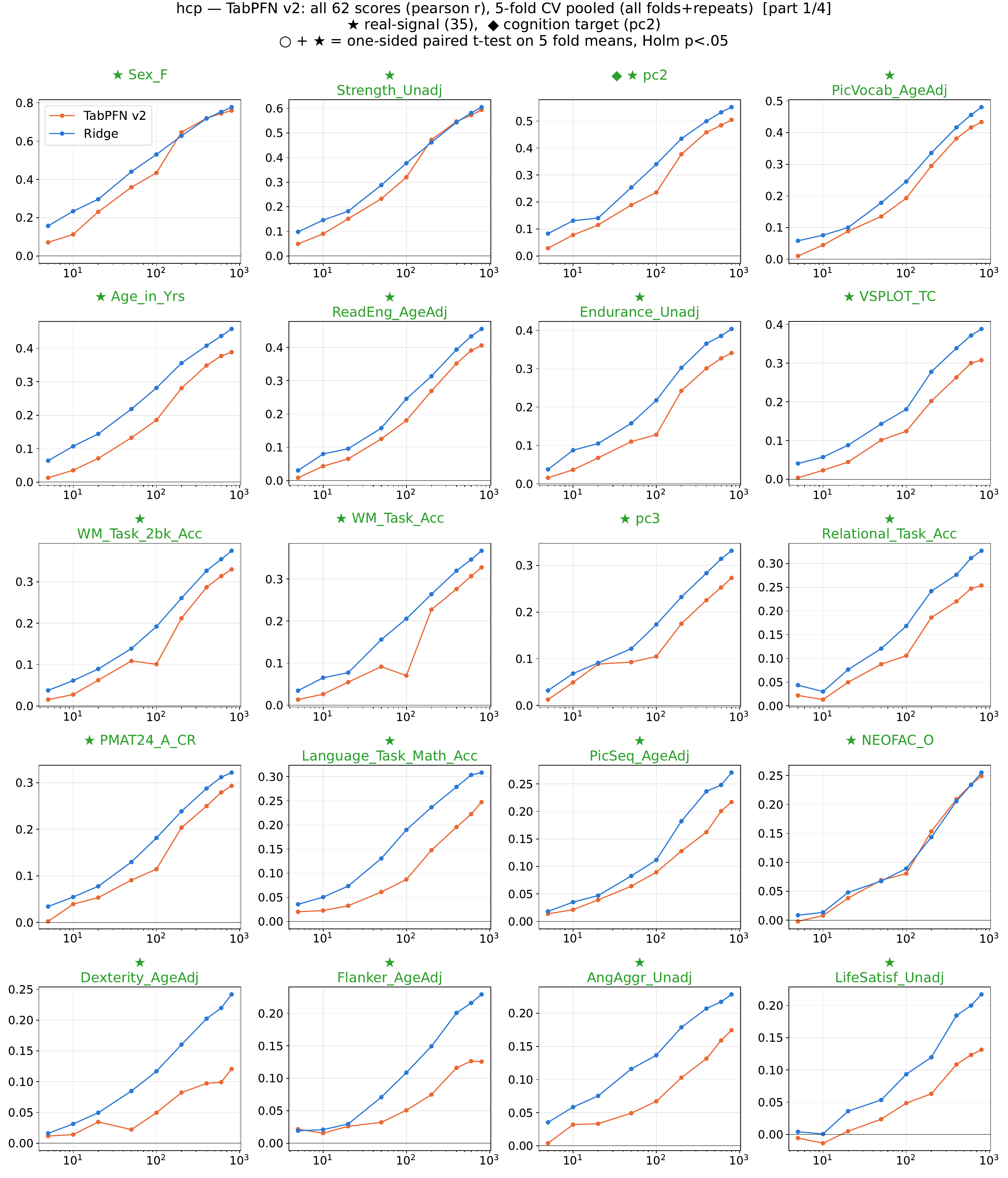}
    \caption{TabPFN v2, all scores, HCP-YA, 5-fold CV (1/3).}
    \label{fig:perscore-tabpfn-hcp-1}
\end{figure}

\begin{figure}
    \centering
    \includegraphics[width=\linewidth]{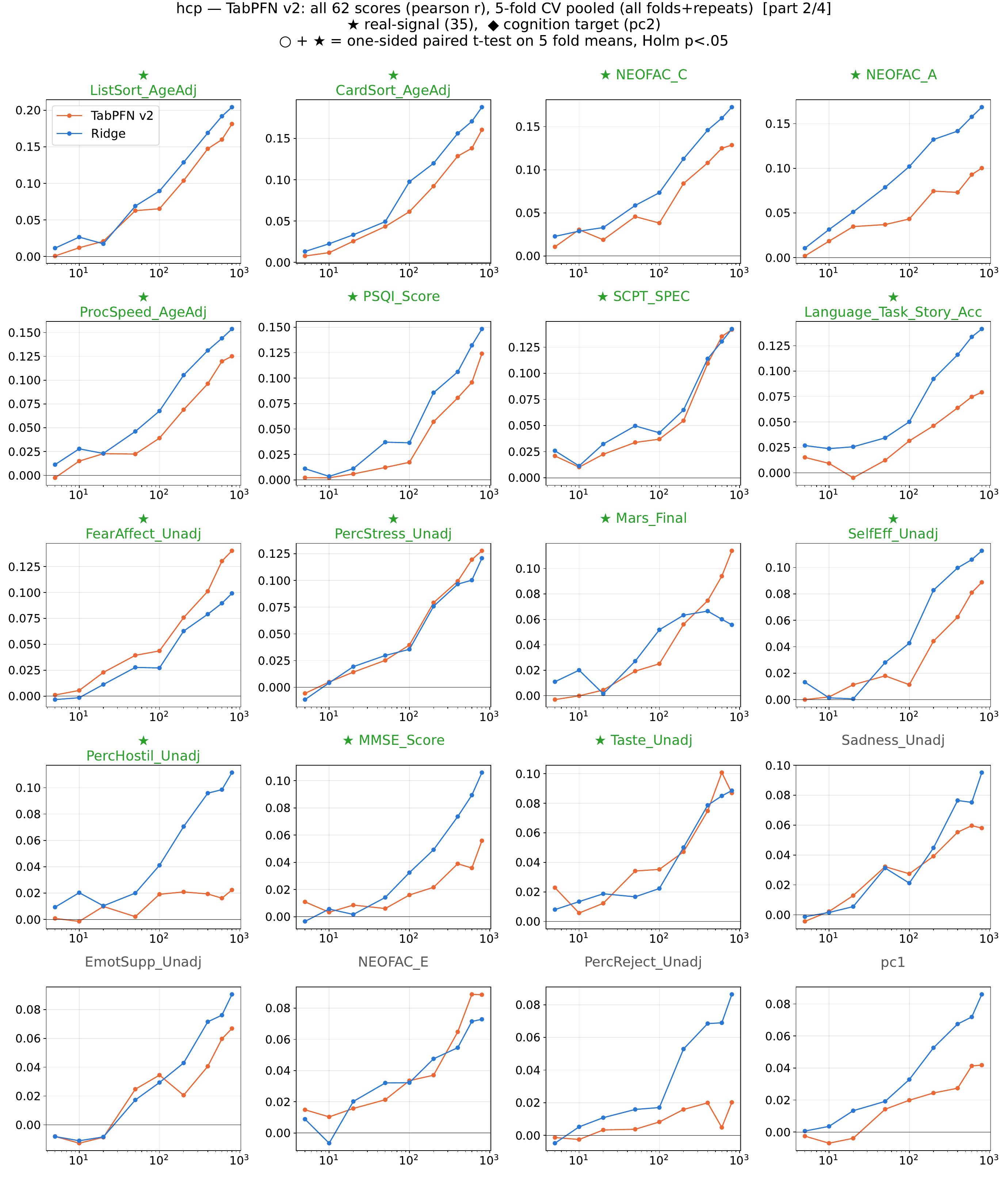}
    \caption{TabPFN v2, all scores, HCP-YA, 5-fold CV (2/3).}
    \label{fig:perscore-tabpfn-hcp-2}
\end{figure}

\begin{figure}
    \centering
    \includegraphics[width=\linewidth]{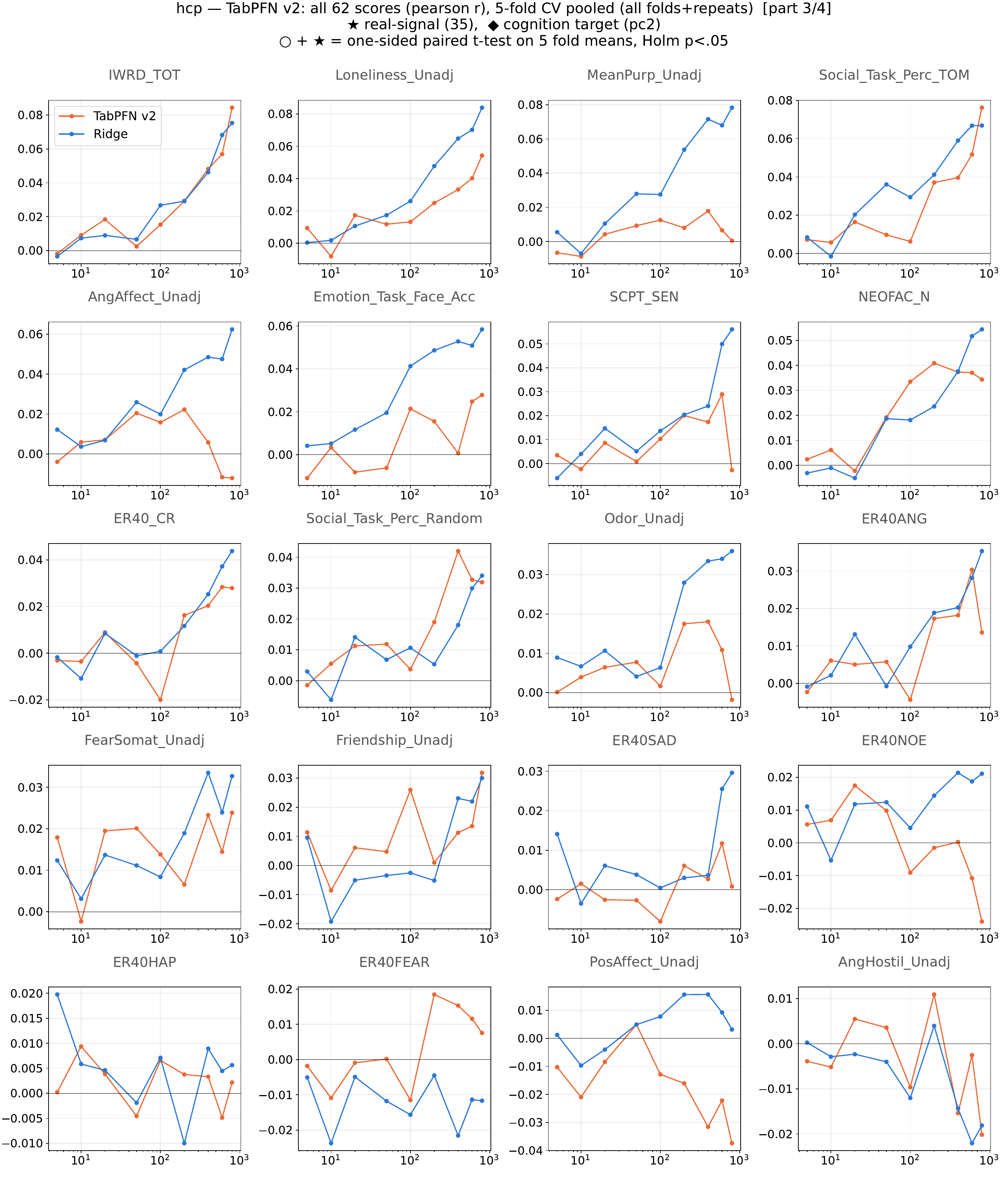}
    \caption{TabPFN v2, all scores, HCP-YA, 5-fold CV (3/3).}
    \label{fig:perscore-tabpfn-hcp-3}
\end{figure}

\begin{figure}
    \centering
    \includegraphics[width=\linewidth]{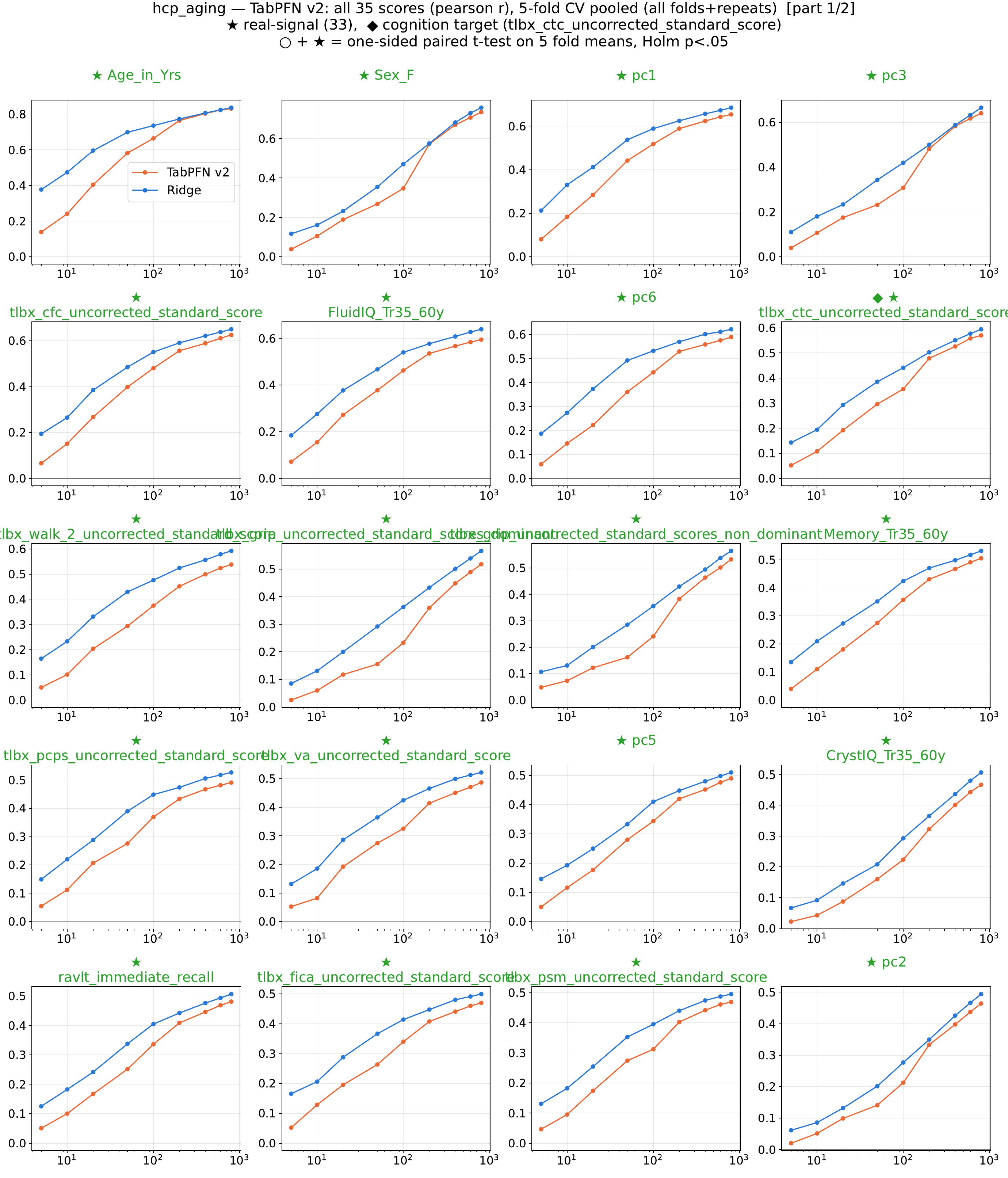}
    \caption{TabPFN v2, all scores, HCP-A, 5-fold CV (1/2).}
    \label{fig:perscore-tabpfn-hcp-aging-1}
\end{figure}

\begin{figure}
    \centering
    \includegraphics[width=\linewidth]{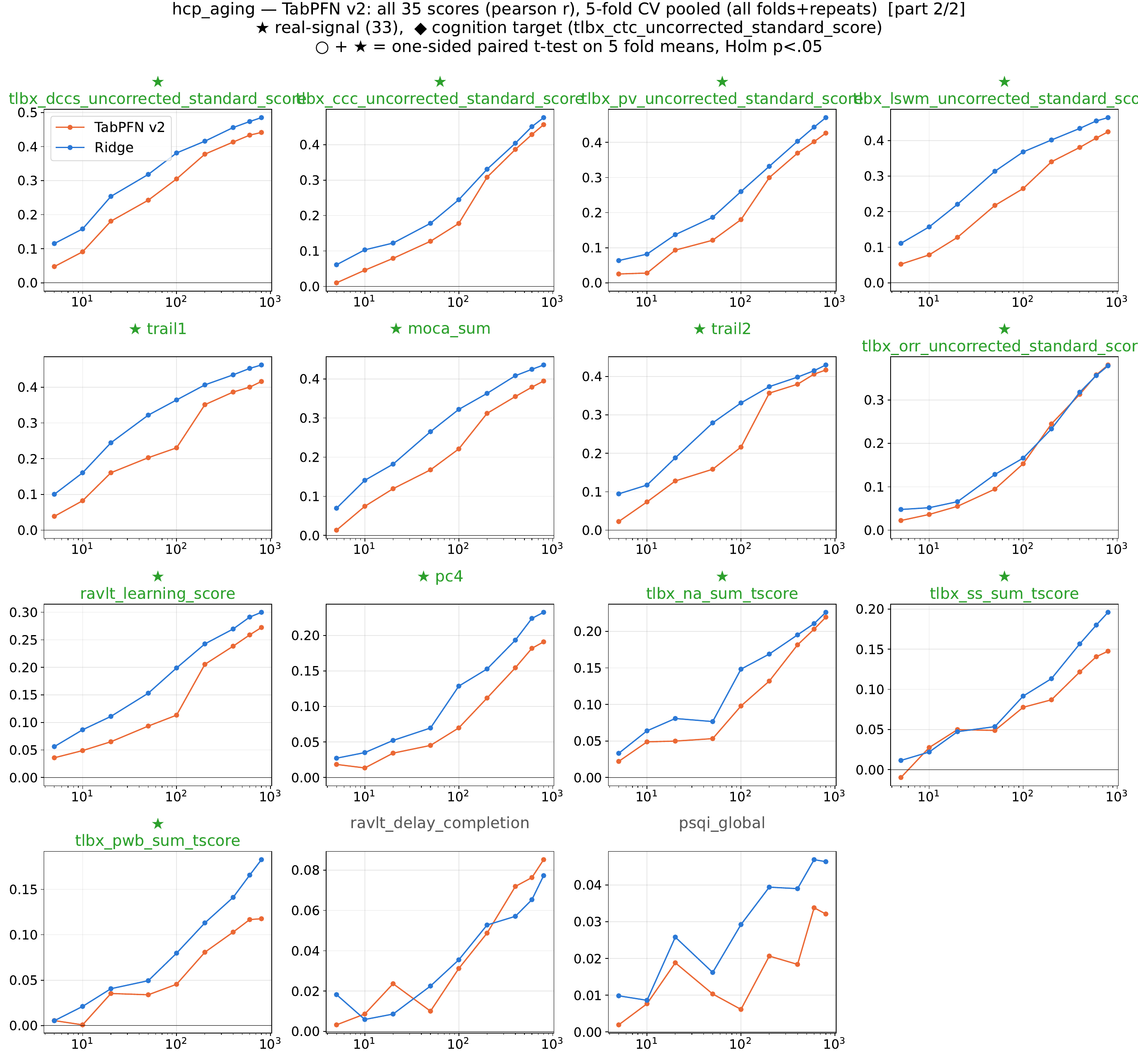}
    \caption{TabPFN v2, all scores, HCP-A, 5-fold CV (2/2).}
    \label{fig:perscore-tabpfn-hcp-aging-2}
\end{figure}

\begin{figure}
    \centering
    \includegraphics[width=\linewidth]{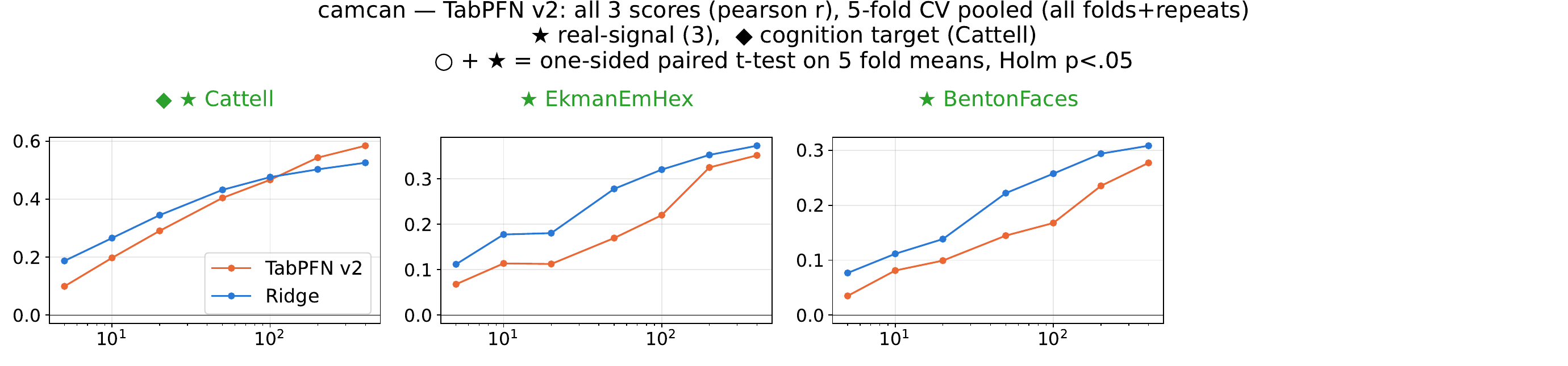}
    \caption{TabPFN v2, all scores, Cam-CAN, 5-fold CV.}
    \label{fig:perscore-tabpfn-camcan}
\end{figure}

\begin{figure}
    \centering
    \includegraphics[width=\linewidth]{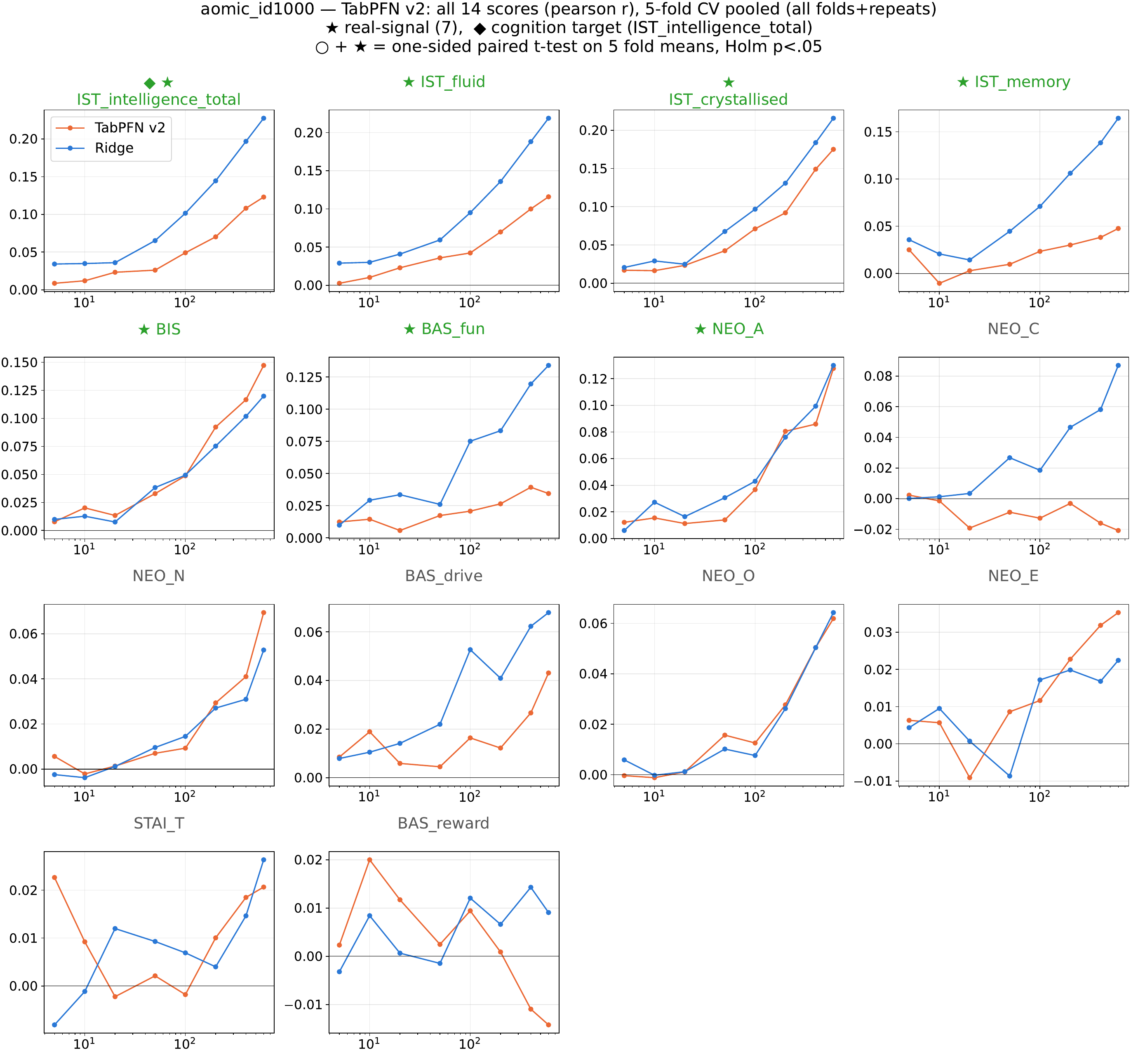}
    \caption{TabPFN v2, all scores, AOMIC-ID1000, 5-fold CV.}
    \label{fig:perscore-tabpfn-aomic-id1000}
\end{figure}

\clearpage

\section{List of datasets}
\label{appendix:dataset-list}

\input{figures/openfmind_datasets_table}

\clearpage

%% file: appendix_ooi.tex
In addition to the individual behavioural scores evaluated throughout the paper, we follow
the dimensionality-reduction approach of \citet{ooi2022comparison} to construct a small set
of interpretable, orthogonal behavioural components for HCP-YA and HCP-Aging, and evaluate
every model on them as additional regression targets. This appendix gives the method, the
resulting components, and the corresponding model comparison.

\begin{figure}[h!]
    \centering
    \includegraphics[width=\linewidth]{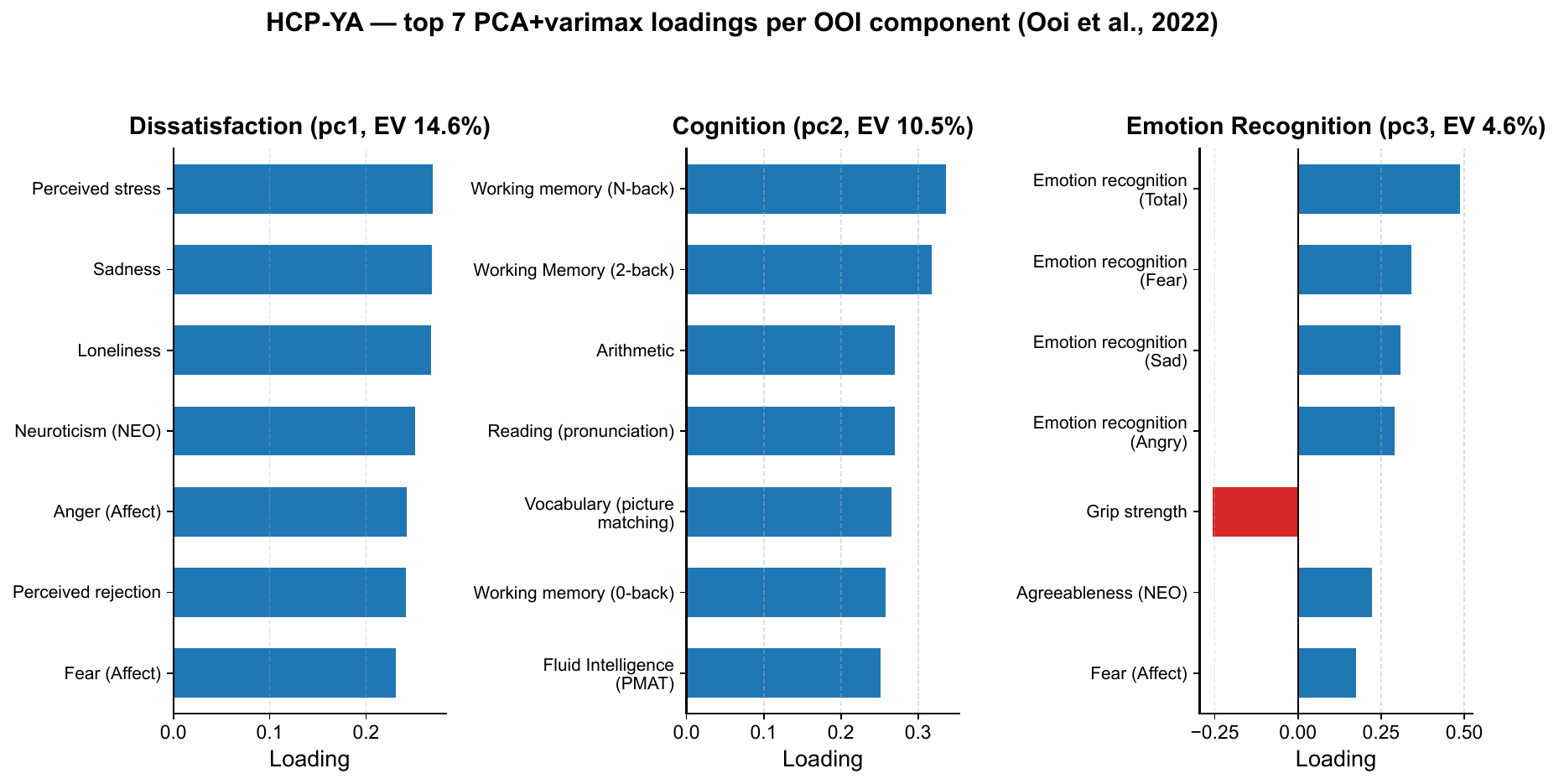}
    \caption{HCP-YA: top-7 PCA+varimax loadings per OOI component.}
    \label{fig:ooi-loadings-hcp}
\end{figure}

\begin{figure}[h!]
    \centering
    \includegraphics[width=\linewidth]{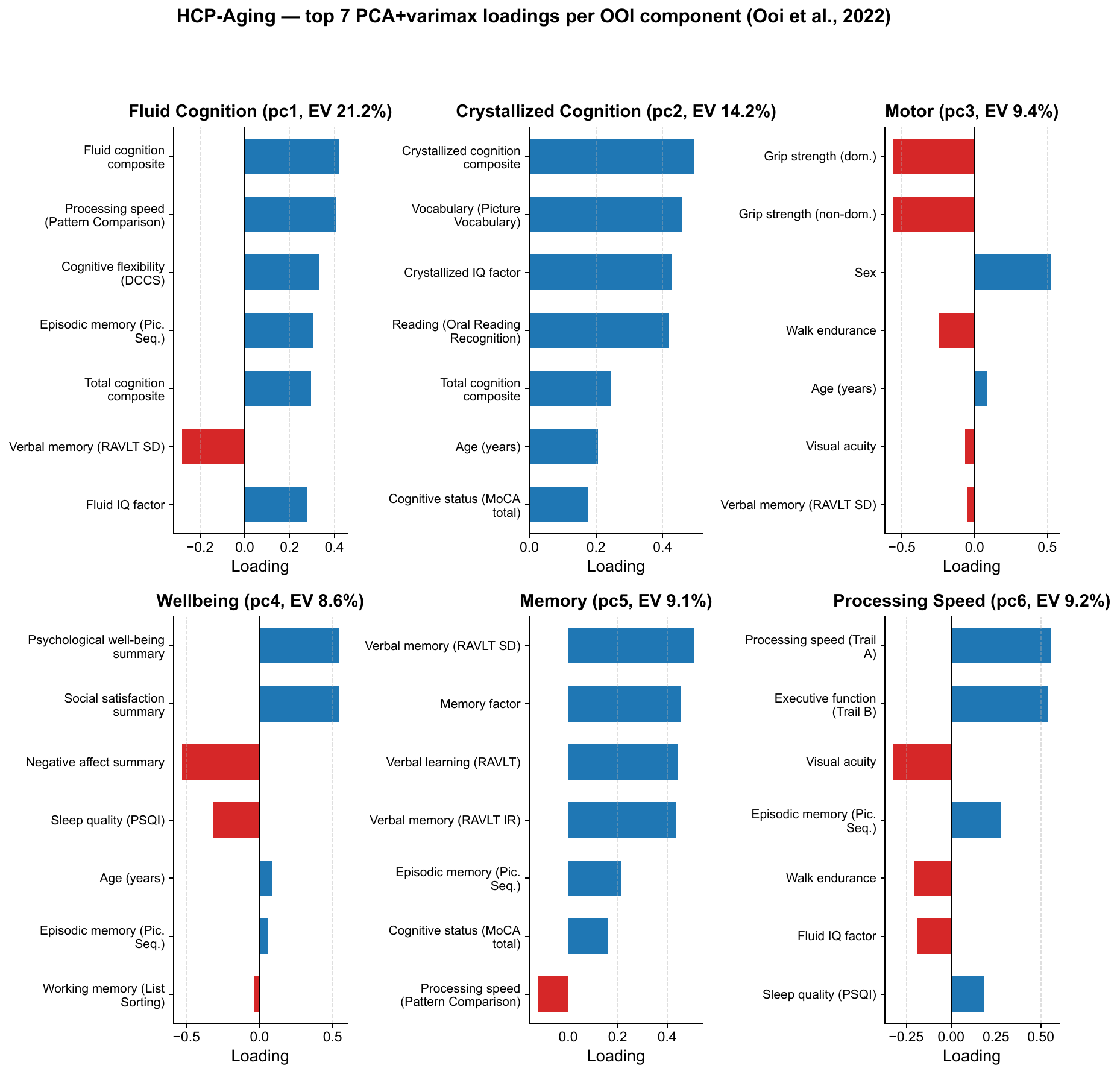}
    \caption{HCP-Aging: top-7 PCA+varimax loadings per OOI component.}
    \label{fig:ooi-loadings-hcp-aging}
\end{figure}

\paragraph{Method.} For each cohort we start from its full pool of behavioural measures --
58 scores for HCP-YA, 29 for HCP-Aging, spanning cognition, motor, and emotion/well-being
domains. Missing entries are imputed with an EM-style
iterative imputer (scikit-learn's \texttt{IterativeImputer}, matching the EM imputation used
by \citet{ooi2022comparison}), the imputed matrix is $z$-scored per column, and we fit PCA
retaining a fixed number of components -- 3 for HCP-YA, 6 for HCP-Aging, chosen to mirror the
component counts reported for these cohorts in the original Ooi et al. analysis. The raw PCA
loadings are then rotated with varimax to sharpen each component
onto a smaller subset of scores, which is what makes the components individually
interpretable (Figures~\ref{fig:ooi-loadings-hcp} and~\ref{fig:ooi-loadings-hcp-aging} show,
for each component, the behavioural scores with the largest rotated loadings). The rotated
per-subject component scores are saved alongside the original behavioural scores and enter
the benchmark as ordinary regression targets, one per component, evaluated
with the exact same 5-fold, context-size-sweep protocol as every other score in this paper.

For HCP-YA, the three components separate into a \emph{cognition} axis (dominated by working
memory, arithmetic, reading, vocabulary, and fluid intelligence), an \emph{emotion
recognition} axis (dominated by task-based emotion-recognition accuracy), and a
\emph{dissatisfaction} axis (dominated by perceived stress, sadness, loneliness, and
neuroticism, loading negatively on positive affect). For HCP-Aging, the six components
separate NIH Toolbox cognition into \emph{fluid} and \emph{crystallized} sub-factors, and add
\emph{motor} (grip strength, sex, walk endurance), \emph{wellbeing} (psychological well-being
and social satisfaction, loading negatively on negative affect and poor sleep),
\emph{memory} (RAVLT verbal learning/recall), and \emph{processing speed} (Trail Making A/B)
components, merged into the per-subject score tables the same way as HCP-YA's three
components above.

\begin{figure}[h!]
    \centering
    \includegraphics[width=\linewidth]{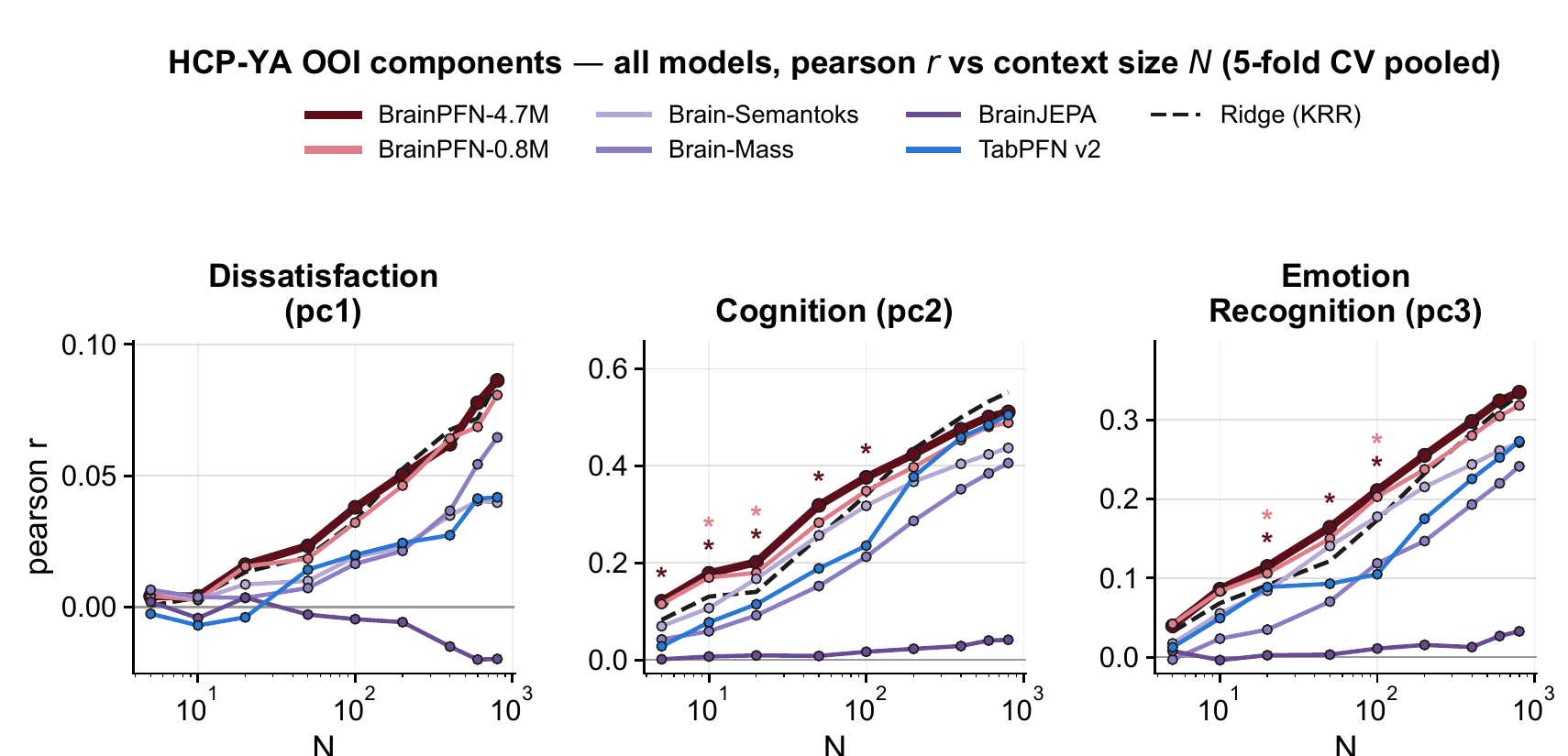}
    \caption{HCP-YA OOI components: pearson $r$ vs. context size $N$ for all 7 models
    (5-fold CV pooled). Dashed black = canonical Ridge reference. ``*'' marks $N$ where a
    model is significantly ahead of its own Ridge baseline (one-sided paired $t$-test on the
    5 fold-level means, Holm-corrected across that panel's $N$-grid).}
    \label{fig:ooi-comparison-hcp}
\end{figure}

\begin{figure}[h!]
    \centering
    \includegraphics[width=\linewidth]{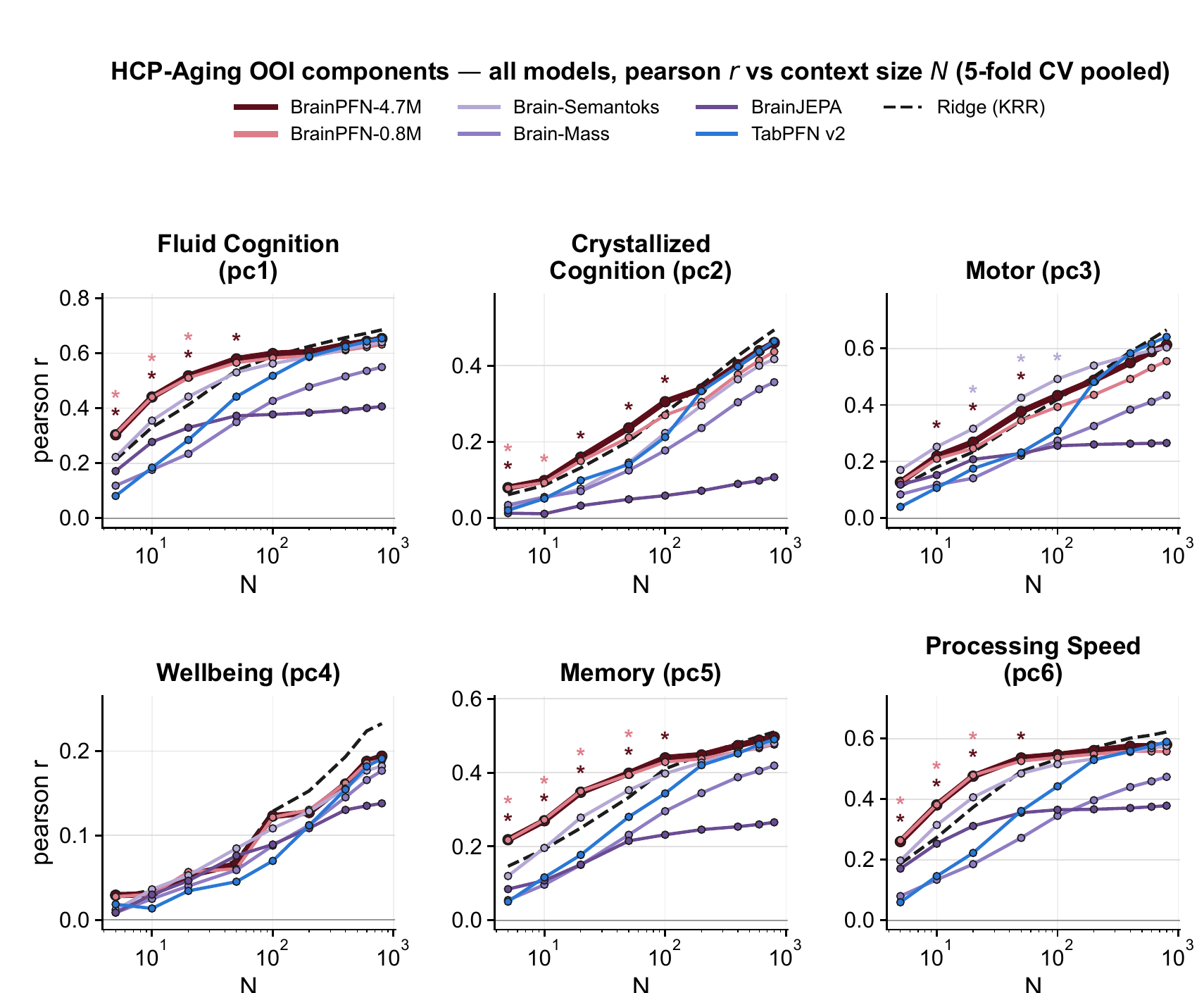}
    \caption{HCP-Aging OOI components: pearson $r$ vs. context size $N$ for all 7 models
    (5-fold CV pooled), same conventions as Figure~\ref{fig:ooi-comparison-hcp}.}
    \label{fig:ooi-comparison-hcp-aging}
\end{figure}

\paragraph{Model comparison.} Figures~\ref{fig:ooi-comparison-hcp}
and~\ref{fig:ooi-comparison-hcp-aging} show pearson $r$ vs. context size $N$ for all 7 models
on every OOI component, and Tables~\ref{tab:ooi-hcp} and~\ref{tab:ooi-hcp-aging} give the
corresponding \logauc{} summary (5-fold CV pooled, same protocol and significance test --
one-sided paired $t$-test over the 5 fold-level \logauc{} values -- as the rest of the paper).
BrainPFN is the most consistent winner on both cohorts: on HCP-Aging it beats Ridge on Fluid
Cognition, Memory, and Processing Speed, and is competitive on Crystallized Cognition and
Motor; on HCP-YA it beats Ridge on Cognition and Emotion Recognition. Wellbeing (HCP-Aging)
and Dissatisfaction (HCP-YA) are the two components no model beats Ridge on -- consistent
with the rest of the paper's finding that affect/self-report targets are the hardest domain
for FC-based prediction regardless of which model does the decoding. BrainJEPA is
consistently the weakest model on both cohorts' OOI components, in line with its performance
on the individual behavioural scores in Appendix~\ref{sec:appendix-all-score-model-comparison}.

\input{figures/ooi_model_comparison_hcp_table.tex}
\input{figures/ooi_model_comparison_hcp_aging_table.tex}

\paragraph{Component loadings.} Figures~\ref{fig:ooi-loadings-hcp}
and~\ref{fig:ooi-loadings-hcp-aging} show the top-7 rotated loadings (by $|{\rm loading}|$)
per component, reproduced from the saved PCA+varimax component loadings computed above.

\clearpage

%% file: figures/ooi_model_comparison_hcp_table.tex
\begin{table}[h!]
\centering
\caption{HCP-YA OOI components: \logauc{} (Pearson $r$ vs. context size $N$, 5-fold CV pooled). \textbf{Bold} = model beats the canonical Ridge row (one-sided paired $t$-test over 5 folds, $p<0.05$).}
\label{tab:ooi-hcp}
\begin{tabular}{lccc}
\toprule
Model & Dissatisfaction & Cognition & Emotion Recognition \\
\midrule
Ridge & 0.032 & 0.298 & 0.162 \\
\midrule
BrainPFN-4.7M & 0.034 & \textbf{0.325} & \textbf{0.186} \\
BrainPFN-0.8M & 0.031 & 0.302 & 0.175 \\
Brain-Semantoks & 0.017 & 0.265 & 0.150 \\
Brain-Mass & 0.017 & 0.194 & 0.099 \\
BrainJEPA & -0.005 & 0.017 & 0.009 \\
TabPFN v2 & 0.014 & 0.241 & 0.124 \\
\bottomrule
\end{tabular}
\end{table}

%% file: figures/ooi_model_comparison_hcp_aging_table.tex
\begin{table}[h!]
\centering
\caption{HCP-Aging OOI components: \logauc{} (Pearson $r$ vs. context size $N$, 5-fold CV pooled). \textbf{Bold} = model beats the canonical Ridge row (one-sided paired $t$-test over 5 folds, $p<0.05$).}
\label{tab:ooi-hcp-aging}
\resizebox{\linewidth}{!}{%
\begin{tabular}{lcccccc}
\toprule
Model & Fluid Cognition & Crystallized Cognition & Motor & Wellbeing & Memory & Processing Speed \\
\midrule
Ridge & 0.512 & 0.246 & 0.375 & 0.107 & 0.347 & 0.462 \\
\midrule
BrainPFN-4.7M & \textbf{0.550} & \textbf{0.256} & 0.384 & 0.095 & \textbf{0.390} & \textbf{0.499} \\
BrainPFN-0.8M & \textbf{0.538} & 0.235 & 0.349 & 0.095 & \textbf{0.383} & \textbf{0.492} \\
Brain-Semantoks & 0.502 & 0.194 & \textbf{0.424} & 0.094 & 0.343 & 0.456 \\
Brain-Mass & 0.356 & 0.164 & 0.243 & 0.079 & 0.246 & 0.290 \\
BrainJEPA & 0.346 & 0.053 & 0.223 & 0.078 & 0.196 & 0.328 \\
TabPFN v2 & 0.427 & 0.207 & 0.313 & 0.075 & 0.292 & 0.366 \\
\bottomrule
\end{tabular}
}
\end{table}

%% file: figures/openfmind_datasets_table.tex
\begin{longtable}{clp{9cm}}
\caption{OpenNeuro datasets included in the meta-training pool.} \label{tab:openfmind-datasets} \\
\toprule
\# & Accession & Name \\
\midrule
\endfirsthead
\toprule
\# & Accession & Name \\
\midrule
\endhead
\bottomrule
\endfoot
1 & \href{https://openneuro.org/datasets/ds000001}{ds000001} & Balloon Analog Risk-taking Task \\
2 & \href{https://openneuro.org/datasets/ds000002}{ds000002} & Classification learning \\
3 & \href{https://openneuro.org/datasets/ds000003}{ds000003} & Rhyme judgment \\
4 & \href{https://openneuro.org/datasets/ds000005}{ds000005} & Mixed-gambles task \\
5 & \href{https://openneuro.org/datasets/ds000006}{ds000006} & ds000006 \\
6 & \href{https://openneuro.org/datasets/ds000007}{ds000007} & Stop-signal task with spoken \& manual responses \\
7 & \href{https://openneuro.org/datasets/ds000008}{ds000008} & Stop-signal task with unconditional and conditional stopping \\
8 & \href{https://openneuro.org/datasets/ds000009}{ds000009} & The generality of self-control \\
9 & \href{https://openneuro.org/datasets/ds000011}{ds000011} & Classification learning and tone-counting \\
10 & \href{https://openneuro.org/datasets/ds000017}{ds000017} & ds000017 \\
11 & \href{https://openneuro.org/datasets/ds000030}{ds000030} & UCLA Consortium for Neuropsychiatric Phenomics LA5c Study \\
12 & \href{https://openneuro.org/datasets/ds000101}{ds000101} & Simon task \\
13 & \href{https://openneuro.org/datasets/ds000102}{ds000102} & Flanker task (event-related) \\
14 & \href{https://openneuro.org/datasets/ds000107}{ds000107} & Word and object processing \\
15 & \href{https://openneuro.org/datasets/ds000108}{ds000108} & Prefrontal-Subcortical Pathways Mediating Successful Emotion Regulation \\
16 & \href{https://openneuro.org/datasets/ds000109}{ds000109} & Social cognitive tasks (animate movement task, moral judgment task, false belief task) \\
17 & \href{https://openneuro.org/datasets/ds000110}{ds000110} & Incidental encoding task (Posner Cueing Paradigm) \\
18 & \href{https://openneuro.org/datasets/ds000113}{ds000113} & Forrest Gump \\
19 & \href{https://openneuro.org/datasets/ds000114}{ds000114} & A test-retest fMRI dataset for motor, language and spatial attention functions. \\
20 & \href{https://openneuro.org/datasets/ds000115}{ds000115} & Working memory in healthy and schizophrenic individuals \\
21 & \href{https://openneuro.org/datasets/ds000116}{ds000116} & Auditory and Visual Oddball EEG-fMRI \\
22 & \href{https://openneuro.org/datasets/ds000119}{ds000119} & Maturational Changes in Anterior Cingulate and Frontoparietal Recruitment Support the Development of Error Processing and Inhibitory Control (Antistate) \\
23 & \href{https://openneuro.org/datasets/ds000120}{ds000120} & Developmental changes in brain function underlying the influence of reward processing on inhibitory control (Slot Reward) \\
24 & \href{https://openneuro.org/datasets/ds000121}{ds000121} & Immaturities in Reward Processing and Its Influence on Inhibitory Control in Adolescence (Ring Reward) \\
25 & \href{https://openneuro.org/datasets/ds000133}{ds000133} & Modafinil alters intrinsic functional connectivity of the right posterior insula: a pharmacological resting state fMRI study \\
26 & \href{https://openneuro.org/datasets/ds000140}{ds000140} & Distinct brain systems mediate the effects of nociceptive input and self-regulation on pain \\
27 & \href{https://openneuro.org/datasets/ds000144}{ds000144} & Preschool Anxiety Disorders \\
28 & \href{https://openneuro.org/datasets/ds000148}{ds000148} & Who can afford self-control? The neural efficiency mechanism explains effective self-regulation of behavior \\
29 & \href{https://openneuro.org/datasets/ds000149}{ds000149} & Scene Perception \\
30 & \href{https://openneuro.org/datasets/ds000157}{ds000157} & Block design food and nonfood picture viewing task \\
31 & \href{https://openneuro.org/datasets/ds000158}{ds000158} & The human Voice Areas: spatial organisation and inter-individual variability in temporal and extra-temporal cortices \\
32 & \href{https://openneuro.org/datasets/ds000164}{ds000164} & Stroop Task \\
33 & \href{https://openneuro.org/datasets/ds000168}{ds000168} & Offline Processing in Associative Learning \\
34 & \href{https://openneuro.org/datasets/ds000171}{ds000171} & Neural Processing of Emotional Musical and Nonmusical Stimuli in Depression \\
35 & \href{https://openneuro.org/datasets/ds000172}{ds000172} & Physiological Contribution in Spontaneous Oscillations: An Approximate Quality - Assurance Index for Resting-State fMRI Signals \\
36 & \href{https://openneuro.org/datasets/ds000200}{ds000200} & Pre-adolescents Exposure to Manganese \\
37 & \href{https://openneuro.org/datasets/ds000205}{ds000205} & Affective Videos \\
38 & \href{https://openneuro.org/datasets/ds000208}{ds000208} & Brain connectivity predicts placebo response across chronic pain clinical trials \\
39 & \href{https://openneuro.org/datasets/ds000210}{ds000210} & Multi-echo fMRI replication sample of autobiographical memory, prospection and theory of mind reasoning tasks \\
40 & \href{https://openneuro.org/datasets/ds000212}{ds000212} & Moral judgments of intentional and accidental moral violations across Harm and Purity domains \\
41 & \href{https://openneuro.org/datasets/ds000213}{ds000213} & Neural mechanism underlying appearance social comparison \\
42 & \href{https://openneuro.org/datasets/ds000214}{ds000214} & EUPD Cyberball \\
43 & \href{https://openneuro.org/datasets/ds000218}{ds000218} & Flavour Pleasantness (Oral Nutritional Supplements) \\
44 & \href{https://openneuro.org/datasets/ds000219}{ds000219} & Flavour Pleasantness (Regular Products) \\
45 & \href{https://openneuro.org/datasets/ds000220}{ds000220} & Cost Analysis TBI \\
46 & \href{https://openneuro.org/datasets/ds000221}{ds000221} & MPI-Leipzig\_Mind-Brain-Body \\
47 & \href{https://openneuro.org/datasets/ds000223}{ds000223} & Magnitude Effect \\
48 & \href{https://openneuro.org/datasets/ds000228}{ds000228} & MRI data of 3-12 year old children and adults during viewing of a short animated film \\
49 & \href{https://openneuro.org/datasets/ds000233}{ds000233} & Neural responses to naturalistic clips of behaving animals in two different task contexts \\
50 & \href{https://openneuro.org/datasets/ds000243}{ds000243} & Washington University 120 \\
51 & \href{https://openneuro.org/datasets/ds000245}{ds000245} & ds000245\_R1.0.0 \\
52 & \href{https://openneuro.org/datasets/ds000249}{ds000249} & Value generalization in human avoidance learning \\
53 & \href{https://openneuro.org/datasets/ds000253}{ds000253} & Female action video game players. \\
54 & \href{https://openneuro.org/datasets/ds000254}{ds000254} & Multiband Multi-Echo Simultaneous ASL/BOLD for task-induced functional MRI \\
55 & \href{https://openneuro.org/datasets/ds000256}{ds000256} & Behavioral interventions for reducing head motion during MRI scans in children \\
56 & \href{https://openneuro.org/datasets/ds000258}{ds000258} & Multi-echo Cambridge \\
57 & \href{https://openneuro.org/datasets/ds001021}{ds001021} & NKI-sample \\
58 & \href{https://openneuro.org/datasets/ds001131}{ds001131} & Milky-Vodka \\
59 & \href{https://openneuro.org/datasets/ds001132}{ds001132} & Sherlock \\
60 & \href{https://openneuro.org/datasets/ds001145}{ds001145} & Twilight Zone Movie Watching Dataset \\
61 & \href{https://openneuro.org/datasets/ds001226}{ds001226} & BTC\_preop \\
62 & \href{https://openneuro.org/datasets/ds001229}{ds001229} & Neural Evidence of the Strategic Choice Between Working Memory and Episodic Memory in Prospective Remembering. \\
63 & \href{https://openneuro.org/datasets/ds001232}{ds001232} & Competition between items in working memory leads to forgetting \\
64 & \href{https://openneuro.org/datasets/ds001235}{ds001235} & Human-Object Interaction \\
65 & \href{https://openneuro.org/datasets/ds001242}{ds001242} & Examining effects of arousal on responses to salient and non-salient stimuli in younger and older adults \\
66 & \href{https://openneuro.org/datasets/ds001297}{ds001297} & Personally familiar faces \\
67 & \href{https://openneuro.org/datasets/ds001302}{ds001302} & Inverse Base Rate \\
68 & \href{https://openneuro.org/datasets/ds001338}{ds001338} & ParanoiaStory \\
69 & \href{https://openneuro.org/datasets/ds001344}{ds001344} & phfMRI\_Control\_Tasks \\
70 & \href{https://openneuro.org/datasets/ds001345}{ds001345} & Decoding of multisensory semantics and memories in low-level visual cortex \\
71 & \href{https://openneuro.org/datasets/ds001353}{ds001353} & Information Filtering in Electronic Networks of Practice: An fMRI Investigation of Expectation [Dis]confirmation \\
72 & \href{https://openneuro.org/datasets/ds001357}{ds001357} & NameFacePriming \\
73 & \href{https://openneuro.org/datasets/ds001379}{ds001379} & Loci case study \\
74 & \href{https://openneuro.org/datasets/ds001399}{ds001399} & multibandACQtests \\
75 & \href{https://openneuro.org/datasets/ds001408}{ds001408} & rsfMRI\_single\_session\_EEG\_NF \\
76 & \href{https://openneuro.org/datasets/ds001419}{ds001419} & thoughtExperiment \\
77 & \href{https://openneuro.org/datasets/ds001430}{ds001430} & Neural Overlap in Item Representations Across Episodes Impairs Context Memory \\
78 & \href{https://openneuro.org/datasets/ds001439}{ds001439} & A role for the medial temporal lobe subsystem in guiding prosociality: the effect of episodic processes on willingness to help others \\
79 & \href{https://openneuro.org/datasets/ds001454}{ds001454} & Human hippocampal replay during rest prioritizes weakly learned information and predicts memory performance \\
80 & \href{https://openneuro.org/datasets/ds001491}{ds001491} & Valence processing differs across stimulus modalities (Multi-echo) \\
81 & \href{https://openneuro.org/datasets/ds001521}{ds001521} & Lingering representations of stimuli influence recall organization \\
82 & \href{https://openneuro.org/datasets/ds001534}{ds001534} & Calorie-labeled food cues \\
83 & \href{https://openneuro.org/datasets/ds001545}{ds001545} & Learning Naturalistic Temporal Structure in the Posterior Medial Network \\
84 & \href{https://openneuro.org/datasets/ds001553}{ds001553} & 100 runs at 3T \\
85 & \href{https://openneuro.org/datasets/ds001554}{ds001554} & Neural pattern change during encoding of a narrative predicts retrospective duration estimates \\
86 & \href{https://openneuro.org/datasets/ds001566}{ds001566} & test \\
87 & \href{https://openneuro.org/datasets/ds001576}{ds001576} & Refresh my memory: Episodic memory reinstatements intrude on working memory maintenance \\
88 & \href{https://openneuro.org/datasets/ds001607}{ds001607} & Reinstated episodic context guides sampling-based decisions for reward \\
89 & \href{https://openneuro.org/datasets/ds001608}{ds001608} & Prestimulus Exp2 \\
90 & \href{https://openneuro.org/datasets/ds001612}{ds001612} & Offline replay supports planning in human reinforcement learning \\
91 & \href{https://openneuro.org/datasets/ds001621}{ds001621} & Neural representations of events arise from temporal community structure \\
92 & \href{https://openneuro.org/datasets/ds001634}{ds001634} & Visual working memory: Study one Task fMRI and Behavioural response \\
93 & \href{https://openneuro.org/datasets/ds001635}{ds001635} & Visual working memory: Study two Task fMRI and Behavioural response \\
94 & \href{https://openneuro.org/datasets/ds001715}{ds001715} & Dissociable neural mechanisms track evidence accumulation for selection of attention versus action \\
95 & \href{https://openneuro.org/datasets/ds001722}{ds001722} & A role for the medial temporal lobe subsystem in guiding prosociality: the effect of episodic processes on willingness to help others (Experiment 2) \\
96 & \href{https://openneuro.org/datasets/ds001728}{ds001728} & Test-GEIER \\
97 & \href{https://openneuro.org/datasets/ds001740}{ds001740} & Brain activity during reciprocal social interaction investigated using conversational robots as control condition \\
98 & \href{https://openneuro.org/datasets/ds001745}{ds001745} & A neural signature of contextually mediated intentional forgetting \\
99 & \href{https://openneuro.org/datasets/ds001747}{ds001747} & Exploring the Resting State Neural Activity of Monolinguals and Late and Early Bilinguals \\
100 & \href{https://openneuro.org/datasets/ds001748}{ds001748} & Differentiation of functional networks during long-term memory retrieval in children and adolescents \\
101 & \href{https://openneuro.org/datasets/ds001762}{ds001762} & Dorsal anterior cingulate and ventromedial prefrontal cortex have inverse roles in both foraging and economic choice \\
102 & \href{https://openneuro.org/datasets/ds001775}{ds001775} & Neural Differentiation Tracks Improved Recall of Competing Memories Following Interleaved Study and Retrieval Practice \\
103 & \href{https://openneuro.org/datasets/ds001796}{ds001796} & Bilingualism and the brain \\
104 & \href{https://openneuro.org/datasets/ds001838}{ds001838} & Handedness and Symbolic Number Representation \\
105 & \href{https://openneuro.org/datasets/ds001840}{ds001840} & Prestimulus Exp1 \\
106 & \href{https://openneuro.org/datasets/ds001847}{ds001847} & A Probability Distribution over Latent Causes, in the Orbitofrontal Cortex \\
107 & \href{https://openneuro.org/datasets/ds001912}{ds001912} & S-cone\_MotionInDepth\_fMRI \\
108 & \href{https://openneuro.org/datasets/ds001921}{ds001921} & Anterior cingulate engagement in a foraging context reflects choice difficulty, not foraging value \\
109 & \href{https://openneuro.org/datasets/ds001923}{ds001923} & Anterior cingulate engagement in a foraging context reflects choice difficulty, not foraging value (2) \\
110 & \href{https://openneuro.org/datasets/ds001927}{ds001927} & The physiological effects of non-invasive brain stimulation fundamentally differ across the human cortex \\
111 & \href{https://openneuro.org/datasets/ds001928}{ds001928} & Functional Connectivity of Music-Induced Analgesia in Fibromyalgia \\
112 & \href{https://openneuro.org/datasets/ds002011}{ds002011} & Locus coeruleus activity strengthens prioritized memories under arousal \\
113 & \href{https://openneuro.org/datasets/ds002014}{ds002014} & Language Production fMRI \\
114 & \href{https://openneuro.org/datasets/ds002041}{ds002041} & Mesolimbic dopamine D2 receptors and neural representations of subjective value \\
115 & \href{https://openneuro.org/datasets/ds002080}{ds002080} & BTC\_postop \\
116 & \href{https://openneuro.org/datasets/ds002149}{ds002149} & Conditional Visual Associative Learning Task \\
117 & \href{https://openneuro.org/datasets/ds002232}{ds002232} & Caltech rsfMRI Dataset \\
118 & \href{https://openneuro.org/datasets/ds002250}{ds002250} & When my wrongs are worse than yours: behavioral and neural asymmetries in first-person and third-person perspectives of accidents \\
119 & \href{https://openneuro.org/datasets/ds002270}{ds002270} & Viewing\_food\_cues \\
120 & \href{https://openneuro.org/datasets/ds002322}{ds002322} & The Alice Dataset: fMRI Dataset to Study Natural Language Comprehension in the Brain \\
121 & \href{https://openneuro.org/datasets/ds002345}{ds002345} & Narratives \\
122 & \href{https://openneuro.org/datasets/ds002363}{ds002363} & Repetition of Computer Security Warnings Results in Differential Repetition Suppression Effects as Revealed with Functional MRI \\
123 & \href{https://openneuro.org/datasets/ds002522}{ds002522} & Contrast Response Functions \\
124 & \href{https://openneuro.org/datasets/ds002547}{ds002547} & SharedStates \\
125 & \href{https://openneuro.org/datasets/ds002603}{ds002603} & MB-EPI-comparison \\
126 & \href{https://openneuro.org/datasets/ds002609}{ds002609} & RPN-signature\_Study2 \\
127 & \href{https://openneuro.org/datasets/ds002643}{ds002643} & Can we have a second helping? A replication study on the neurobiological mechanisms underlying self-control \\
128 & \href{https://openneuro.org/datasets/ds002711}{ds002711} & Smiley \\
129 & \href{https://openneuro.org/datasets/ds002715}{ds002715} & AVScenes\_Blind \\
130 & \href{https://openneuro.org/datasets/ds002717}{ds002717} & Unexplained Repeated Pregnancy Loss is Associated with Altered Perceptual and Brain Responses to Men’s Body-Odor \\
131 & \href{https://openneuro.org/datasets/ds002731}{ds002731} & Multiple interactive memory representations underlie the induction of false memory \\
132 & \href{https://openneuro.org/datasets/ds002732}{ds002732} & Unnamed Dataset \\
133 & \href{https://openneuro.org/datasets/ds002734}{ds002734} & Evidence Accumulation in Value-Based decisions \\
134 & \href{https://openneuro.org/datasets/ds002737}{ds002737} & multibandCFtests \\
135 & \href{https://openneuro.org/datasets/ds002748}{ds002748} & Resting state with closed eyes for patients with depression and healthy participants \\
136 & \href{https://openneuro.org/datasets/ds002750}{ds002750} & YanDataBIDS \\
137 & \href{https://openneuro.org/datasets/ds002785}{ds002785} & AOMIC-PIOP1 \\
138 & \href{https://openneuro.org/datasets/ds002790}{ds002790} & AOMIC-PIOP2 \\
139 & \href{https://openneuro.org/datasets/ds002793}{ds002793} & ToM following strong vs weak priors \\
140 & \href{https://openneuro.org/datasets/ds002797}{ds002797} & eLife39497 \\
141 & \href{https://openneuro.org/datasets/ds002835}{ds002835} & Prospection \\
142 & \href{https://openneuro.org/datasets/ds002837}{ds002837} & Naturalistic Neuroimaging Database \\
143 & \href{https://openneuro.org/datasets/ds002841}{ds002841} & Intuitive physics with fMRI \\
144 & \href{https://openneuro.org/datasets/ds002843}{ds002843} & Cognitive Training \\
145 & \href{https://openneuro.org/datasets/ds002848}{ds002848} & mriquality \\
146 & \href{https://openneuro.org/datasets/ds002872}{ds002872} & Aging \\
147 & \href{https://openneuro.org/datasets/ds002896}{ds002896} & Resting State fMRI study of non bothersome tinnitus \\
148 & \href{https://openneuro.org/datasets/ds002938}{ds002938} & Task-Dependent Effort-Induced Connectivity \\
149 & \href{https://openneuro.org/datasets/ds003007}{ds003007} & Two sessions of resting state with closed eyes for patients with depression in treatment course (NFB, CBT or No treatment groups) \\
150 & \href{https://openneuro.org/datasets/ds003011}{ds003011} & Social Processes Initiative in Neurobiology of the Schizophrenia(s) Traveling Human Phantoms \\
151 & \href{https://openneuro.org/datasets/ds003012}{ds003012} & action-in-auctions \\
152 & \href{https://openneuro.org/datasets/ds003037}{ds003037} & SUDMEX\_TMS \\
153 & \href{https://openneuro.org/datasets/ds003076}{ds003076} & Brain Correlates of Deductive Reasoning in Adults \\
154 & \href{https://openneuro.org/datasets/ds003083}{ds003083} & Brain Correlates of Math Processing in Adults \\
155 & \href{https://openneuro.org/datasets/ds003085}{ds003085} & Temporal Dynamics of Emotional Music \\
156 & \href{https://openneuro.org/datasets/ds003096}{ds003096} & Valuation of peers' safe choices is associated with substance-naivete in adolescents' \\
157 & \href{https://openneuro.org/datasets/ds003103}{ds003103} & Consensus-seeking and conflict-resolving: an fMRI study on college couples’ shopping interaction \\
158 & \href{https://openneuro.org/datasets/ds003114}{ds003114} & How ovarian hormones influence the behaviroal activation and inhibition system through the dopamine pathway \\
159 & \href{https://openneuro.org/datasets/ds003126}{ds003126} & Reading-related functional activity in children with isolated spelling deficits and dyslexia \\
160 & \href{https://openneuro.org/datasets/ds003146}{ds003146} & vocal task \\
161 & \href{https://openneuro.org/datasets/ds003171}{ds003171} & Modeling an auditory stimulated brain under altered states of consciousness using the generalized ising model \\
162 & \href{https://openneuro.org/datasets/ds003346}{ds003346} & SUDMEX\_CONN: The Mexican dataset of cocaine use disorder patients. \\
163 & \href{https://openneuro.org/datasets/ds003354}{ds003354} & EmpaToM with PPI-R, TAS, and EQ \\
164 & \href{https://openneuro.org/datasets/ds003381}{ds003381} & Estimation of in-scanner head pose changes during structural MRI using a convolutional neural network trained on eye tracker video \\
165 & \href{https://openneuro.org/datasets/ds003382}{ds003382} & Monash vis-fPET-fMRI \\
166 & \href{https://openneuro.org/datasets/ds003425}{ds003425} & Mental imagery can generate and regulate acquired differential fear conditioned reactivity \\
167 & \href{https://openneuro.org/datasets/ds003427}{ds003427} & Effects of Phase Regression on High-Resolution Functional MRI of the Primary Visual Cortex \\
168 & \href{https://openneuro.org/datasets/ds003436}{ds003436} & Agreeableness personality trait and social information encoding \\
169 & \href{https://openneuro.org/datasets/ds003459}{ds003459} & Component processes of word reading in adults and children \\
170 & \href{https://openneuro.org/datasets/ds003466}{ds003466} & Haptic three-dimensional curved surface exploration fMRI dataset \\
171 & \href{https://openneuro.org/datasets/ds003469}{ds003469} & Speech disfluencies: Neurophysiological aspect in normal population \\
172 & \href{https://openneuro.org/datasets/ds003495}{ds003495} & dStreamUpgrade \\
173 & \href{https://openneuro.org/datasets/ds003500}{ds003500} & Response inhibition and selective attention in adults and children with and without ADHD \\
174 & \href{https://openneuro.org/datasets/ds003542}{ds003542} & Adult language learners \\
175 & \href{https://openneuro.org/datasets/ds003545}{ds003545} & Adolescent language learners \\
176 & \href{https://openneuro.org/datasets/ds003548}{ds003548} & Emotion Category and Face Perception Task Optimized for Multivariate Pattern Analysis \\
177 & \href{https://openneuro.org/datasets/ds003550}{ds003550} & Visser, Scholte, Beemsterboer, \& Kindt (2013) Nature Neuroscience \\
178 & \href{https://openneuro.org/datasets/ds003553}{ds003553} & Visser, Scholte \& Kindt (2011) Journal of Neuroscience \\
179 & \href{https://openneuro.org/datasets/ds003554}{ds003554} & Visser et al. 2015 Psychoneuroendocrinology \\
180 & \href{https://openneuro.org/datasets/ds003574}{ds003574} & Reward biases spontaneous neural reactivation during sleep \\
181 & \href{https://openneuro.org/datasets/ds003592}{ds003592} & Neurocognitive aging data release with behavioral, structural, and multi-echo functional MRI measures \\
182 & \href{https://openneuro.org/datasets/ds003688}{ds003688} & Open multimodal iEEG-fMRI dataset from naturalistic stimulation with a short audiovisual film \\
183 & \href{https://openneuro.org/datasets/ds003709}{ds003709} & NIMH-CompyPsych MMI \\
184 & \href{https://openneuro.org/datasets/ds003763}{ds003763} & Interoception during aging: The heartbeat detection task \\
185 & \href{https://openneuro.org/datasets/ds003770}{ds003770} & Depression Treatment by Medial Prefrontal Real-Time fMRI Neurofeedback \\
186 & \href{https://openneuro.org/datasets/ds003772}{ds003772} & Functionally dissociable influences on learning rate in a dynamic environment \\
187 & \href{https://openneuro.org/datasets/ds003789}{ds003789} & Dynamic changes in neural representations underlie the repetition eﬀect on false memory \\
188 & \href{https://openneuro.org/datasets/ds003831}{ds003831} & Cognitive Control Theoretic Mechanisms of Real-time fMRI-Guided Neuromodulation (CTM) \\
189 & \href{https://openneuro.org/datasets/ds003835}{ds003835} & PE-Update \\
190 & \href{https://openneuro.org/datasets/ds003849}{ds003849} & Individual differences in frontoparietal plasticity in humans \\
191 & \href{https://openneuro.org/datasets/ds003858}{ds003858} & Monetary Incentive Delay task \\
192 & \href{https://openneuro.org/datasets/ds003871}{ds003871} & Resting-state for 34 younger and 28 older adults \\
193 & \href{https://openneuro.org/datasets/ds003892}{ds003892} & Loss and enhancement of layer-selective signals in geniculostriate and corticotectal pathways of adult human amblyopia \\
194 & \href{https://openneuro.org/datasets/ds003965}{ds003965} & NI-edu-data-minimal \\
195 & \href{https://openneuro.org/datasets/ds003972}{ds003972} & Body Motor \\
196 & \href{https://openneuro.org/datasets/ds003990}{ds003990} & Fast Event-Related Mapping of Population Fingertip Tuning Properties in Human Sensorimotor Cortex at 7T \\
197 & \href{https://openneuro.org/datasets/ds003999}{ds003999} & Pre-Post rehabilitation fMRI data of post-stroke patients. \\
198 & \href{https://openneuro.org/datasets/ds004021}{ds004021} & Visuomotor Rotation Adaptation Experiment \\
199 & \href{https://openneuro.org/datasets/ds004101}{ds004101} & Resting-state fMRI data acquired before and after a course of infra-low frequency neurofeedback \\
200 & \href{https://openneuro.org/datasets/ds004128}{ds004128} & Neural representations of others' traits predict social decisions \\
201 & \href{https://openneuro.org/datasets/ds004134}{ds004134} & Experience Sampling in Resting State \\
202 & \href{https://openneuro.org/datasets/ds004141}{ds004141} & Targeting Dynamic Facial Processing Mechanisms in Superior Temporal Sulcus Using fMRI Neurofeedback \\
203 & \href{https://openneuro.org/datasets/ds004142}{ds004142} & Exploring the Neural Correlates of Feedback-Related Reward Saliency and Valence During Real-Time fMRI-Based Neurofeedback \\
204 & \href{https://openneuro.org/datasets/ds004144}{ds004144} & A behavioral, clinical and brain imaging dataset with focus on emotion regulation of females with fibromyalgia \\
205 & \href{https://openneuro.org/datasets/ds004182}{ds004182} & Magic, Memory, and Curiosity (MMC) fMRI Dataset \\
206 & \href{https://openneuro.org/datasets/ds004196}{ds004196} & Bimodal dataset on Inner speech \\
207 & \href{https://openneuro.org/datasets/ds004213}{ds004213} & BrainBeats \\
208 & \href{https://openneuro.org/datasets/ds004219}{ds004219} & Litebook\_Alertness\_Study \\
209 & \href{https://openneuro.org/datasets/ds004302}{ds004302} & Brain correlates of speech perception in schizophrenia patients with and without auditory hallucinations \\
210 & \href{https://openneuro.org/datasets/ds004392}{ds004392} & Parkinson's disease, functional connectivity, and cognition \\
211 & \href{https://openneuro.org/datasets/ds004400}{ds004400} & Number comparison task \\
212 & \href{https://openneuro.org/datasets/ds004440}{ds004440} & Stanford Child and Adult Checkerboard Retinotopy Dataset \\
213 & \href{https://openneuro.org/datasets/ds004450}{ds004450} & XCP\_WALKTHROUGH \\
214 & \href{https://openneuro.org/datasets/ds004467}{ds004467} & MRI data from 20 adults in response to videos of dialogue and monologue from Sesame Street \\
215 & \href{https://openneuro.org/datasets/ds004469}{ds004469} & Temporal Lobe Epilepsy - UNAM \\
216 & \href{https://openneuro.org/datasets/ds004498}{ds004498} & Perinatal Stroke \\
217 & \href{https://openneuro.org/datasets/ds004544}{ds004544} & An fMRI Investigation of the Neurocognitive Processing of Strategies and Measures \\
218 & \href{https://openneuro.org/datasets/ds004557}{ds004557} & The role of superstition of cognitive control during neurofeedback training - part 2 \\
219 & \href{https://openneuro.org/datasets/ds004562}{ds004562} & MVPA for motor adaptation \\
220 & \href{https://openneuro.org/datasets/ds004564}{ds004564} & Demo \\
221 & \href{https://openneuro.org/datasets/ds004594}{ds004594} & Brain surface fingerprints \\
222 & \href{https://openneuro.org/datasets/ds004604}{ds004604} & Cerebrovascular Reactivity Normative Dataset \\
223 & \href{https://openneuro.org/datasets/ds004627}{ds004627} & Characterization and Treatment of Adolescent Depression (CAT-D) \\
224 & \href{https://openneuro.org/datasets/ds004648}{ds004648} & Sympathovagal quotient and functional connectivity of control networks are related to gut Ruminococcaceae abundance in healthy men \\
225 & \href{https://openneuro.org/datasets/ds004656}{ds004656} & Neural reactivity to visual food stimuli is reduced in some areas of the brain during evening hours compared to morning hours: an fMRI study in women \\
226 & \href{https://openneuro.org/datasets/ds004662}{ds004662} & Denoising task-correlated head motion from motor-task fMRI data with multi-echo ICA \\
227 & \href{https://openneuro.org/datasets/ds004697}{ds004697} & Food and Brain Study \\
228 & \href{https://openneuro.org/datasets/ds004718}{ds004718} & Le Petit Prince Hong Kong: Naturalistic fMRI and EEG dataset from older Cantonese speakers \\
229 & \href{https://openneuro.org/datasets/ds004720}{ds004720} & Descriptive name for this dataset \\
230 & \href{https://openneuro.org/datasets/ds004737}{ds004737} & CS-DSI \\
231 & \href{https://openneuro.org/datasets/ds004765}{ds004765} & language fMRI \\
232 & \href{https://openneuro.org/datasets/ds004775}{ds004775} & Vicarious Punishment of Moral Violations in Naturalistic Drama Narratives Predicts Cortical Synchronization \\
233 & \href{https://openneuro.org/datasets/ds004783}{ds004783} & Functional MRI dataset for evaluating brain synchrony among dance spectators \\
234 & \href{https://openneuro.org/datasets/ds004798}{ds004798} & fMRI data for: Multimodal single-neuron, intracranial EEG, and fMRI brain responses during movie watching in human patients \\
235 & \href{https://openneuro.org/datasets/ds004835}{ds004835} & Processing of frequency and location in human subcortical auditory structures \\
236 & \href{https://openneuro.org/datasets/ds004848}{ds004848} & Game of Thrones - A naturalistic viewing dataset \\
237 & \href{https://openneuro.org/datasets/ds004873}{ds004873} & BOLD signal changes can oppose oxygen metabolism across the human cortex \\
238 & \href{https://openneuro.org/datasets/ds004956}{ds004956} & Braille letters - fMRI \\
239 & \href{https://openneuro.org/datasets/ds005027}{ds005027} & Michigan Longitudinal Study \\
240 & \href{https://openneuro.org/datasets/ds005040}{ds005040} & Political Moralization (PMM) \\
241 & \href{https://openneuro.org/datasets/ds005069}{ds005069} & test \\
242 & \href{https://openneuro.org/datasets/ds005072}{ds005072} & test \\
243 & \href{https://openneuro.org/datasets/ds005075}{ds005075} & Simultaneous Brain and Spinal Cord at rest: an fMRI dataset \\
244 & \href{https://openneuro.org/datasets/ds005128}{ds005128} & Deception Signaling Task \\
245 & \href{https://openneuro.org/datasets/ds005134}{ds005134} & Sequence test for multiband DWI and magnetization transfer sequence \\
246 & \href{https://openneuro.org/datasets/ds005166}{ds005166} & A framework for leveraging technology to probe mechanism: A proof of concept study of inhibitory control \\
247 & \href{https://openneuro.org/datasets/ds005230}{ds005230} & Reinforcement-Learning Generalization \\
248 & \href{https://openneuro.org/datasets/ds005239}{ds005239} & motor-fmri \\
249 & \href{https://openneuro.org/datasets/ds005270}{ds005270} & BOLD variability during cognitive control for an adult lifespan sample \\
250 & \href{https://openneuro.org/datasets/ds005339}{ds005339} & Yale\_reading\_dataset \\
251 & \href{https://openneuro.org/datasets/ds005355}{ds005355} & BSC\_BIDS\_071724 \\
252 & \href{https://openneuro.org/datasets/ds005365}{ds005365} & Neurobiological substrates of Altered State of Consciousness Induced by High Ventilation Breathwork \\
253 & \href{https://openneuro.org/datasets/ds005418}{ds005418} & Cerebrovascular Reactivity: CO2 inhalation and resting state comparisons \\
254 & \href{https://openneuro.org/datasets/ds005454}{ds005454} & fMRIPrep boot camp -- 7T training dataset \\
255 & \href{https://openneuro.org/datasets/ds005455}{ds005455} & An fMRI dataset for investigating language control and cognitive control in bilinguals \\
256 & \href{https://openneuro.org/datasets/ds005479}{ds005479} & Monetary Incentive Delay task - structural and functional images of 37 men; study of associations between circadian characteristics (eveningness, distinctness) and affective processing \\
257 & \href{https://openneuro.org/datasets/ds005504}{ds005504} & Psychosocial adversity and inhibitory control: an fMRI study of children growing up in extreme poverty \\
258 & \href{https://openneuro.org/datasets/ds005518}{ds005518} & Deciphering the neural responses to a naturalistic persuasive message \\
259 & \href{https://openneuro.org/datasets/ds005525}{ds005525} & Circadian misalignment and energy balance \\
260 & \href{https://openneuro.org/datasets/ds005530}{ds005530} & Depotentiation of emotional reactivity using TMR during REM sleep \\
261 & \href{https://openneuro.org/datasets/ds005533}{ds005533} & High-quality multimodal MRI with simultaneous EEG using conductive ink and polymer-thick film nets \\
262 & \href{https://openneuro.org/datasets/ds005573}{ds005573} & A comparison of resting state functional magnetic resonance imaging to invasive electrocortical stimulation for sensorimotor mapping in pediatric patients \\
263 & \href{https://openneuro.org/datasets/ds005588}{ds005588} & SoDec - Responsibility fMRI experiment dataset \\
264 & \href{https://openneuro.org/datasets/ds005596}{ds005596} & A naturalistic fMRI dataset in response to puclic speaking \\
265 & \href{https://openneuro.org/datasets/ds005598}{ds005598} & Visuomotor Rotation Learning and Reward-based motor learning \\
266 & \href{https://openneuro.org/datasets/ds005603}{ds005603} & Mitchell\_Hacker\_2013 \\
267 & \href{https://openneuro.org/datasets/ds005623}{ds005623} & test \\
268 & \href{https://openneuro.org/datasets/ds005625}{ds005625} & Connectomix test dataset 1 \\
269 & \href{https://openneuro.org/datasets/ds005699}{ds005699} & Connectomix test dataset 2 \\
270 & \href{https://openneuro.org/datasets/ds005754}{ds005754} & Multivariate Assessment of Inhibitory Control in Youth: Links with Psychopathology and Brain Function Dataset \\
271 & \href{https://openneuro.org/datasets/ds005783}{ds005783} & Representation of navigational affordances and ego-motion in the occipital place area - dataset \\
272 & \href{https://openneuro.org/datasets/ds005795}{ds005795} & MULTI-CLARID (Multimodal Category Learning and Resting-state Imaging Data) \\
273 & \href{https://openneuro.org/datasets/ds005850}{ds005850} & Spoken narrative comprehension in Chinese \\
274 & \href{https://openneuro.org/datasets/ds005882}{ds005882} & Integration of contextual reward motivation with category representations \\
275 & \href{https://openneuro.org/datasets/ds005883}{ds005883} & CoSpine Database\_pain\_dataset \\
276 & \href{https://openneuro.org/datasets/ds005884}{ds005884} & CoSpine Database\_motor\_dataset \\
277 & \href{https://openneuro.org/datasets/ds005892}{ds005892} & Resting State MRI data from healthy control (HC), Parkinson's disease with normal cognition (PD-NC), and Parkinson's disease with mild cognitive impairment (PD-MCI) cohorts \\
278 & \href{https://openneuro.org/datasets/ds006105}{ds006105} & Skewed Gambling Task: Deconstructing neural predictors of risky choice \\
279 & \href{https://openneuro.org/datasets/ds006156}{ds006156} & Executive Functioning-Related Neural Processes in Emerging Adults with Type 1 Diabetes: A Functional Magnetic Resonance Imaging Study \\
280 & \href{https://openneuro.org/datasets/ds006193}{ds006193} & PAFIN: PennLINC AFfective INstability (Test Dataset) \\
281 & \href{https://openneuro.org/datasets/ds006211}{ds006211} & Brain bases for navigating acoustic features - fMRI dataset \\
282 & \href{https://openneuro.org/datasets/ds006559}{ds006559} & Attitude Change Study 1 Persuade \\
283 & \href{https://openneuro.org/datasets/ds006568}{ds006568} & Attitude Change Study 2 Debate \\
284 & \href{https://openneuro.org/datasets/ds006572}{ds006572} & RSfMRI\_Vision \\
285 & \href{https://openneuro.org/datasets/ds006592}{ds006592} & The First Comprehensive Study of Early-Emerging Prosopometamorphopsia \\
286 & \href{https://openneuro.org/datasets/ds006594}{ds006594} & RSfMRI\_Function \\
287 & \href{https://openneuro.org/datasets/ds006729}{ds006729} & R01\_Neuromuscular\_Signature\_Pilot \\
288 & \href{https://openneuro.org/datasets/ds006893}{ds006893} & The interplay of central insulin and menstrual cycle on functional brain networks and neural food cue reactivity in women \\
289 & \href{https://openneuro.org/datasets/ds006944}{ds006944} & One-person corticospinal fMRI dataset for Stanford Spinal Cord workshop 2025 \\
290 & \href{https://openneuro.org/datasets/ds006945}{ds006945} & Dataset: T1-Weighted Structural MRI and fMRI of Participants Viewing Self-Avatar Exoskeleton Walking (11 SWS Cycles) \\
291 & \href{https://openneuro.org/datasets/ds006952}{ds006952} & PSYCH-REST \\
292 & \href{https://openneuro.org/datasets/ds006971}{ds006971} & Shoulder abduction motor-task fMRI in healthy participants \\
293 & \href{https://openneuro.org/datasets/ds007048}{ds007048} & Whole-Brain Auditory Mapping with 3T Multi-Echo fMRI in healthy participants \\
294 & \href{https://openneuro.org/datasets/ds007115}{ds007115} & Neural Mechanisms of Cognitive Conflict: Processing COVID-19 Vaccine Misinformation \\
295 & \href{https://openneuro.org/datasets/ds007126}{ds007126} & An open fMRI dataset examining the effects of online social and non-social information distraction on global-local attention \\
296 & \href{https://openneuro.org/datasets/ds007135}{ds007135} & Investigating Neuroepigenetic Alterations in Chronic Low Back Pain with PET/MR \\
297 & \href{https://openneuro.org/datasets/ds007185}{ds007185} & Neural Responses to Curiosity in Young Children \\
298 & \href{https://openneuro.org/datasets/ds007313}{ds007313} & Brain and spinal cord fMRI and qMRI - Single participant \\
299 & \href{https://openneuro.org/datasets/ds007345}{ds007345} & Multimodal Brain MRI and Clinical Data in Olfactory Groove Meningioma: A Prospective Data Report \\
300 & \href{https://openneuro.org/datasets/ds007441}{ds007441} & Human Dignity \\
301 & \href{https://openneuro.org/datasets/ds007534}{ds007534} & Neuroil \\
\end{longtable}